\documentclass{article}

\usepackage[utf8]{inputenc}
\usepackage[T1]{fontenc}
\usepackage[english]{babel} 
\usepackage{anyfontsize} 
\usepackage{authblk} 

\usepackage[title,titletoc]{appendix}

\usepackage{amsmath}
\usepackage{amssymb}
\usepackage{amsthm}

\theoremstyle{plain}
\newtheorem{theorem}{Theorem}[section]

\theoremstyle{definition}

\newtheorem{assumption}{Assumption}[section]

\usepackage{graphicx}
\usepackage{subcaption} 
\usepackage{booktabs}   
\usepackage{float}      
\usepackage{comment}

\usepackage[margin=2.5cm]{geometry}

\usepackage{hyperref}
\hypersetup{
    colorlinks=true,
    linkcolor=blue,
    filecolor=magenta,      
    urlcolor=cyan,
    citecolor=blue,
}

\usepackage{titling}

\pretitle{\centering\Huge\bfseries} 
\posttitle{\par\vskip 0.5em} 

\title{Emulating spatially extended systems using Fourier Neural Operator}
\author{Gabriele Poccianti$^{1,2}$, Duccio Fanelli$^{1,3}$}
\affil{1. LNCA UMR 7364,  University of Strasbourg, Strasbourg, France, \\
2. Department of Physics and Astronomy, University of Florence, Sesto Fiorentino, Italy \\ 3. INFN, Florence, Italy}

\begin{document}
\maketitle

\begin{abstract}
Operator Learning (OL) algorithms approximate underlying mathematical operators by mapping entire classes of input functions to their corresponding solutions. For instance, architectures such as the Fourier Neural Operator (FNO) have been trained on numerical datasets generated by isolated, stiff ionic models to provide a data-driven alternative to traditional numerical integration schemes. However, real-world biological and physical systems typically consist of interconnected spatial units arranged in complex, heterogeneous networks. Motivated by this observation, we generalize this data-driven framework to spatially extended systems defined on diverse graph topologies, ranging from Erdős–Rényi to Barabási–Albert networks. Furthermore, we evaluate the proposed approach on empirical data, demonstrating its ability to accurately and efficiently model task-oriented neural activity.
\end{abstract}

\section{Introduction}
Scientific Machine Learning (SciML) has emerged as a rapidly developing field that bridges the versatility of data-driven techniques, designed to recognize complex patterns from observations, with the rigor of traditional modeling based on explicit mathematical formulations, such as partial differential equations (PDEs). According to recent categorizations \cite{boulle2024mathematical}, the SciML landscape can be broadly divided into three main areas: (i) Equation Discovery, which focuses on inverse problems to uncover governing equations from data; (ii) Neural ODE/ PDE Solvers, which leverage neural networks to approximate the solution of specific DE instances; (iii) Operator Learning, the most recent development, aimed at learning mappings between infinite-dimensional function spaces. 
The second of the above mentioned areas, related to direct problems, has proven highly effective in accelerating traditional solvers. A significant breakthrough in this domain was the introduction of Physics-Informed Neural Networks (PINNs) and their variants \cite{raissi2019physics}. These architectures privilege a physics-oriented approach, embedding physical laws directly into the optimization process rather than solely relying on data. However, PINNs are typically trained to solve a \textit{single} problem instance. Since a solution often depends on specific input parameters or boundary conditions, modifying these inputs often requires retraining the network, thereby limiting their applicability in many-query scenarios. To address this fundamental limitation, the third branch of recent developments, the so-called Operator Learning approach as listed above, has gained conceptual and applied relevance 
\cite{lu2021learning, li2020fourier}. Unlike traditional neural solvers, Operator Learning algorithms aim to approximate the underlying mathematical operator, by mapping an entire class of input functions to their corresponding solutions. This approach has been successfully applied to complex PDE systems involving parametric dependencies and forcing terms, as well as in various industrial applications \cite{osorio2022forecasting}. Among the emerging approaches, two architectures, based on different universal approximation theorems for operators, stand out: the Deep Operator Network (DeepONet) \cite{lu2021learning} and Neural Operators \cite{kovachki2023neural}. Within the latter class, integral kernel-based methods such as the Fourier Neural Operator (FNO) \cite{li2020fourier} have demonstrated remarkable capabilities in learning highly non-linear operators on regular domains. In this work, we expand upon the results presented in \cite{centofanti2024comconf} and further developed in \cite{centofanti2024cma} and \cite{pellegrini2025mlcse}. In fact, in these prior studies, Operator Learning architectures, with a particular focus on the FNO (detailed in Section \ref{Chp:3}), were employed to learn the dynamics of isolated neuronal systems governed by stiff ionic models, such as the FitzHugh-Nagumo and Hodgkin-Huxley models. To generate the training data, the underlying stiff systems of ordinary differential equations (ODEs) were numerically integrated. Early studies utilized robust solvers (e.g., MATLAB's \texttt{ode15s}) over fixed timespans \cite{centofanti2024comconf, centofanti2024cma}, while subsequent extensions to higher-dimensional systems, such as the 41-variable O'Hara-Rudy model, employed customized Runge-Kutta schemes \cite{pellegrini2025mlcse}. In all cases, the systems were perturbed using time-dependent applied currents, typically square pulses of varying intensities and durations, to capture the threshold-dependent dynamics and complex oscillatory behaviors characteristic of excitable cells. Using datasets ranging from 2,000 to 3,750 simulated trajectories, FNOs were trained to learn the non-linear mapping from the applied input current to the system's state variables (e.g., transmembrane potential, recovery, and gating variables). By leveraging the Fast Fourier Transform (FFT) on uniform discretization grids, the FNO architecture computes the integral kernel operator directly in the frequency domain. This approach proved highly effective in capturing both global dependencies and multiscale temporal behaviors of the stiff ionic models. The optimized FNOs consistently achieved low relative $L^2$ test errors: approximately 0.8\% for the FitzHugh-Nagumo model, 2.0\% for the Hodgkin-Huxley model, and roughly 2.0\% for the highly complex O'Hara-Rudy model \cite{pellegrini2025mlcse}.  Furthermore, these findings highlighted the remarkable scalability of FNOs, maintaining bounded errors and linear scaling in training time despite substantial increases in system dimensionality \cite{pellegrini2025mlcse}.
\newline
While previous work has firmly established the capability of Operator Learning to emulate isolated neural dynamics, real-world biological and physical phenomena are generally governed by spatially extended networks of interacting units. Building on these observations, the primary aim of the present work is to investigate whether Operator Learning architectures can effectively emulate the dynamics of such spatially extended systems. Crucially, we treat these systems entirely as black-box models: the FNO will be trained solely on macroscopic input-output data, without any explicit information regarding the internal constituents, local physical laws, or exact network topology. By evaluating this data-driven methodology across different network topologies, including synthetic graphs such as Erdős-Rényi and Barabási-Albert networks, and further applying it to real-world neural datasets, we aim to demonstrate the expressive power and flexibility of Neural Operators in capturing emergent macroscopic behaviors in complex systems. The paper is organized as follows: in Section \ref{Chp:2} we will introduce the celebrated FitzHugh-Nagumo reaction scheme that defines the reference model for the subsequent analysis. Moving from the isolated formulation of the model we will turn to considering its spatially extended version, by enforcing diffusive coupling among the variables hosted on different locations of the underlying network support. Section \ref{Chp:3} is devoted to revisiting the Operator Learning theory, including the relevant universal approximation theorems. In the following Section \ref{Chp:4} we apply the Operator Learning algorithm to different datasets obtained by integrating the aforementioned FitzHugh-Nagumo reaction scheme deployed on networks of increasing complexity. The adequacy of the proposed algorithm will be challenged under different conditions as by varying the network size. Then in Section \ref{Chp:5} we will turn to considering an application to real data  featuring electrophysiological recordings from a macaque monkey ({\it Macaca mulatta}) performing a delayed center-out reaching task. Finally, we will sum up and conclude.

\section{The FitzHugh-Nagumo model}
\label{Chp:2}
The FitzHugh-Nagumo (FHN) model \cite{centofanti2024comconf, fitzhugh1961impulses, nagumo1962active} describes a prototype of an excitable system, such as a neuron. More specifically, it is a two-dimensional simplification of the more well-known Hodgkin-Huxley (HH) model \cite{hodgkin1952quantitative}, while preserving the essential characteristics of excitability. The dynamical equations for the fast variable $V$ (potential) and the slow variable $w$ (recovery) are:

\begin{equation}
    \begin{aligned}
        \frac{dV}{dt} &= bV(V-\beta)(\delta-V) - cw + I_{app} \\
        \frac{dw}{dt} &= \epsilon(V - \gamma w)
    \end{aligned}
    \label{eq:FHN}
\end{equation}

The parameters used in all the FHN simulations carried out throughout this article are $b = 5$, $c = 1$, $\beta = 0.1$, $\delta = 1$, $\epsilon = 0.1$, $\gamma = 0.25$ \cite{centofanti2024comconf}.

\begin{figure}[H]
    \centering
    \begin{subfigure}[b]{0.58\textwidth}
        \centering
        \includegraphics[width=\textwidth]{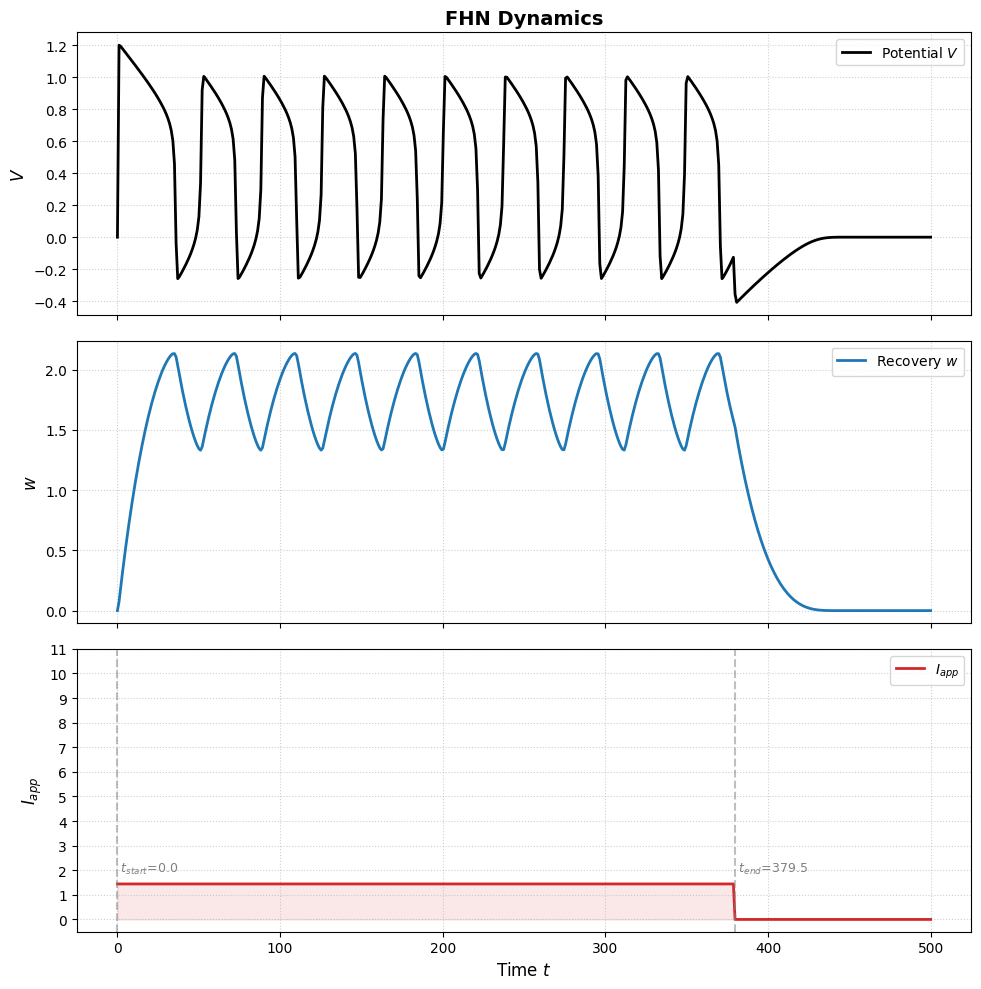}
        \caption{Time evolution of $V$ and $w$ (top) and applied current (bottom).}
        \label{fig:fhn_time_series}
    \end{subfigure}
    \hfill
    \begin{subfigure}[b]{0.38\textwidth}
        \centering
        \includegraphics[width=\textwidth]{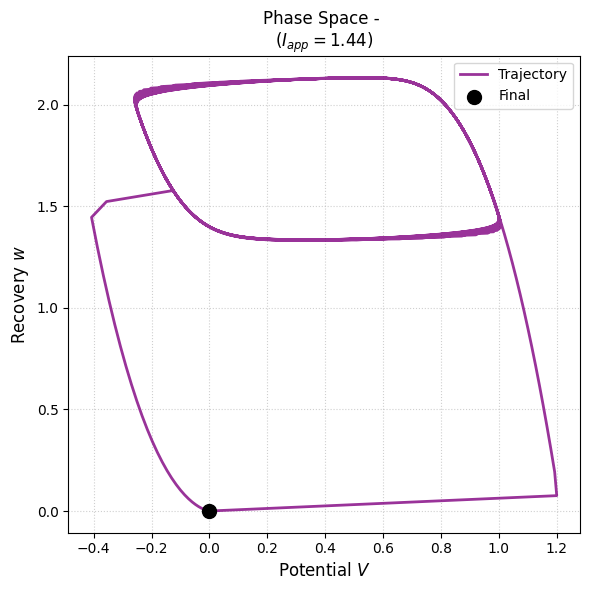}
        \caption{Trajectory in the phase plane $(V, w)$.}
        \label{fig:fhn_phase_space}
    \end{subfigure}
    \caption{Dynamics of the FitzHugh-Nagumo model in response to a square-wave current of amplitude $I_{app} \approx 1.44$. While the current is on, after a transient the system settles onto a stable limit cycle; once the current is switched off, it returns to its only stable equilibrium $(0,0)$.}
    \label{fig:fhn_dynamics}
\end{figure}

In Figure \ref{fig:fhn_dynamics}, and as a representative example, we display a response of the system to a square wave perturbation. In particular, we observe an oscillatory dynamics while the external current is on. Then, when the perturbation is turned off the system recovers the unperturbed state. This is the setting relevant for the subsequent study. Since the operator-learning approach that we shall thoroughly discuss in the following is based on mapping  functions (perturbations) into functions (responses), we have chosen to deal with the limited sub-class of square wave perturbations as it provides an easy-to-study three-parameter family of input functions (onset, duration, amplitude). Starting from these premises, and before focusing on the employed ML scheme, we will turn to discussing the spatial version of the FHN model and analyze its response to a square wave perturbation localized in space (i.e. applied to a single node).

\subsection{From single neuron to spatially extended systems}
\label{sec:extended_network}
\begin{figure}[htbp]
    \centering
    \begin{subfigure}[b]{1.0\textwidth}
        \centering
        \includegraphics[width=\textwidth]{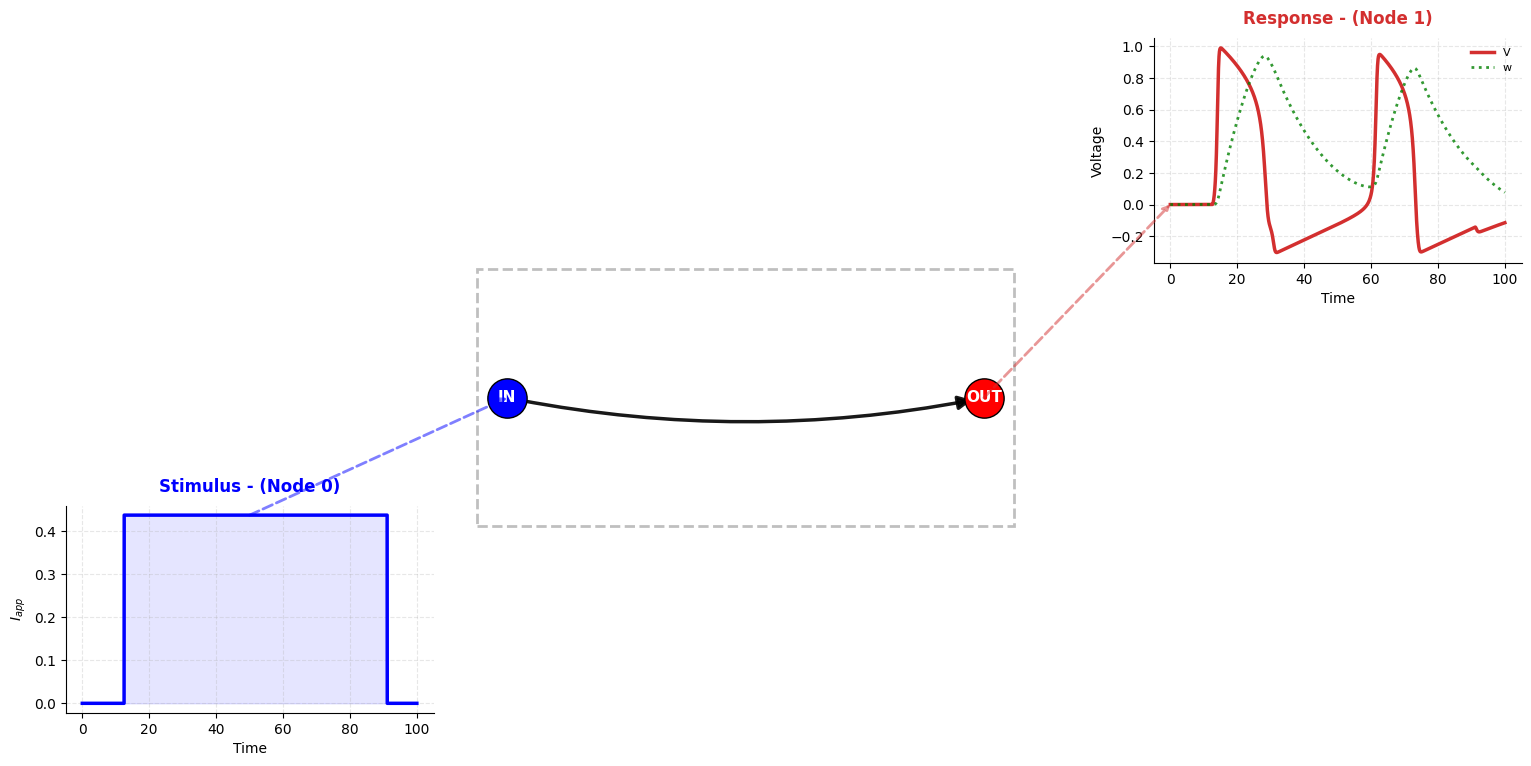}
        \caption{Minimal setup: Input $\rightarrow$ Output}
        \label{fig:fh_output}
    \end{subfigure}
    
    \vspace{0.5cm} 
    
    \begin{subfigure}[b]{1.0\textwidth}
        \centering
        \includegraphics[width=\textwidth]{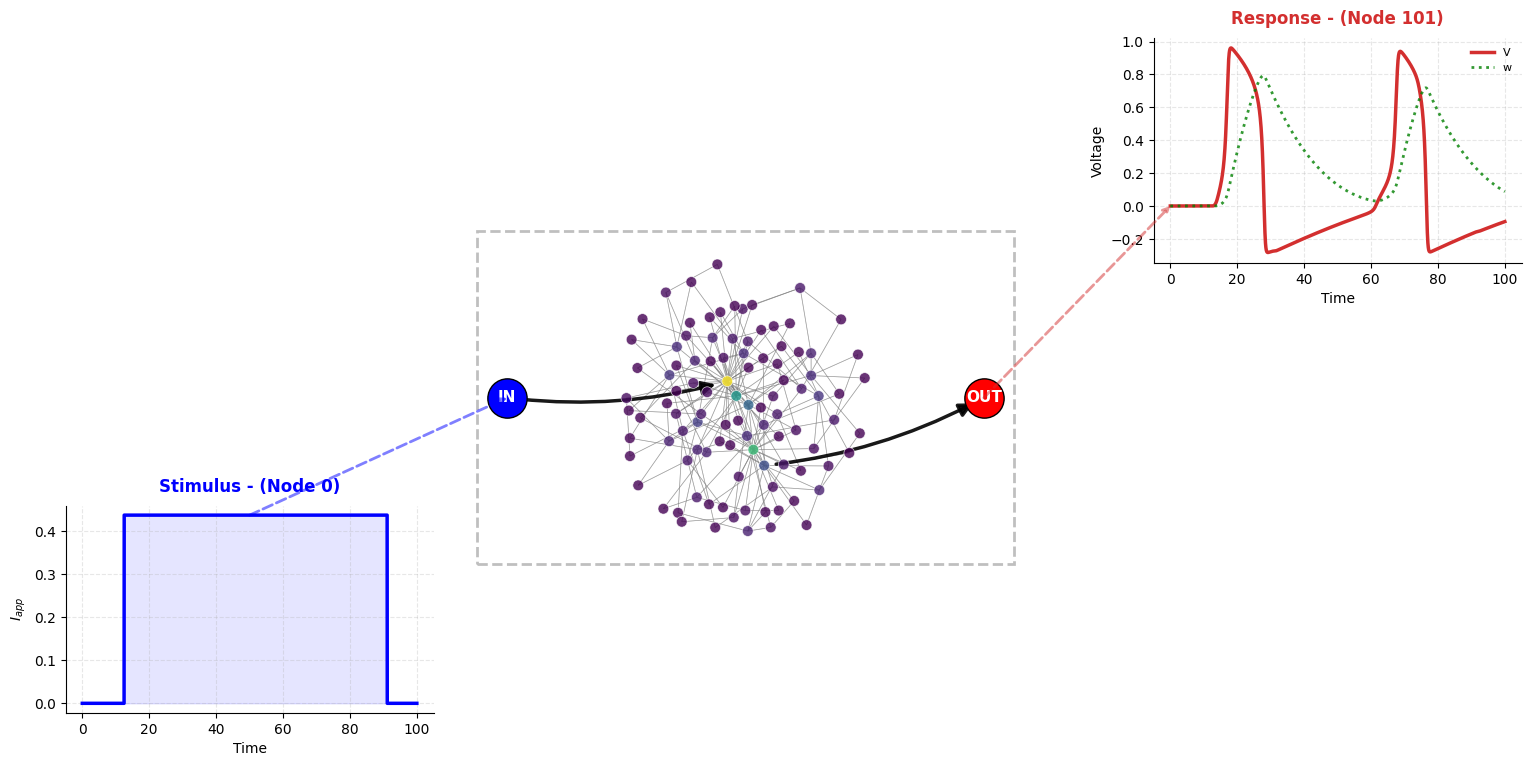}
        \caption{Extended topology: Input $\rightarrow$ Network $\rightarrow$ Output}
        \label{fig:fh_dynamics}
    \end{subfigure}
    
    \caption{Schematic representation of the network setup. (a) A minimal configuration illustrating the direct signal transmission from an input to an output node. (b) A more complex topology utilizing a Barabási-Albert (BA) scale-free network ($N=100$, $m=2$): the input stimulus propagates through the network before reaching the output node.}
    \label{fig:fh_comparison}
\end{figure}

The framework discussed thus far has focused on the dynamics of an isolated single node. In order to transition from a microscopic to a spatially extended (mesoscopic) description of neural dynamics, we embed the FHN model within a network topology. Our primary objective, to which we shall return in the following, is to investigate whether the resulting high-dimensional system of ordinary differential equations (ODEs) can be effectively approximated by a single neural operator trained via deep learning techniques. To this end, we will consider a spatially extended system characterized by a specific input-output mapping, which generalizes beyond the setting considered for the local reaction. More specifically, an external stimulus, modeled as a square-wave current, will be injected into a designated {\textbf{input node}} (pictured in blue in all relevant drawings). The signal then propagates through the network via diffusive coupling, while each node undergoes local reaction dynamics governed by the FHN equations. Finally, the response of the system is recorded by extracting the membrane potential of a designated {\textbf{output node}} (depicted in red). This setup is illustrated by two examples in Fig. \ref{fig:fh_comparison}.

Formally, let $\mathcal{G} = (\mathcal{V}, \mathcal{E})$ be the unweighted graph that hosts the dynamics. It consists of the hidden network, whose $N$ nodes are linked by undirected edges, and of the input and output nodes, which are attached to it through two directed edges: the input node drives the node of the hidden network it is attached to, without receiving any feedback from it, while the output node is driven by the node of the hidden network it is attached to, without acting back on it. Hence $|\mathcal{V}| = N + 2$.

The connectivity is encoded in the $|\mathcal{V}| \times |\mathcal{V}|$ adjacency matrix $\mathbf{A}$, whose element $A_{ij}$ equals one if node $i$ receives a signal from node $j$, and zero otherwise. Within the hidden network $A_{ij} = A_{ji}$, whereas the two directed edges break this symmetry, so that $\mathbf{A}$ is not symmetric. The diffusive coupling is described by the (in-degree) graph Laplacian $\mathbf{L} = \mathbf{D}_{\mathrm{in}} - \mathbf{A}$, where $\mathbf{D}_{\mathrm{in}}$ is the diagonal matrix of the in-degrees $k_i^{\mathrm{in}} = \sum_{l} A_{il}$:
\begin{equation}
L_{ij} =
\begin{cases}
k_i^{\mathrm{in}} & \text{if } i = j \\
-A_{ij} & \text{if } i \neq j
\end{cases}
\end{equation}

The dynamics of the coupled system is therefore governed by the following set of equations for each node $i \in \mathcal{V}$:
\begin{equation}
\label{eq:coupled_FHN}
\begin{aligned}
\frac{dV_i}{dt} &= b V_i (V_i - \beta)(\delta - V_i) - c w_i + I_{app}(t)\,\delta_{i,\mathrm{in}} - \sigma \sum_{j \in \mathcal{V}} L_{ij} V_j \\
\frac{dw_i}{dt} &= \epsilon (V_i - \gamma w_i)
\end{aligned}
\end{equation}
where $\delta_{i,\mathrm{in}}$ is the Kronecker delta (the current is applied to the input node only) and the coefficient $\sigma$ modulates the global coupling strength (i.e., the intensity of diffusion) across the network. 


In Figure \ref{fig:fh_comparison} two examples are reported to illustrate
the spatially extended setting that defines our reference platform for subsequent investigations. In panel \ref{fig:fh_output} the network is solely composed of two interacting nodes. The perturbation is injected in the input (blue) node and extracted on the (red) node coupled to the former. In panel \ref{fig:fh_dynamics} the setting is generalized to a more complex network architecture of the Barabasi-Albert type. 

Having established the mathematical setting for the reference coupled system that we shall aim at exploring, we turn in the next Section to introduce  the principles SciML. Specifically, we will focus on the operator learning techniques that we will consistently employ throughout the second part of the work to emulate the input-output mapping of the spatially extended system under inspection. This amounts to predicting the continuous temporal evolution observed at the output node's location as a direct consequence of a specific perturbation applied on the input node, for different pairs of input/output nodes. 

\clearpage

\section{Operator Learning} 
\label{Chp:3}

This Section is devoted to reviewing the basic elements of the Operator Learning theory. The techniques here discussed will be applied in Section \ref{Chp:4} to emulate the dynamical behavior of the FHN model deployed on a network, as illustrated above.

\subsection{Universal approximation of operators}
A key theoretical foundation of Artificial Neural Networks (ANNs) is the \textit{Universal Approximation Theorem} ~\cite{cybenko1989approximation, hornik1989multilayer}. This theorem states that a feedforward network with at least one hidden layer and a suitable (e.g., sigmoidal) activation function can approximate any continuous \textbf{function} on a compact set to arbitrary accuracy, given a sufficient number of neurons.

However, many physical systems and partial differential equations (PDEs) require mapping an infinite-dimensional space to another infinite-dimensional space (e.g., mapping a parametric initialization or a boundary condition to a complete solution field over time). To address this, the operator learning paradigm extends the concept of universal approximation to operators, a result first established by Chen and Chen~\cite{Chen_Chen_1995}.

Kovachki et al. \cite{kovachki2023neural, kovachki2021universal} provide a comprehensive approximation theory for Neural Operators, proving that an architecture based on alternating integral kernel operators and pointwise non-linearities is a universal approximator for \textbf{operators} between general Banach spaces.

\subsubsection{Discretization invariance and resolution independence}

A defining characteristic of Neural Operators, which distinguishes them from classical deep learning architectures such as Convolutional Neural Networks (CNNs), is the property of \textit{discretization invariance}. In standard architectures, model parameters are intrinsically tied to the specific discretization of the input data (e.g., the number of pixels in an image or the grid size of a simulation). Conversely, Neural Operators are defined as mappings between infinite-dimensional function spaces, meaning the learned parameters are independent of the underlying mesh used during training \cite{kovachki2023neural}.

Here we will not delve into the mathematical definition of such property.

\subsubsection{Zero-Shot Super-Resolution}

The immediate consequence of discretization invariance is the ability to perform \textit{zero-shot super-resolution}. The neural operator is mesh-invariant, so it can be trained on a lower resolution and evaluated at a
higher resolution, without seeing any higher resolution data. 

As demonstrated in the numerical experiments for parametric PDEs in article \cite{li2020fourier}, the relative error of a discretization-invariant model remains consistent across different resolutions. This is formally expressed by the property that, as the discretization becomes finer ($n \to \infty$), the discrete approximation converges to the continuous operator:
\[ 
\lim_{n \to \infty} \hat{\mathcal{G}}^{(n)}_{\theta}(a_n) = \mathcal{G}_{\theta}(a) 
\]
This feature is particularly valuable in neuroscience and complex systems modeling, where neural field dynamics or networked ODEs may be simulated at different spatial/ temporal scales. A Neural Operator trained on a low-resolution brain atlas or a certain temporal discretization can, in principle, generalize its learned dynamics to higher-resolution representations or finer temporal scales without requiring new training data.

\subsection{Fourier Neural Operator (FNO)}
\label{FNO}
The Fourier Neural Operator (FNO), proposed by Li et al. \cite{li2020fourier}, tackles the approximation problem by parameterizing the integral operator directly in the frequency domain. Unlike standard networks, FNO learns a mapping between infinite-dimensional function spaces, guaranteeing discretization independence (\textit{resolution-invariance}).

\begin{figure}[H]
    \centering
    \includegraphics[width=0.9\textwidth]{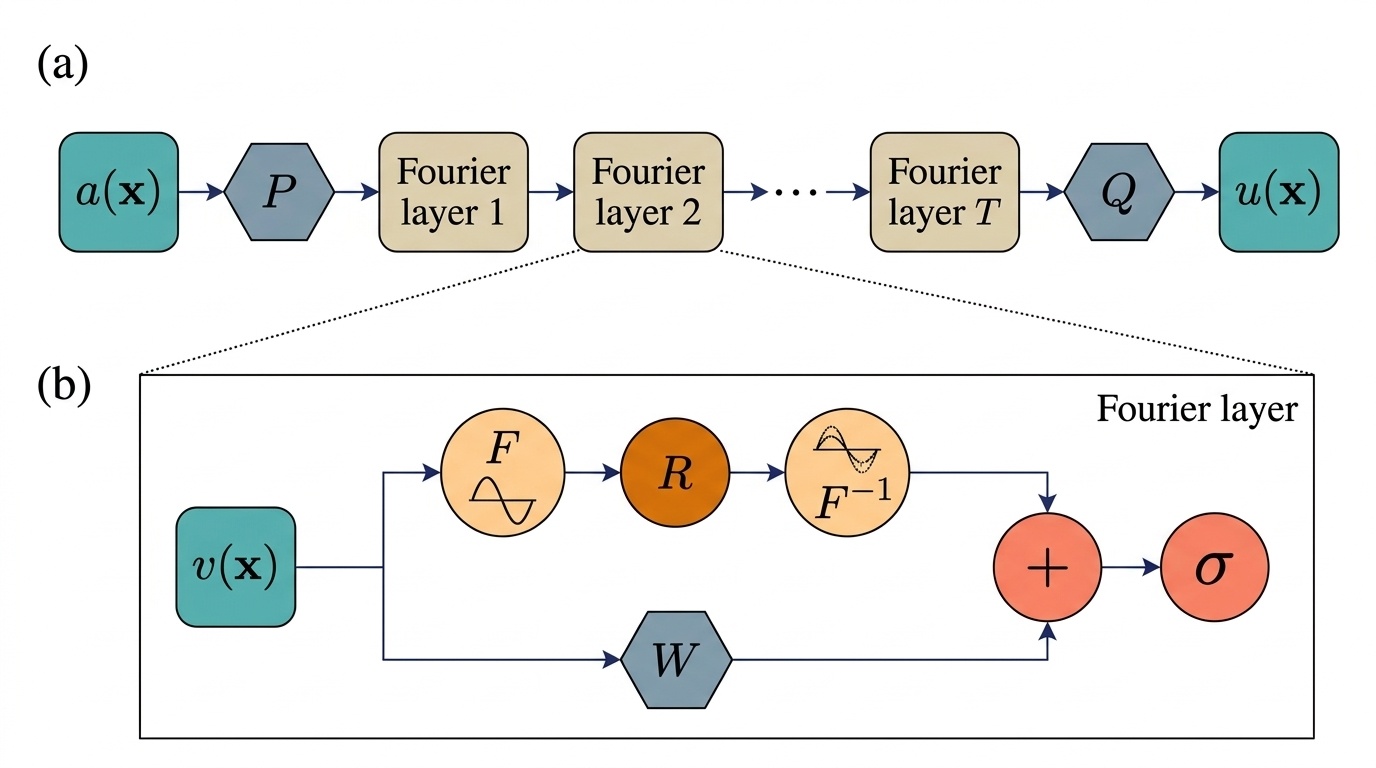}
    \caption{Schematic of the FNO architecture. (a) General structure with lifting ($P$), iterative layers ($T$), and projection ($Q$). (b) Detail of a single Fourier layer. Adapted from \cite{li2020fourier}.}
    \label{fig:fno_schema}
\end{figure}

In the following, $D\subset\mathbb{R}^d$ denotes the bounded domain on which the input and output functions are defined (in our case a time interval, $d=1$).
As shown in Figure \ref{fig:fno_schema}, the architecture follows three main phases:
\begin{enumerate}
    \item \textbf{Lifting:} The input function $a(x)$ is lifted to a higher-dimensional hidden representation $v_0(x)$ via a local transformation $P$ (typically a shallow fully connected network with activation $\sigma_1$): \\
    $v_0(x) = P(a(x)) = \sigma_1(W_{\text{in}} a(x) + b_{\text{in}}), \quad x \in D$
    
    \item \textbf{Iterative Kernel Updates:} For $t = 0, \dots, T-1$ (in Section \ref{Chp:4} we chose $T=3$), the hidden representation is updated using a pointwise linear operator $W_t$, an integral kernel operator with parameterized kernel $\kappa_t(x,y)=\kappa_t(x-y)$\footnote{In our case we applied an MLP after the kernel operation in each layer.}, and the non-linear activation function $\sigma_2$: \\
    $\displaystyle v_{t+1}(x) = \sigma_2 \left( W_t v_t(x) + \int_{D} \kappa_t(x,y) v_t(y) \, dy + b_t \right), \quad x \in D$ \\
    where $v_t : D \to \mathbb{R}^{d_v}$ ($d_v = 32$ in our case), $W_t \in \mathbb{R}^{d_v \times d_v}$, $\kappa_t : D \times D \to \mathbb{R}^{d_v \times d_v}$ and $b_t \in \mathbb{R}^{d_v}$
    
    \item \textbf{Projection:} The final hidden representation $v_T$ is projected back to the target space defined on $D$ using another local transformation $Q$ 
    (with activation $\sigma_3$), for instance: \\
    $\mathcal{G}(a)(x) = Q(v_T(x)) = W_{out,2}\, \sigma_3\big(W_{out,1} v_T(x) + b_{out,1}\big) + b_{out,2}, \quad x \in D$
\end{enumerate}

In our implementation \footnote{For the experiments of Section~\ref{Chp:4} we used the FNO implementation of~\cite{centofanti2024cma} by E.~Centofanti and M.~Ghiotto, available at \url{https://github.com/edoardo100/DON_HH}.} $\sigma_1 = \mathrm{Id}$ (linear lifting) and $\sigma_2 = \sigma_3 = \mathrm{ReLU}$.
\newline
In the specific case of the Fourier Neural Operator (FNO) \cite{li2020fourier}, the integral kernel is parameterized directly in the frequency domain \footnote{Other choices for the kernel are valid and lead to different architectures.}. By leveraging the convolution theorem, the continuous integral operation is efficiently resolved by computing the Fourier transform of the input, performing a pointwise multiplication with a learnable complex-valued weight tensor over a truncated set of low-frequency modes (we kept $16$ modes), and finally applying the inverse Fourier transform. Mathematically, the integral term in the iterative update is replaced by:
\[ 
\int_{D} \kappa_t(x,y) v_t(y) \, dy = \mathcal{F}^{-1} \left( R_t \cdot \mathcal{F}(v_t) \right)(x) 
\]
where $\mathcal{F}$ and $\mathcal{F}^{-1}$ denote the forward and inverse Fourier transforms, respectively, and $R_t$ is the learnable complex tensor acting as the Fourier kernel at layer $t$. The truncation of higher frequencies acts as a low-pass filter, ensuring both discretization invariance and high computational efficiency. A schematic layout of the FNO architecture is provided in Figure \ref{fig:fno_schema}.

\clearpage
\section{Data-driven FNO approximation of spatially extended FHN systems on different network topologies}
\label{Chp:4}

We now turn to evaluate the performance of the FNO data-driven approximation scheme on the class of FHN models hosted on graphs, as introduced in Section \ref{Chp:2}. To this end, and for the different network topologies considered, we will generate a large dataset of output responses (on the node of extraction), as produced upon injection (at the input node) of square-wave perturbations with different structural characteristics (onset, time duration and amplitude). Recall that the trained FNO has no access to detailed information about the underlying network's architecture. As a further technical notation, and unless otherwise specified, the input node (depicted in blue) is connected via a directed link to a random node of the (undirected and unweighted) network where the dynamics takes place. Further, a node randomly selected among the $N$ nodes that compose the underlying network is attached via a directed link to the designated output node (colored in red), where extraction is performed. For details about the numerical schemes adopted for data generation, refer to Appendix \ref{app:B}.

\subsection{First minimal test: input and output nodes are short-circuited}

As a preliminary test of the explored numerical setting, we consider just two nodes (with no actual network in between), connected by a directed edge from the input to the output node, see Figure \ref{fig:fh_output}.

To evaluate the performance of the Fourier Neural Operator applied to this minimal (though spatially extended) configuration, we analyzed both the global error metrics (MSE\footnote{Notice that even if the authors of \cite{centofanti2024comconf} used relative $L^2$ errors, we preferred to use an absolute metric to avoid exploding gradients when the solution remains too close to the equilibrium point at $0$.}) and (qualitatively) the network's ability to accurately capture the spiking behavior of the potential, typically activated when the input current exceeds a certain threshold. Figure \ref{fig:1to2_peaks} illustrates some instances of the test set. The trained FNO correctly predicts the action potential peaks for the vast majority of the test samples, which is a crucial requirement for excitable models.

Furthermore, Figure \ref{fig:1to2_l2histo} reports the distribution of the MSE evaluated across the test set. The histogram confirms that the network maintains a highly concentrated and low error profile, indicating that the operator learning approach remains robust and reliable in this first spatial extension step.

\subsection{Extending to a toy network setting}
\label{subsec: Toy_Networks}
\begin{figure}[p]
    \centering

    \begin{subfigure}[b]{0.64\textwidth}
        \centering
        \includegraphics[height=5.4cm,width=\linewidth,keepaspectratio]{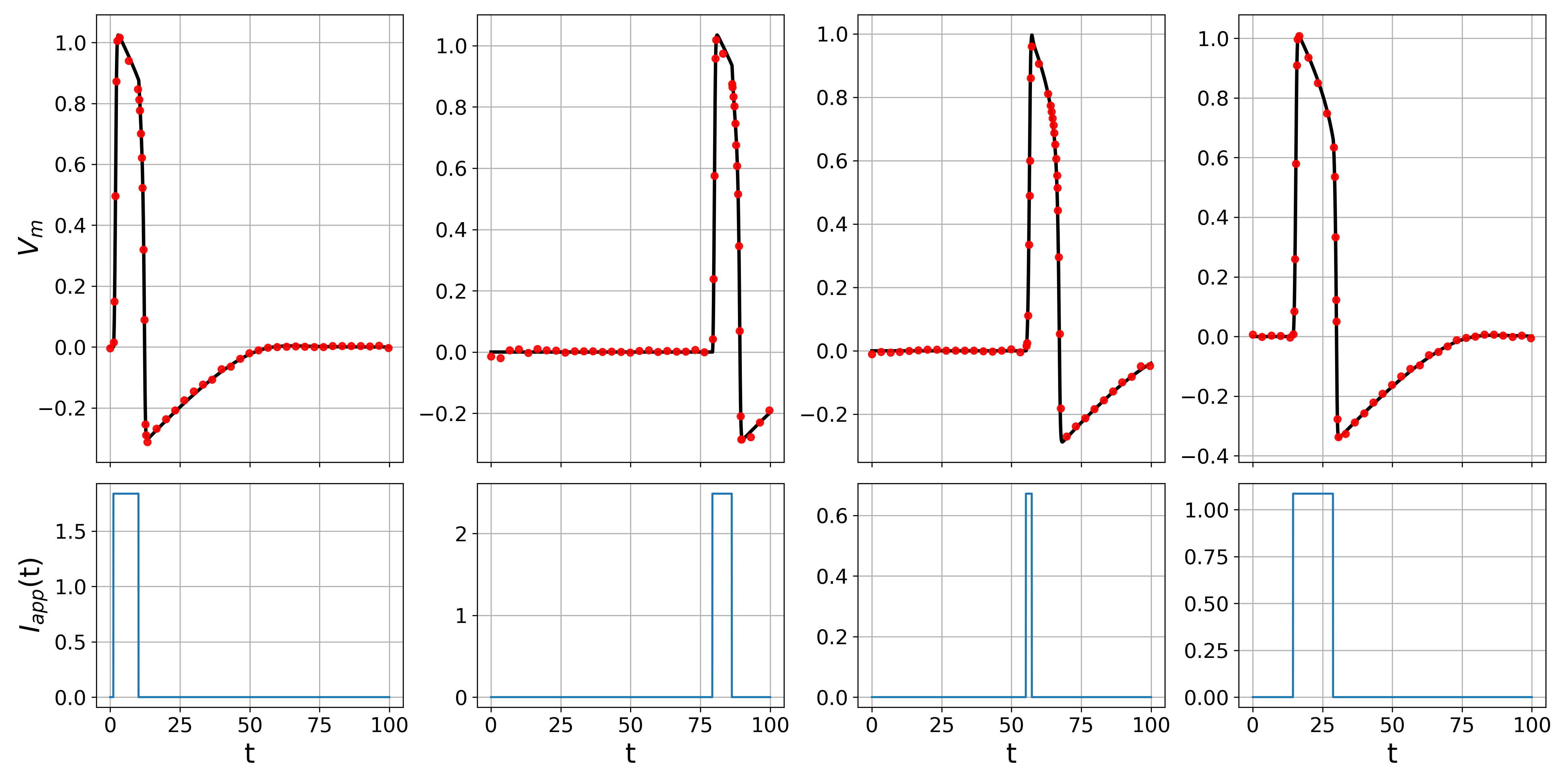}
        \caption{1to2: test set instances. }
        \label{fig:1to2_peaks}
    \end{subfigure}
    \hfill
    \begin{subfigure}[b]{0.34\textwidth}
        \centering
        \includegraphics[height=5.4cm,width=\linewidth,keepaspectratio]{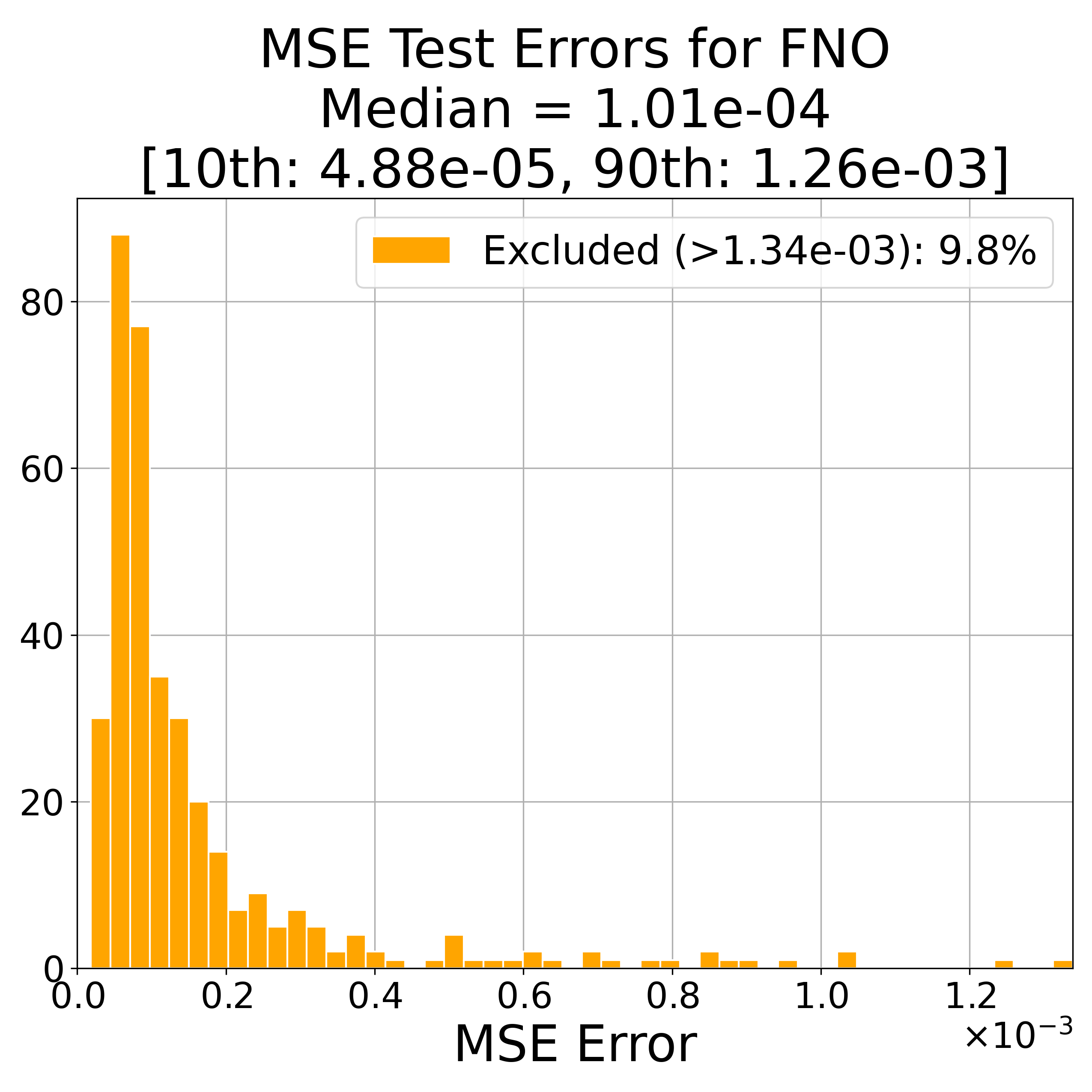}
        \caption{1to2: MSE distribution.}
        \label{fig:1to2_l2histo}
    \end{subfigure}

    \vspace{0.4cm}

    \begin{subfigure}[b]{\textwidth}
        \centering
        \includegraphics[height=4.5cm,width=\linewidth,keepaspectratio]{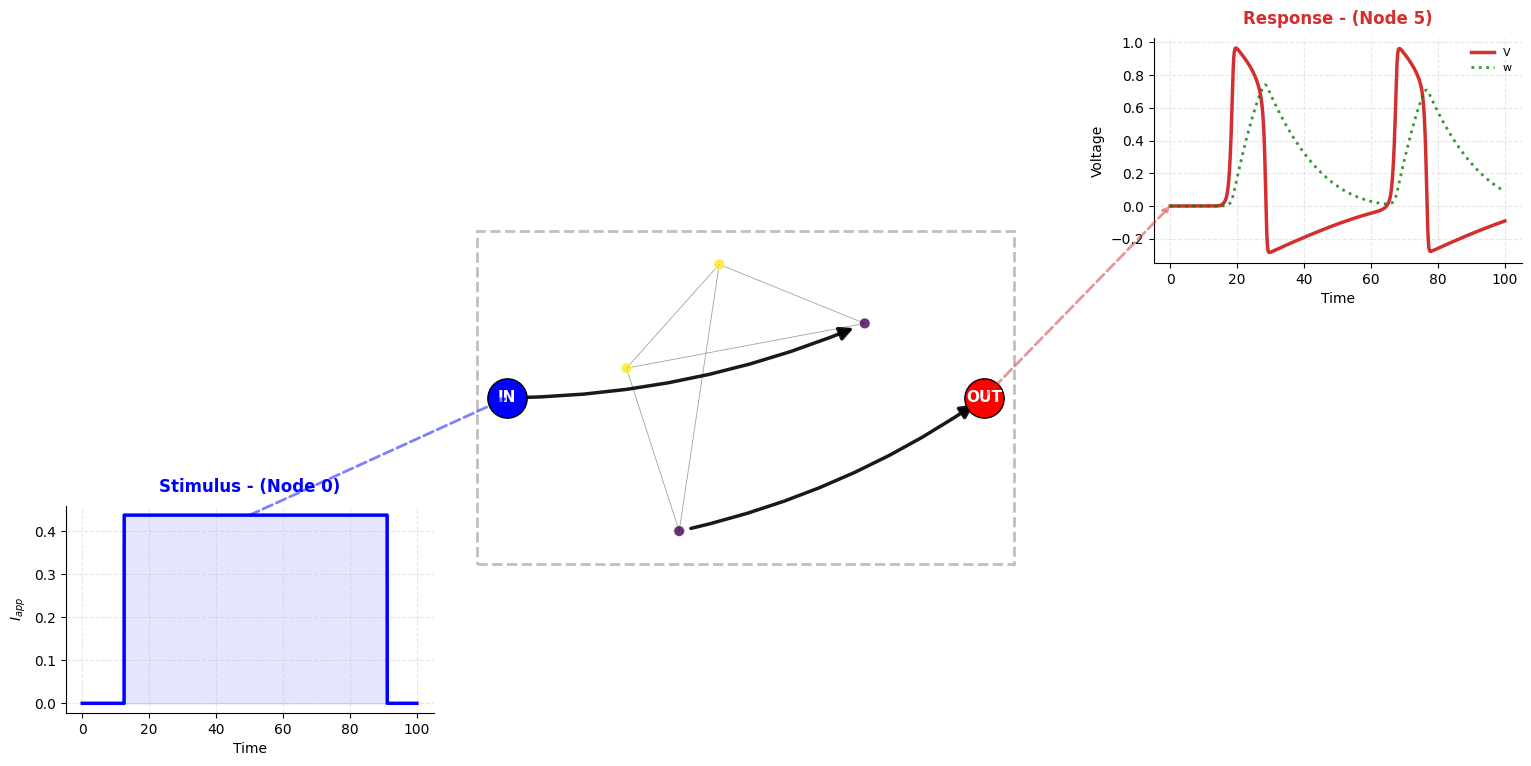}
        \caption{Simple cycle: network schematic setup.}
        \label{fig:cycle_setup}
    \end{subfigure}

    \vspace{0.4cm}

    \begin{subfigure}[b]{0.64\textwidth}
        \centering
        \includegraphics[height=5.4cm,width=\linewidth,keepaspectratio]{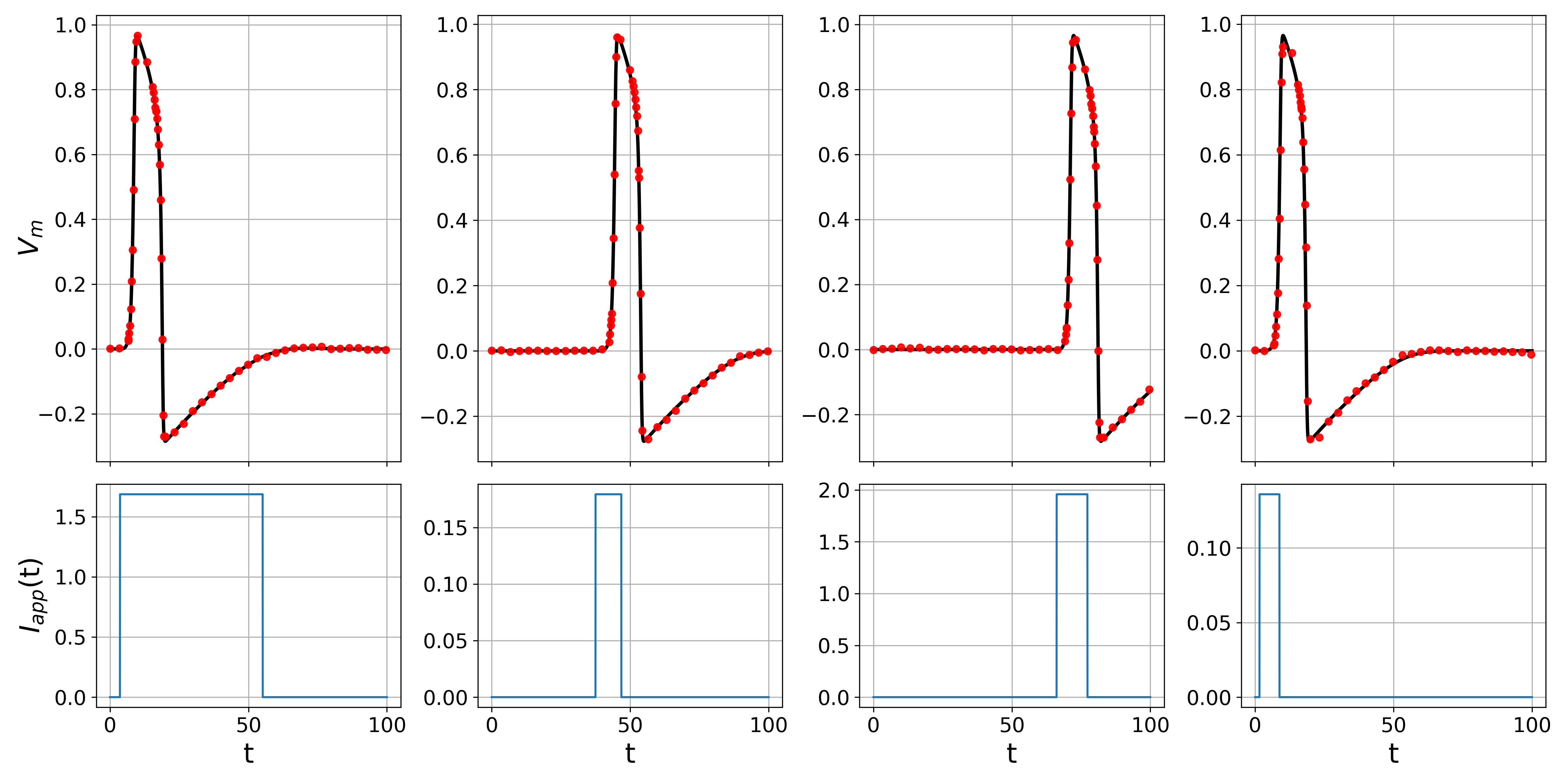}
        \caption{Simple cycle: test set instances.}
        \label{fig:cycle_peaks}
    \end{subfigure}
    \hfill
    \begin{subfigure}[b]{0.34\textwidth}
        \centering
        \includegraphics[height=5.4cm,width=\linewidth,keepaspectratio]{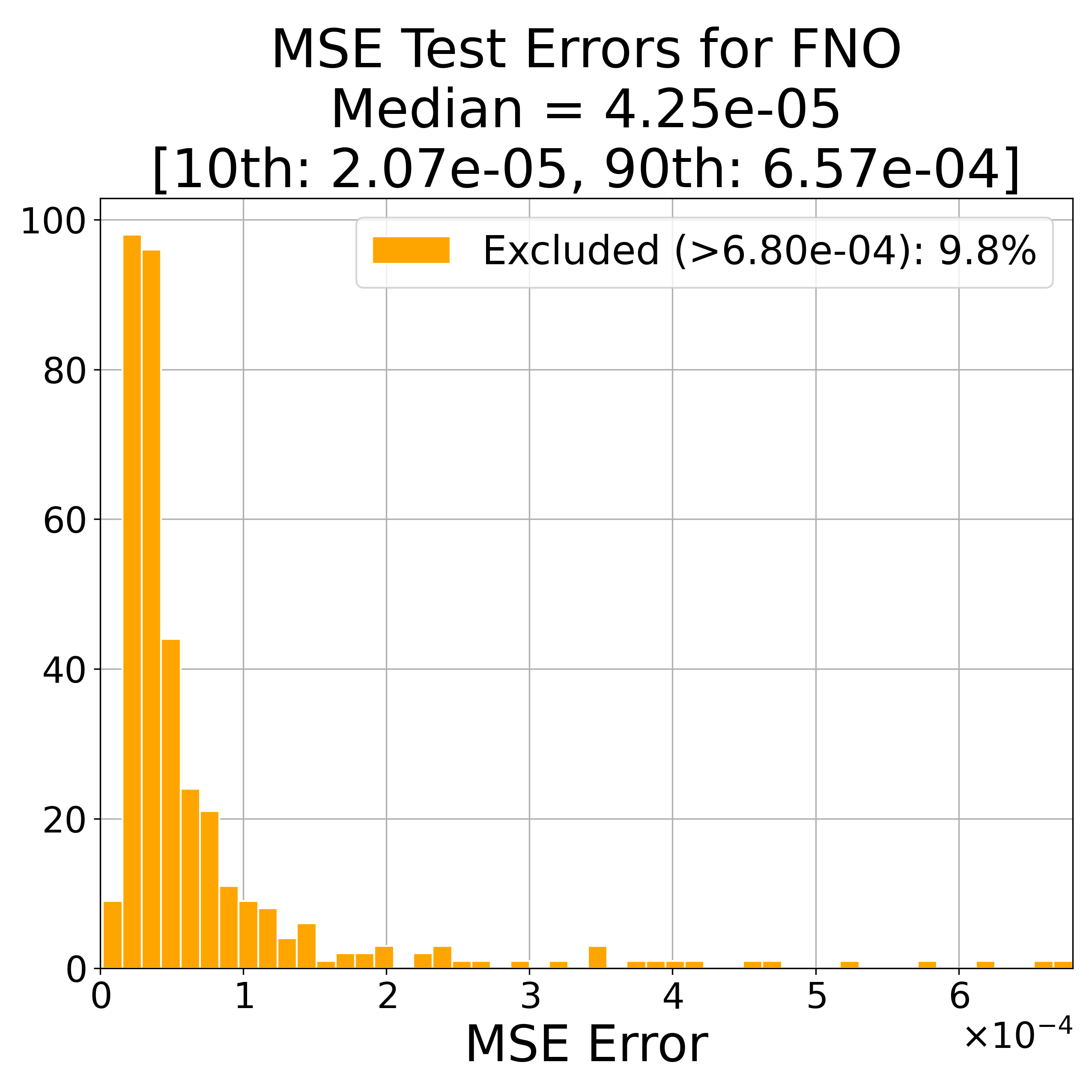}
        \caption{Simple cycle: MSE distribution.}
        \label{fig:cycle_l2histo}
    \end{subfigure}

    \caption{Results for the minimal 1to2 configuration (top row) and for the simple (cyclical) hidden network (middle and bottom rows). \textbf{(a)}, \textbf{(d)} Four instances of the test set for the 1to2 and the simple cycle configuration, respectively. For each column we can see the applied input current (bottom) with its corresponding potential response (top): the continuous lines represent the result of the simulations, whereas the red dots indicate the output of the NN. \textbf{(b)}, \textbf{(e)} Histogram of the MSEs evaluated on the test set. As descriptors, we reported the median, 10th and 90th percentile. \textbf{(c)} Schematic representation of the network setup in the case of a simple (cyclical) hidden network.}
    \label{fig:small_networks_results}
\end{figure}

Following the same approach, we evaluated the FNO performance on a more complex spatial configuration, introducing a simple cycle topology (see Figure \ref{fig:cycle_setup}). 

Consistent with the findings of the previous Section, the network is found capable of capturing the underlying threshold dynamics. Figure \ref{fig:cycle_peaks} illustrates some instances of the test set for this new configuration. As we can see, the predicted dynamics closely follows the numerical simulations, successfully reproducing the spiking behavior and the delay introduced by the diffusion over the cycle.

The MSE distribution across the test set (Figure \ref{fig:cycle_l2histo}) remains concentrated at low values, indicating that the operator learning approach is robust and its accuracy does not degrade when dealing with diffusion over a small ($4$ nodes) network.


\newpage
\subsection{The case of an Erd\H{o}s-R\'enyi (ER) network}
\label{Sec:ER}

We now move on to account for larger computing networks. In particular, we want to check if the performance of FNO degrades with the growing size ($N$) of the hidden network.
To conduct this experiment we performed our analysis for different sizes of the ER network $N= 10,100,200,300,400,500,600,700,800,900,1000,2000$ (with connection probability $p = \frac{3}{N}$ \footnote{We verified that for our instances a path Input $\rightarrow$ Output always existed, even though the hidden network is not always connected.}).
Since the connection nodes for the input and output nodes are randomly chosen we repeated each experiment with three different seeds  
to distinguish if some effects are only due to chance of the particular topology/ sample of input/ output. For considerations about the choice of the  constant $\sigma$ as the size $N$ of the network is being varied, we refer to Appendix \ref{appA}.

\begin{figure}[htbp]
    \centering
    \includegraphics[width=0.8\textwidth]{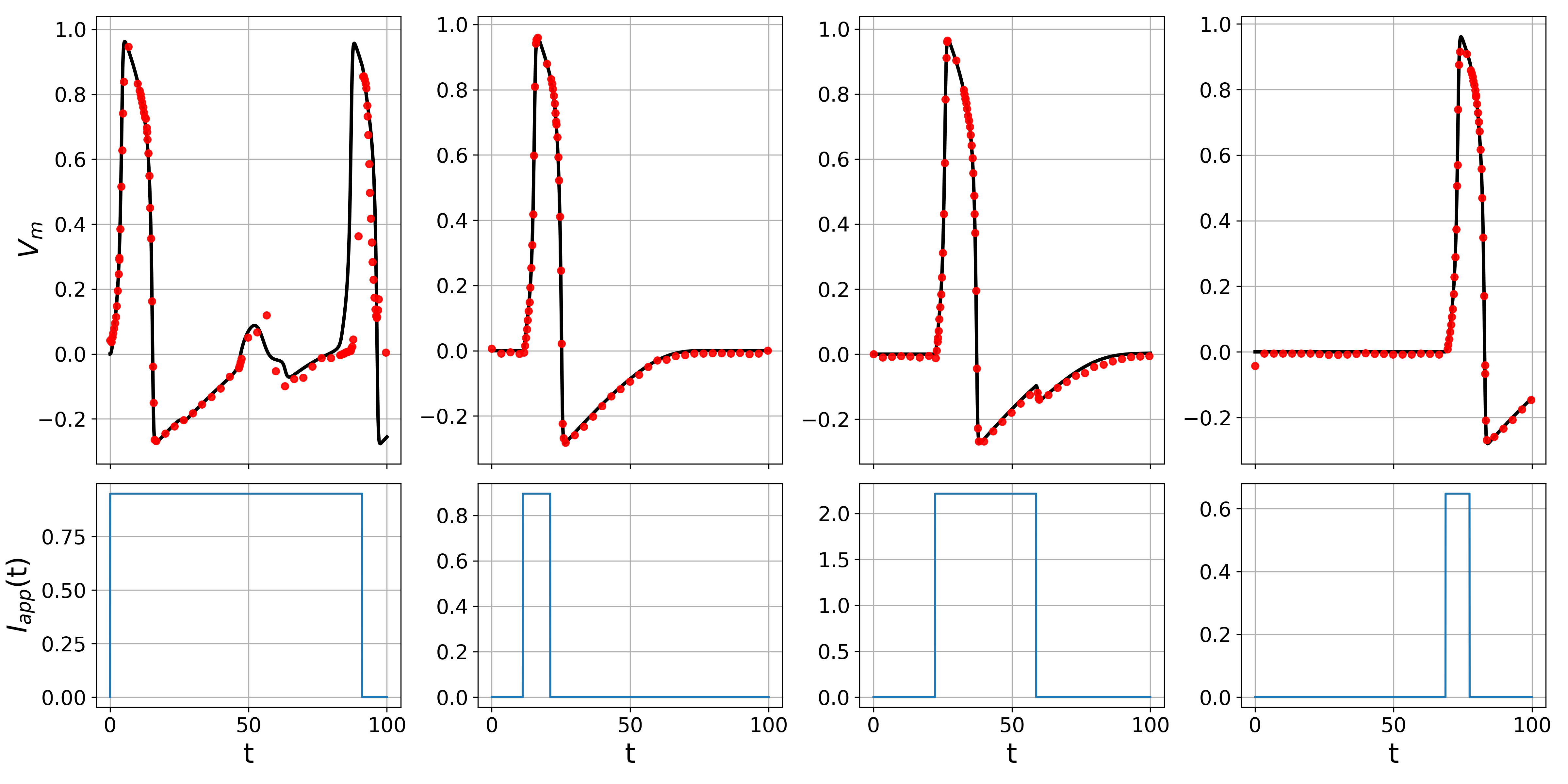}
    \caption{Some instances of the test set for a hidden ER network of size $N=500$. We can see the applied input current (bottom) with its corresponding potential response (top): the continuous line represents the result of the simulations, whereas the red dots indicate the output of the NN.}
    \label{fig:er_peaks_N500}
\end{figure}

\begin{figure}[htbp]
    \centering
    \includegraphics[width=0.8\textwidth]{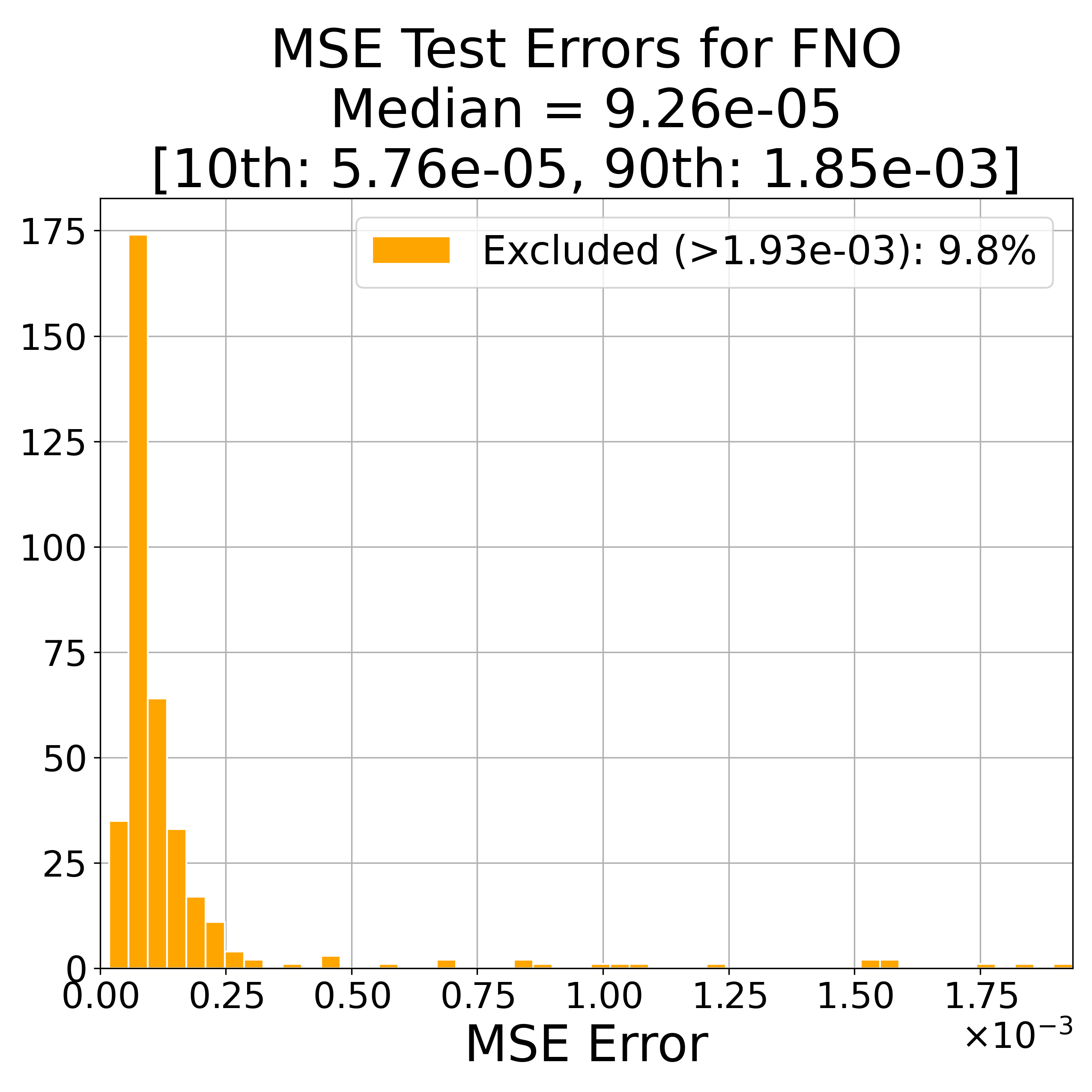}
    \caption{MSE distributions across the test set for a hidden ER network of size $N=500$.}
    \label{fig:er_histo_N500}
\end{figure}

Figure~\ref{fig:er_peaks_N500} shows some instances of the test set for a representative network of size $N = 500$. As in the smaller configurations of Section~\ref{subsec: Toy_Networks}, the FNO reproduces the timing and the shape of the action potentials recorded at the output node. The corresponding distribution of the MSE over the test set (Figure~\ref{fig:er_histo_N500}) is concentrated at low values.
\newline
The same behavior is observed for all the network sizes considered. Test instances, MSE distributions and training/test loss curves for each value of $N$ are reported in Appendix~\ref{app:C} (Figures~\ref{fig:er_peaks_grid}, \ref{fig:er_histo_grid} and~\ref{fig:er_loss_grid}, respectively): neither the accuracy of the predictions nor the convergence of the training appears to deteriorate as $N$ increases. A quantitative comparison across sizes is given in Section~\ref{sec: performances_over_N}.

\clearpage
\subsection{The case of a Barabasi-Albert (BA) network}
\label{Sec:BA}

Before detailing the results, it is worth briefly recalling that a Barab\'asi-Albert (BA) network is a widely used model for generating random scale-free networks. It relies on a preferential attachment mechanism, meaning that new nodes are more likely to establish connections with existing nodes that already have a high degree.
Due to the particular structure of this family of networks, we defined the hubs as the nodes whose degree exceeds the 95th percentile (representing the top 5\% most connected nodes in the graph). Consequently, we opted to randomly connect the input and output nodes exclusively to one of these hubs.
In this way we reduced the variability of the dynamics due to the seed.

Following the same rationale as the previous section, we proceeded to observe if the performance of FNO degraded with the growing size ($N$) of the hidden network. To conduct this experiment we performed our analysis for different sizes of the BA network: $N= 10,100,200,300,400,500,600,700,800,900,1000,2000$. 
Since the hubs to which the input and output nodes are attached are randomly chosen, we repeated each experiment with three different seeds 
to distinguish whether specific effects are intrinsic to the topology or merely due to the chance of the particular sampling of input/output nodes.

\begin{figure}[htbp]
    \centering
    \includegraphics[width=0.8\textwidth]{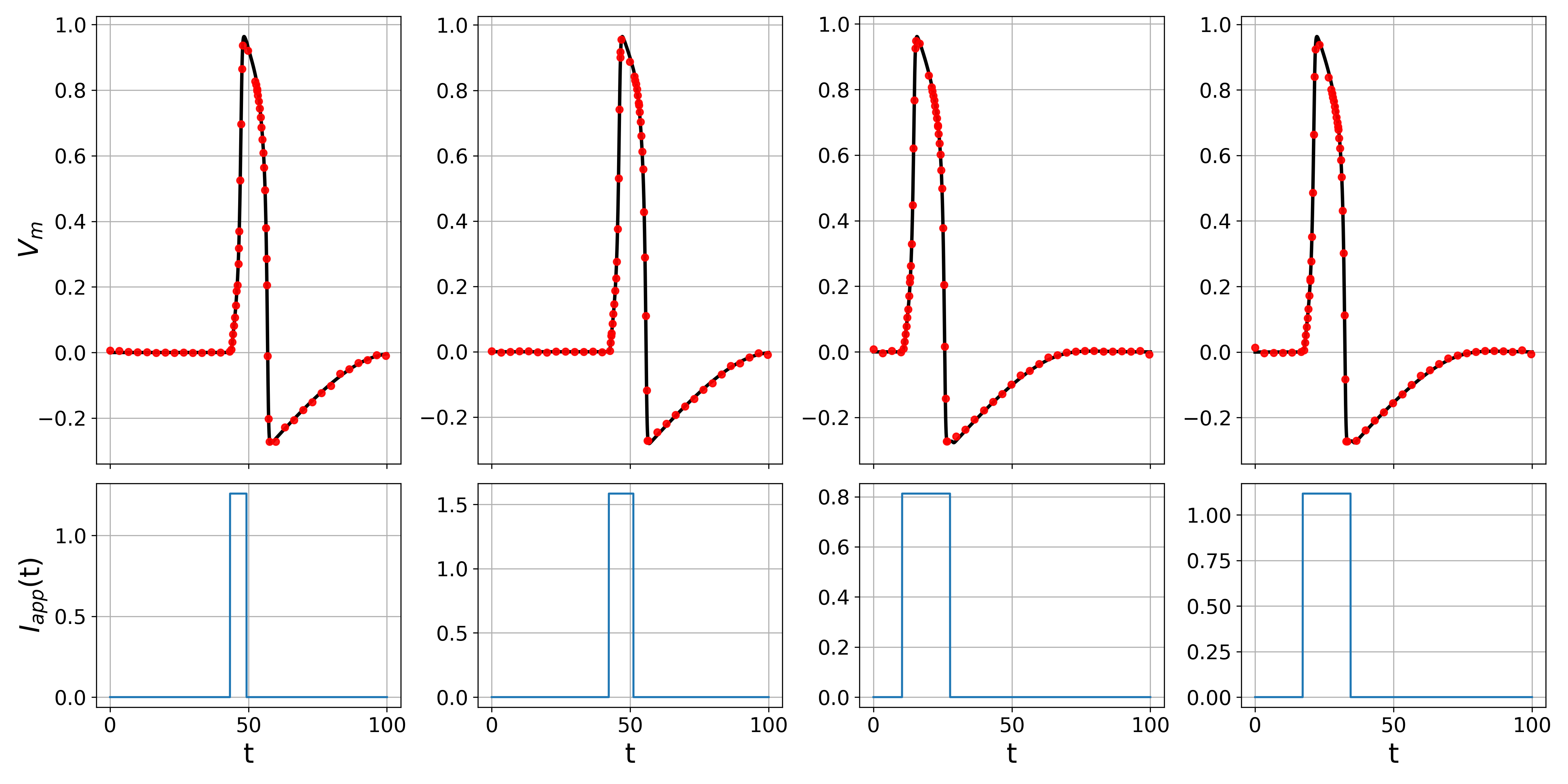}
    \caption{Some instances of the test set for a hidden BA network of size $N=500$. We can see the applied input current (bottom) with its corresponding potential response (top): the continuous line represents the result of the simulations, whereas the red dots indicate the output of the NN.}
    \label{fig:ba_peaks_N500}
\end{figure}

\begin{figure}[htbp]
    \centering
    \includegraphics[width=0.8\textwidth]{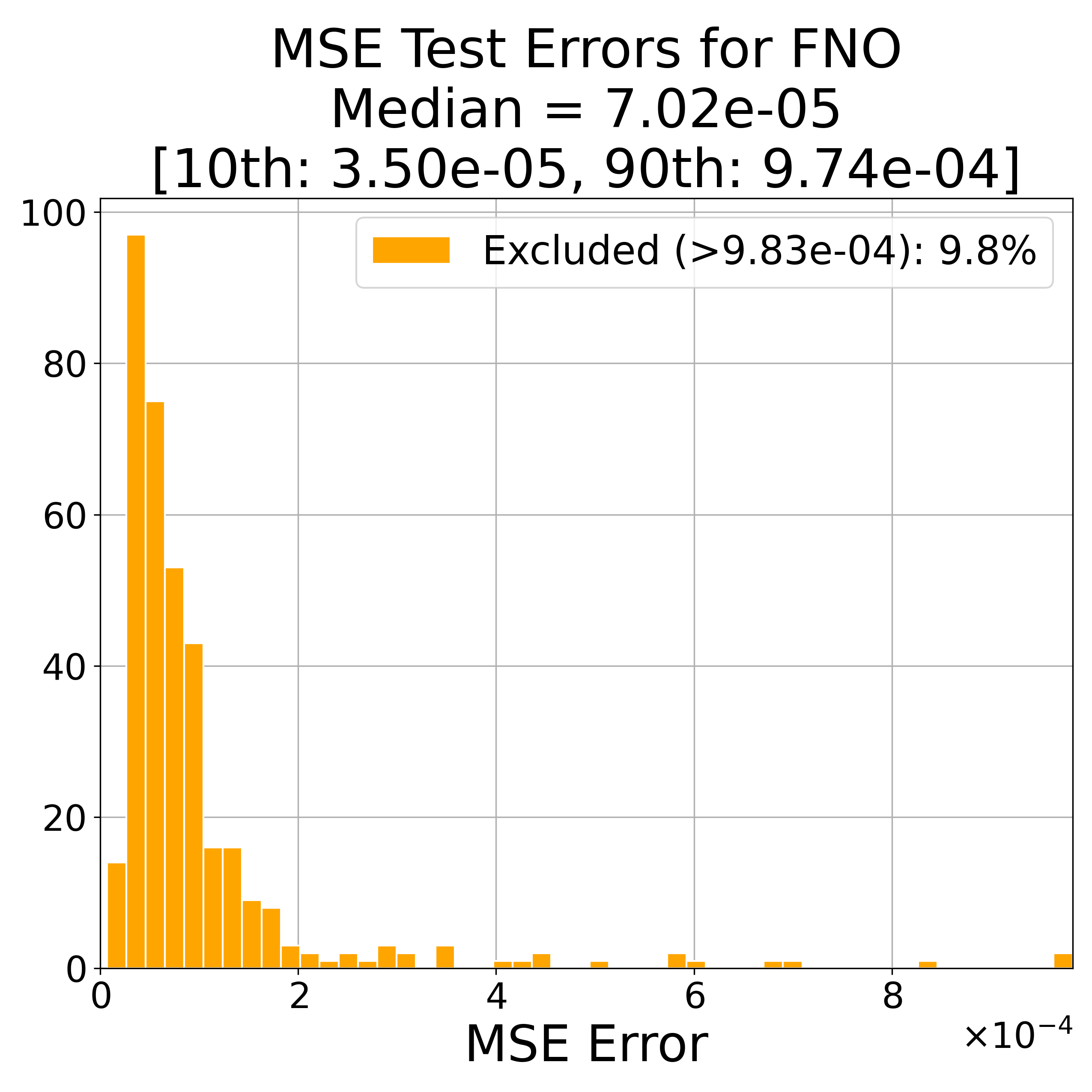}
    \caption{MSE distributions across the test set for a hidden BA network of size $N=500$.}
    \label{fig:ba_histo_N500}
\end{figure}

Figure~\ref{fig:ba_peaks_N500} shows some instances of the test set for a representative network of size $N = 500$. As for the ER network (Section~\ref{Sec:ER}), the FNO reproduces the timing and the shape of the action potentials recorded at the output node, and the corresponding distribution of the MSE over the test set (Figure~\ref{fig:ba_histo_N500}) is concentrated at low values.

The results for all the network sizes considered are reported in Appendix~\ref{app:C}: test instances, MSE distributions and training/test loss curves are shown in Figures~\ref{fig:ba_peaks_grid}, \ref{fig:ba_histo_grid} and~\ref{fig:ba_loss_grid}, respectively. As in the ER case, the performance of the FNO does not seem to degrade with the growing network size $N$, and the spiking response of the system to the external stimuli is generally well captured. A quantitative comparison across sizes is given in Section~\ref{sec: performances_over_N}.

\clearpage
\subsection{Performance across different sizes of the hidden network: general conclusion}
\label{sec: performances_over_N}

\begin{figure}[htbp]
    \centering
    \includegraphics[width=1.0\textwidth]{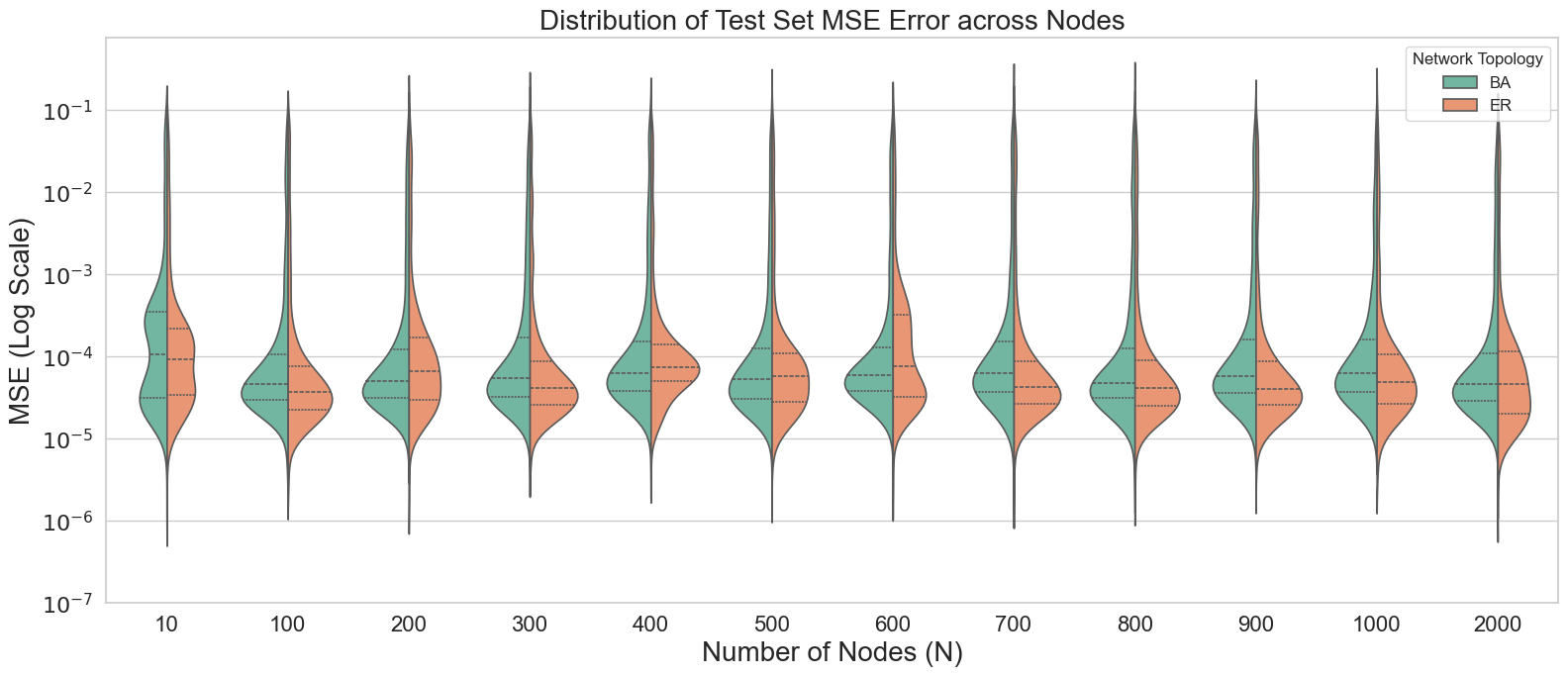}
    \caption{Violin plots of the FNO error distribution across different network sizes ($N$). The dotted lines represent (respectively, from bottom to top) the first quartile, median and third quartile of each distribution. The plot compares the error scaling for both Erd\H{o}s-R\'enyi (ER) and Barab\'asi-Albert (BA) topologies, demonstrating the robustness and scalability of the architecture.}
    \label{fig:fno_scaling_error}
\end{figure}

As we could see in the previous graphs (at least for the sizes considered) the FNO generalizes well to the deterministic dynamics. 
These results are strengthened and summarized in the violin plots reported in Figure \ref{fig:fno_scaling_error}, for the distribution of $\log(\mathrm{MSE})$, as network size and type vary. To gain statistical relevance we aggregated data for the three different seeds employed.
\newline
As the network size $N$ increases, we do not observe a significant error distribution spread toward higher values.
To quantify this observation, we computed the Spearman rank correlation coefficient $\rho$ between $N$ and two descriptors of the MSE distribution, namely the median and the third quartile (Q3), using one value per network size (12 sizes, with data aggregated over the three seeds). No significant monotonic trend was found for either topology: for the BA networks, $\rho = -0.09$ ($p = 0.78$) for the median and $\rho = -0.05$ ($p = 0.88$) for Q3; for the ER networks, $\rho = -0.24$ ($p = 0.46$) and $\rho = -0.11$ ($p = 0.73$), respectively.
\newline
A future extension could be to investigate the robustness of the proposed scheme to external noise injection.

\newpage
\section{Testing the spatially extended FNO approximation scheme against real data}
\label{Chp:5}

After validating the Operator Learning approach on synthetic data generated by deterministic models and on complex networks, we now extend the analysis to real data, more specifically electrophysiological data.

The transition from the simulated to the biological domain introduces significant challenges. In synthetic models, the cause-and-effect relationship is well-defined: a known input current $I_{app}(t)$ produces a deterministic (or stochastic with controlled noise) voltage response. 

In real experiments, however, the observed neuronal response is the result of the integration of multiple signals: not only the sensory stimulus controlled by the experimenter, but also feedback signals, the internal state of the animal, attention, preparatory motor activity, and intrinsic background noise. 

Consequently, the fundamental assumption of Operator Learning---namely, the existence of an operator $G: u \to v$ that uniquely maps a single input to the observed response---is a simplification. In this work, we will nevertheless assume this framework to be valid, treating the unobserved variables as stochastic noise that the neural network must filter out in order to learn the average dynamics induced by the stimulus.

\subsection{The dataset}
The data analyzed in this section is sourced from the Neural Latents Benchmark (NLB), a standardized suite of datasets and an associated public challenge originally established to evaluate machine learning models on neural population activity \cite{neurallatents, PeiYe2021NeuralLatents}. The primary objective of the NLB challenge was to infer underlying low-dimensional neural dynamics to accurately reconstruct or forecast held-out discrete spike trains across various brain regions and experimental paradigms \cite{neurallatents}.

\begin{figure}[htbp]
    \centering
    \includegraphics[width=0.8\textwidth]{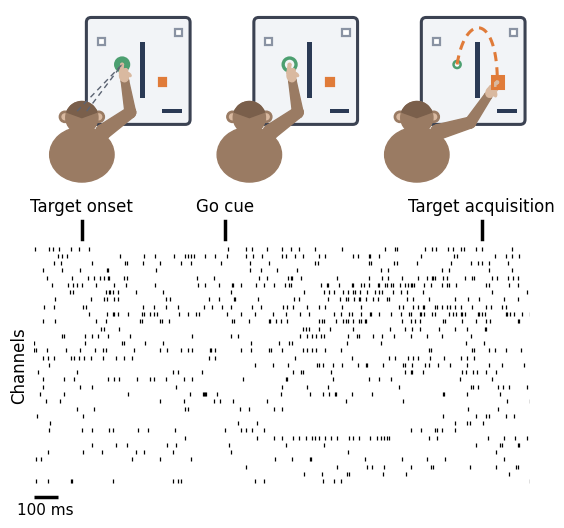}
    \caption{Overview of the \texttt{mc\_maze} (adapted from \cite{PeiYe2021NeuralLatents}) dataset and the behavioral decoding task. The upper panel illustrates the delayed center-out reaching task, displaying a 2D spatial trajectory of the macaque's hand navigating around obstacles to reach specific targets. The lower panels display the corresponding neural population activity (spike raster plots) recorded from the M1 and PMd cortices.}
    \label{fig:mc_maze_dataset}
\end{figure}

For our specific analysis, we selected the \texttt{mc\_maze} dataset \cite{churchland2010cortical, neurallatents}. This dataset features electrophysiological recordings from a macaque monkey (\textit{Macaca mulatta}) performing a delayed center-out reaching task \cite{churchland2010cortical}. During the experiment, the animal was trained to move its hand to specific target locations while navigating around virtual obstacles (forming a "maze", see Figure \ref{fig:mc_maze_dataset}). The neural activity was recorded using multielectrode Utah arrays implanted in the primary motor cortex (M1) and the dorsal premotor cortex (PMd)---areas heavily involved in the planning and execution of voluntary movements. Alongside the neural recordings, the dataset provides rich continuous behavioral metrics, most notably the continuous hand kinematics (e.g., position and velocity over time).

It is important to note how our methodology positions itself relative to the original scope of the NLB challenge. First, our analysis is strictly confined to the subset of data designated as the official "train" split, without utilizing the test split reserved for the benchmark's competitive leaderboard. Second, while the primary focus of the NLB is often the point-process estimation of discrete spike trains, the benchmark also includes a behavioral decoding phase. In the official challenge, decoding continuous hand kinematics from neural activity is performed using a simple linear baseline, namely Ridge Regression.

In our work, we replaced this standard linear decoder with our Operator Learning framework, which naturally operates on continuous function spaces. Specifically, we defined the input function $u(t)$ as the smoothed, continuous neural firing rates, and the target output function $v(t)$ as the continuous hand velocity of the macaque. By training the neural network to learn this operator (with FNO) $G: u \to v$, we inferred the continuous velocity profile directly from the underlying neural dynamics. Finally, the predicted velocity was numerically integrated over time to reconstruct the full 2D reaching trajectory (position) of the hand. This setup allowed us to rigorously test the neural operator's ability to capture the complex, nonlinear mapping between cortical activity and physical behavior, moving beyond the limitations of simple linear regression.

\clearpage
\subsection{Pre-processing}
As part of our pre-processing pipeline, the time series data was rigorously aligned and resampled at a 5 ms resolution to provide a strictly uniform temporal grid. The resulting dataset consists of 2295 individual reaching trajectories.
 We defined the input function $u(t)$ as the continuous neural firing rates from $182$ neural units, which have been smoothed using a Gaussian kernel ($\sigma = 50$ ms) to reduce point-process noise. The target output function $v(t)$ was defined as the continuous hand velocity of the macaque along the $x$ and $y$ axes. To account for the physiological delay required for signals to travel from the motor cortex to the muscles, we applied an $80$ ms lag to the kinematic data relative to the neural activity. We trained the FNO to learn this operator $G: u \to v$. Finally, the predicted velocity was numerically integrated over time to reconstruct the full 2D reaching trajectory (position) of the hand.

\subsection{Training}
We divided our dataset (trajectories) into $80\%$ for training, $10\%$ for validation, and $10\%$ for testing. During the training phase, the continuous neural and kinematic data were processed in mini-batches of $32$ trials. The FNO architecture was configured to retain $16$ Fourier modes with a hidden channel width of $32$ in each area-specific stream and $64$ in the shared layers. 
To exploit the known anatomical separation between the two recorded cortical areas, we adopted a dual-stream design: PMd and M1 firing rates were processed by two independent streams of 2 Fourier layers each, whose outputs were concatenated and passed through 2 additional shared Fourier layers modelling inter-area interactions; the analogous Long Short Term Memory (LSTM) baseline \footnote{One might argue that this comparison is not fair, since FNO acts on the entire time sequence all at once, while LSTM operates only in a sequential manner.  In Appendix \ref{app:C} we added the results relative to a BiLSTM benchmark (with the same hyperparameters as our LSTM), which do not change our conclusions.} (see below) used 1 independent layer per area followed by 1 shared layer.

To optimize the network parameters, we employed the Adam optimizer with an initial learning rate of $1 \times 10^{-3}$ and a weight decay of $1 \times 10^{-4}$ to introduce $L_2$ regularization. The network was trained to minimize the Mean Squared Error (MSE) loss function, quantifying the discrepancy between the predicted continuous hand velocities and the actual recorded kinematics. 

The training loop was executed for a total of $300$ epochs. To mitigate the risk of overfitting and ensure optimal generalization to unseen data, we implemented a model checkpointing strategy based on the validation set performance. At the end of each epoch, the model was evaluated on the validation split: whenever the validation loss reached a new historical minimum, the model's weight configuration was saved. Upon completion of the training process, the weights corresponding to the best validation performance were restored and utilized for the final inference and evaluation on the test set.

To rigorously evaluate the performance of our framework, we additionally trained two baseline models: a recurrent neural network and a standard linear model. 

\begin{figure}[htbp]
    \centering
    \begin{subfigure}{0.48\textwidth}
        \centering
        \includegraphics[width=\textwidth]{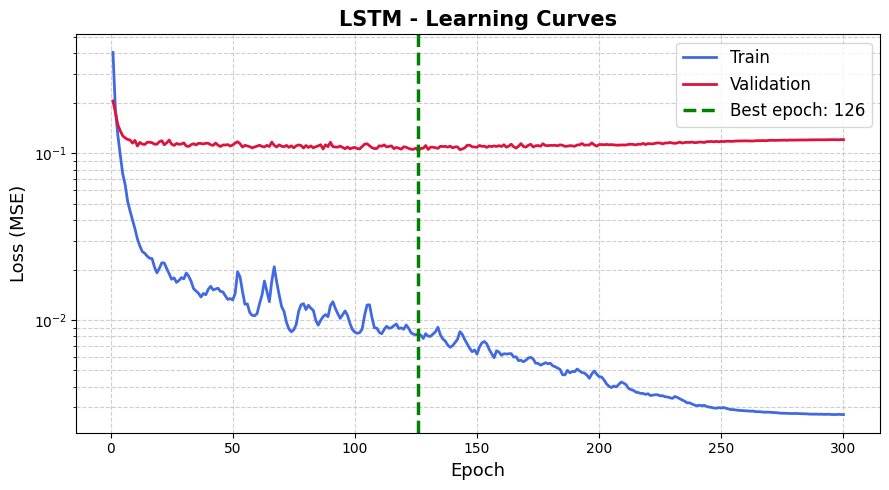}
        \caption{LSTM Training Dynamics}
    \end{subfigure}\hfill
    \begin{subfigure}{0.48\textwidth}
        \centering
        \includegraphics[width=\textwidth]{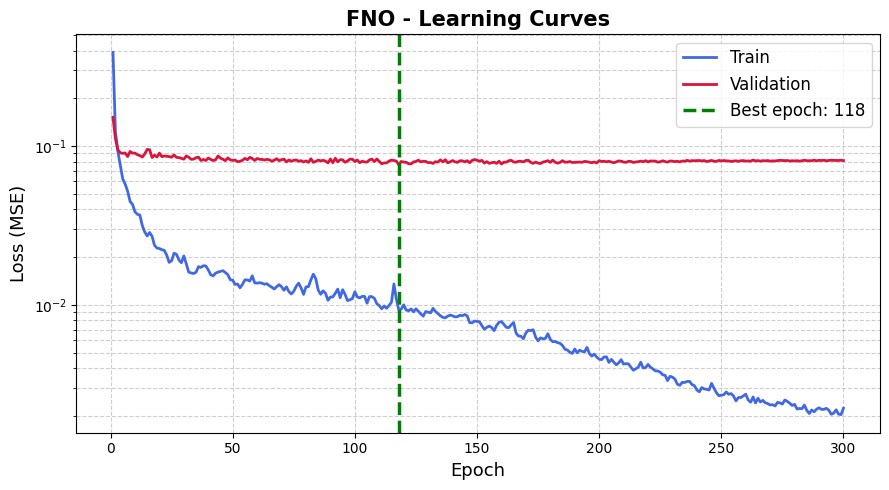}
        \caption{FNO Training Dynamics}
    \end{subfigure}
    \caption{Training and validation loss curves over 300 epochs (MSE on standardized velocities with statistics computed on the training set only). Both non-linear models converge stably, while the dashed line marks the epoch with the lowest validation loss, whose weights were retained.}
    \label{fig:loss_curves}
\end{figure}

For the non-linear recurrent baseline, we implemented a Long Short Term Memory (LSTM) network. The architecture consisted of the previously described LSTM layers with a hidden dimension of $128$, followed by a 2 layer MLP readout to map the recurrent hidden states to the 2D kinematic outputs. 
The network was trained using the identical optimization scheme as the FNO: an Adam optimizer with a learning rate of $1 \times 10^{-3}$ and a weight decay of $1 \times 10^{-4}$, minimizing the Mean Squared Error (MSE) over $300$ epochs. Furthermore, the same validation-based checkpointing strategy was employed, saving the model weights only when the validation loss reached a new minimum.
In Figure \ref{fig:loss_curves} we can see the training and validation curves for both types of architectures with the corresponding best epoch. 

As a standard linear baseline, we implemented a Ridge Regression model. 
To ensure optimal generalization, the $L_2$ regularization hyperparameter ($\alpha$) was selected via an exhaustive grid search combined with 5-fold cross-validation executed strictly on the training set. The grid search evaluated $100$ logarithmically spaced candidate values ranging from $10^{-4}$ to $10^{0}$, automatically selecting the most robust linear mapping between the neural firing rates and the macaque's hand velocity \footnote{The selected $\alpha=0.91$.}.

\subsection{Results}

The evaluation of the Operator Learning framework was conducted on a strictly held-out test set comprising 230 unseen reaching trials. To quantitatively assess the predictive capabilities of the models, we computed the Global $R^2$ score, the individual $R^2$ scores for the $X$ and $Y$ velocity components, the Mean Squared Error (MSE), and the Relative $L_2$ Error. The results are summarized and compared against the baseline models in Table \ref{tab:test_performance}.

\begin{table}[htbp]
\centering
\resizebox{\textwidth}{!}{
\begin{tabular}{l c c c c c}
\toprule
\textbf{Model} & \textbf{Global $R^2$} & \textbf{X / Y Velocity $R^2$} & \textbf{Mean Rel. $L_2$} & \textbf{Median Rel. $L_2$} & \textbf{MSE $(mm/s)^2$} \\
\midrule
Ridge Regression & 0.65 & 0.70 / 0.60 & 0.59 $\pm$ 0.14 & 0.58 & $\left(3.4 \pm 1.8\right)10^{4}$ \\
LSTM & 0.88 & 0.91 / 0.85 & 0.33 $\pm$ 0.12 & 0.31 & $\left(1.2 \pm 1.0\right) 10^{4}$ \\
BiLSTM & 0.90 & 0.91 / 0.88 & 0.31 $\pm$ 0.11 & 0.29 & $\left(9.9 \pm 8.1\right) 10^{3}$ \\
\textbf{FNO} & \textbf{0.92} & \textbf{0.93} / \textbf{0.90} & \textbf{0.27 $\pm$ 0.09} & \textbf{0.26} & $\mathbf{\left(7.8 \pm 6.0\right) 10^{3}}$ \\
\bottomrule
\end{tabular}
}
\caption{Quantitative performance comparison of the decoding models on the unseen \texttt{mc\_maze} test set (230 trials). The FNO achieves the highest $R^2$ scores and the lowest relative $L_2$ errors, demonstrating superior decoding.}
\label{tab:test_performance}
\end{table}

The quantitative analysis clearly highlights the limitations of the linear baseline. Ridge regression yields a Global $R^2$ of $0.65$ and struggles significantly with the Y-axis kinematics ($R^2 = 0.60$). Introducing non-linearity through the LSTM network provides a performance leap, elevating the Global $R^2$ to $0.88$ and reducing the relative $L_2$ error. 

However, FNO outperforms all baselines. By treating the neural dynamics and the resulting kinematics as continuous functions, the FNO reaches a Global $R^2$ of $0.92$, proving highly reliable in predicting both the $X$ ($0.93$) and $Y$ ($0.90$) velocity components. The FNO also boasts the lowest and most stable error margin, with a Median Relative $L_2$ error of $0.26$.

\begin{figure}[htbp]
    \centering
    \begin{subfigure}{1.0\textwidth}
        \centering
        \includegraphics[width=\textwidth]{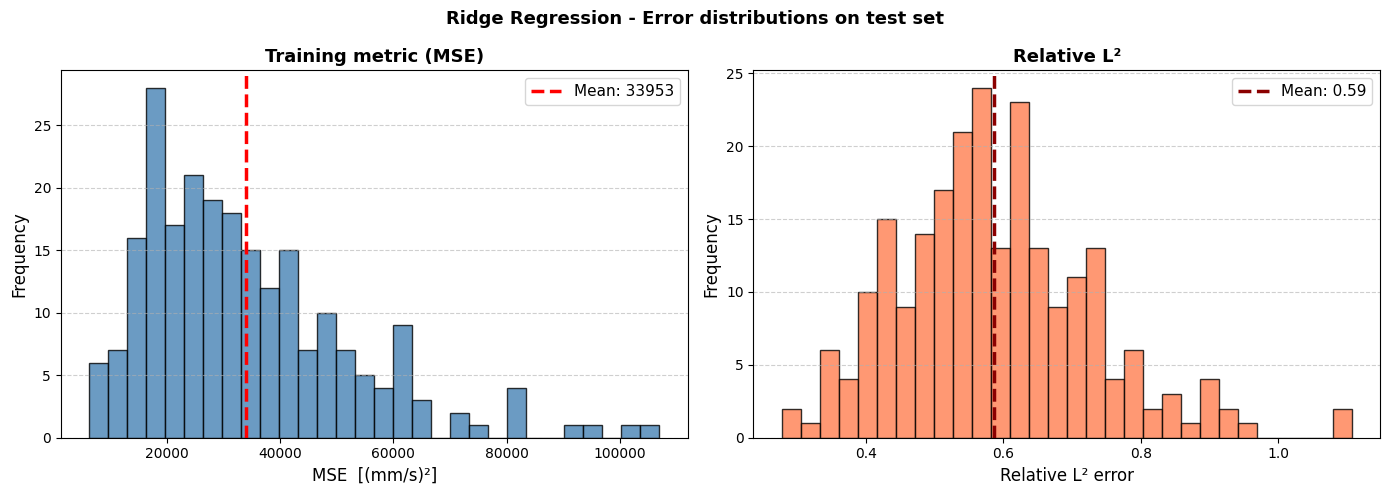}
        \caption{Ridge Regression}
    \end{subfigure}
    
    \vspace{0.4cm}
    \begin{subfigure}{1.0\textwidth}
        \centering
        \includegraphics[width=\textwidth]{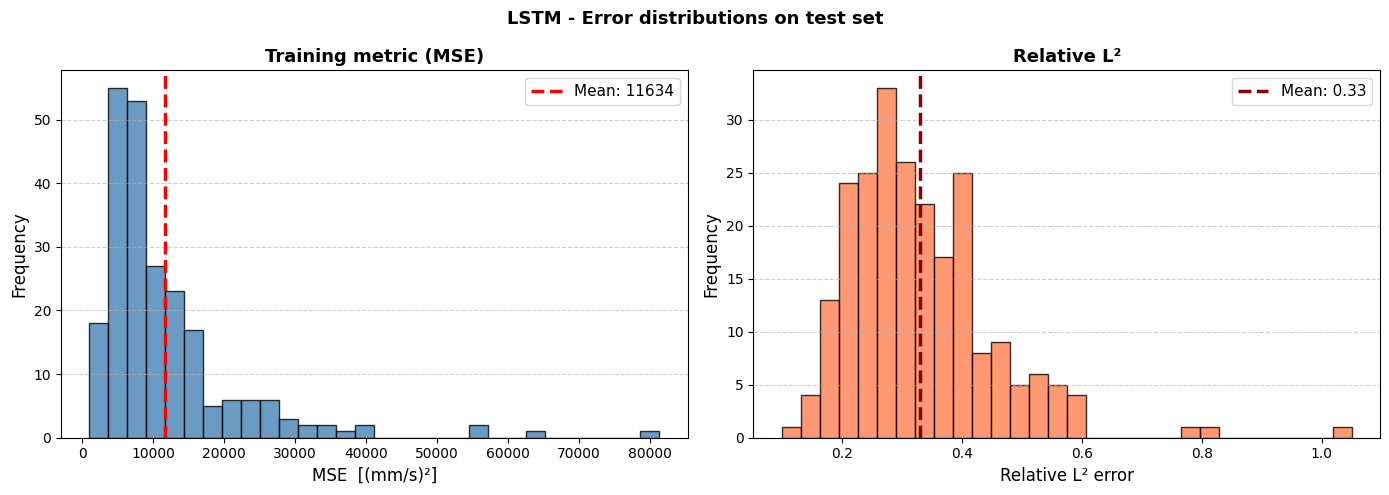}
        \caption{LSTM}
    \end{subfigure}
    
    \vspace{0.4cm}
    \begin{subfigure}{1.0\textwidth}
        \centering
        \includegraphics[width=\textwidth]{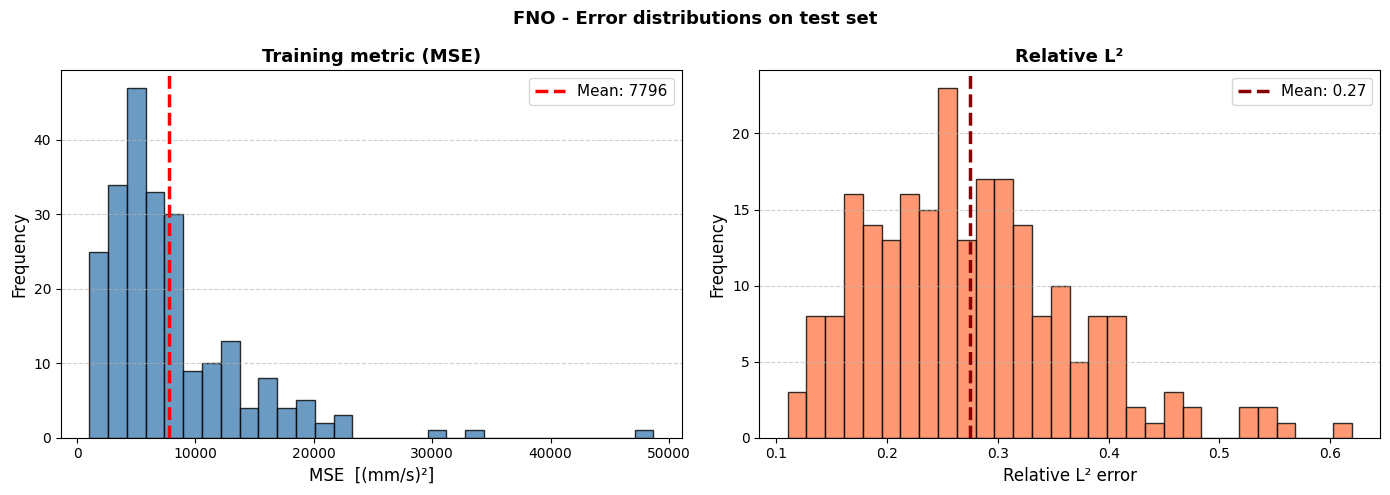}
        \caption{FNO}
    \end{subfigure}
    \caption{Histograms of the Relative $L_2$ and MSE errors across the test set. The FNO (c) demonstrates a narrower and lower error distribution, indicating greater consistency across different reaching trials.}
    \label{fig:histograms}
\end{figure}

To further investigate the error distribution across the 230 test trials, we analyzed the histograms of the relative $L_2$ errors for each model (Figure \ref{fig:histograms}). The FNO distribution is highly skewed towards lower errors, exhibiting a tighter variance compared to the LSTM and a drastic shift compared to Ridge Regression.

Finally, qualitative assessments were performed by numerically integrating the predicted 2D velocities to reconstruct the physical hand trajectories of the macaque monkey. The results are visualized across the following figures for Ridge Regression (Figure \ref{fig:qual_ridge} in Appendix \ref{app:C}), the LSTM baseline (Figure \ref{fig:qual_lstm}, Appendix \ref{app:C}), and our FNO model (Figure \ref{fig:qual_fno}, Appendix \ref{app:C}). The linear predictions produced by Ridge Regression often fail to capture the curvature and subtle velocity shifts of the reaches. In contrast, both non-linear models—the LSTM baseline and the FNO model successfully capture the complex, non-linear dynamics of the biological movement. Visually, there is no drastic qualitative difference between the LSTM and FNO reconstructions; both neural networks reliably reproduce the ground truth trajectories, navigating sharp turns and complex maze configurations with fidelity. This visual similarity aligns with our quantitative findings: while the FNO maintains a slight statistical advantage (lower relative $L_2$ error and higher global $R^2$), both architectures demonstrate a robust capacity to decode complex kinematics.

\begin{figure}[p]
    \centering
    \includegraphics[width=\textwidth, height=0.9\textheight, keepaspectratio]{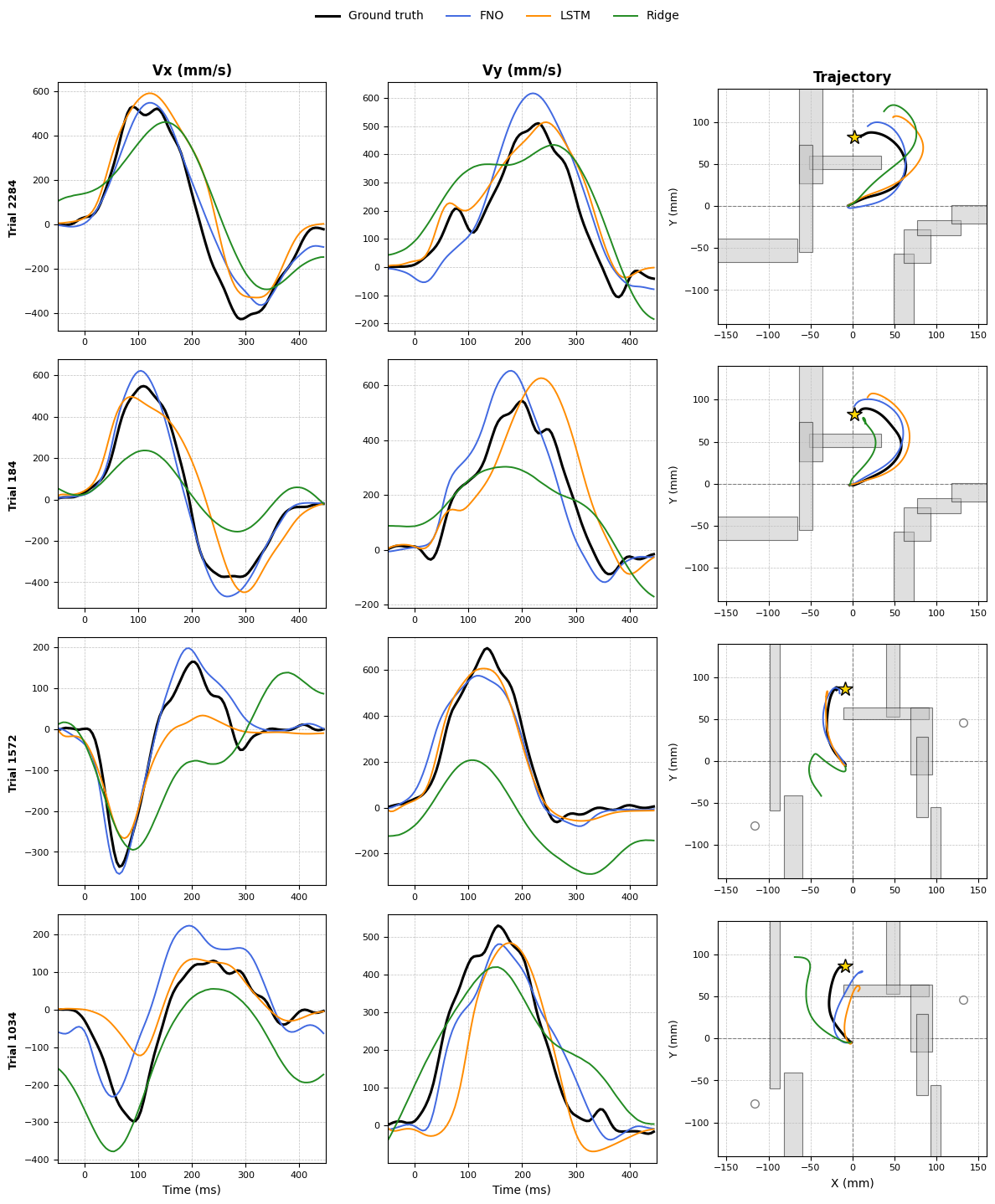}
    \caption{Comparison on some instances of the test set of the three models proposed for two particular maze types. Further tests are reported in Appendix \ref{app:C}.}
    \label{fig:maze_models}
\end{figure}
\clearpage

\section{Conclusions}

In this work, we have extended the application of Scientific Machine Learning (SciML), specifically Operator Learning via FNO, from isolated dynamical systems to spatially extended analogues hosted on heterogeneous networks. Building upon previous studies that established the efficacy of FNO in emulating isolated stiff ionic models, we demonstrated that neural operators can effectively capture the macroscopic dynamics of interconnected systems without requiring explicit knowledge of their internal constituents, exact coupling topologies, or local physical laws.

By framing these spatially extended systems as strictly black-box models, our framework relies entirely on macroscopic input-output data. Numerical experiments using the diffusively coupled FitzHugh-Nagumo reaction scheme demonstrated the high expressivity and robustness of the methodology across diverse synthetic network architectures, including Erd\H{o}s-R\'enyi and scale-free Barabasi–Albert graphs. The architecture maintained its accuracy and generalization capabilities across varying network sizes. Furthermore, the practical utility of this data-driven paradigm was validated on real-world biological data, where the operator successfully learned emergent macro-level dynamics from electrophysiological recordings of a macaque monkey (\textit{Macaca mulatta}) performing a delayed center-out reaching task. In this respect, the FNO scheme proved more capable, as compared to the baseline models (Ridge Regression and (Bi)LSTM), in grasping and consequently resolving the complex nature of analyzed data.  

Overall, these findings highlight the power of Operator Learning as a flexible, scalable surrogate model for complex networked systems where internal governing equations are either partially unknown or computationally intractable to simulate directly. Future work will focus on integrating graph-based spatial operators in the predicting scheme, exploring the framework's resilience to missing or noisy macroscopic data, and extending these surrogate models toward real-time feedback control and intervention in complex biological networks.
\clearpage

\begin{appendices}

\section{Dependence on the Diffusion Coefficient $\sigma$}
\label{appA}

\begin{figure}[htbp]
    \centering
    \includegraphics[width=0.8\textwidth]{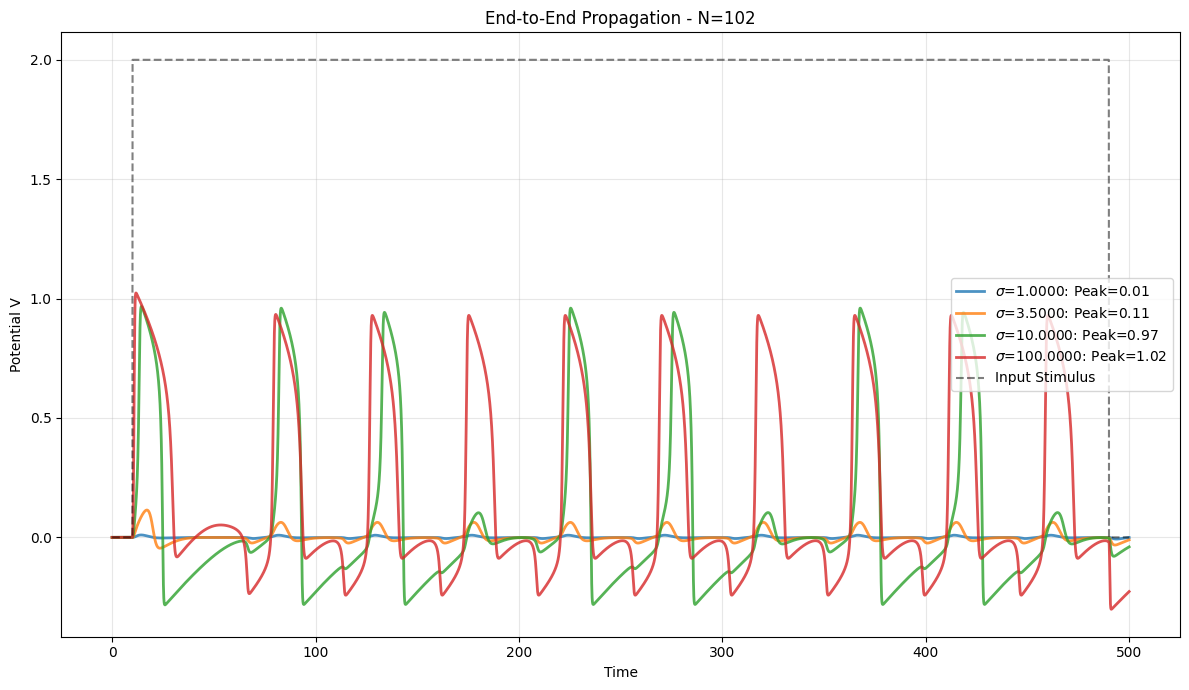}
    \caption{Dynamical transition of the neural network as the diffusion coefficient $\sigma$ varies. For low values of $\sigma$, the signal fails to propagate efficiently, whereas a critical threshold allows for coherent signal transmission. Here $|\mathcal{V}| = 102$ represents the size of the hidden network plus the input and output neurons.}
    \label{fig:diffusion_transition}
\end{figure}

During the experiments presented in the previous sections, it became essential to understand how to appropriately tune the diffusion coefficient $\sigma$ as a function of the hidden network size $N$. In fact, if we maintain $\sigma$ fixed while $N$ increases, the signal amplitude dissipates and the output signal progressively attenuates until it completely dies out. An initial numerical investigation suggested that the correct scaling to preserve the macroscopic dynamical regime is $\sigma = \mathcal{O}(N)$. In Figure \ref{fig:diffusion_transition} we can see that for $\sigma$ several orders of magnitude smaller than N (here $N=100$) the signal does not seem to flow through the network.

This empirical observation is supported by the numerical study of the spectral gap of the network's directed Laplacian, specifically the first non-zero real part of the eigenvalues, $\text{Re}(\lambda_2)$. In networked dynamical systems, the global synchronization and the ability of a perturbation to travel across the graph 
 are governed by the effective macroscopic diffusion rate, which is proportional to the product $\sigma \text{Re}(\lambda_2)$ \cite{barrat2008dynamical}. 

\begin{figure}[htbp]
    \centering
    \begin{minipage}[t]{0.49\textwidth}
        \vspace{0pt}
        \centering
        \includegraphics[width=\textwidth]{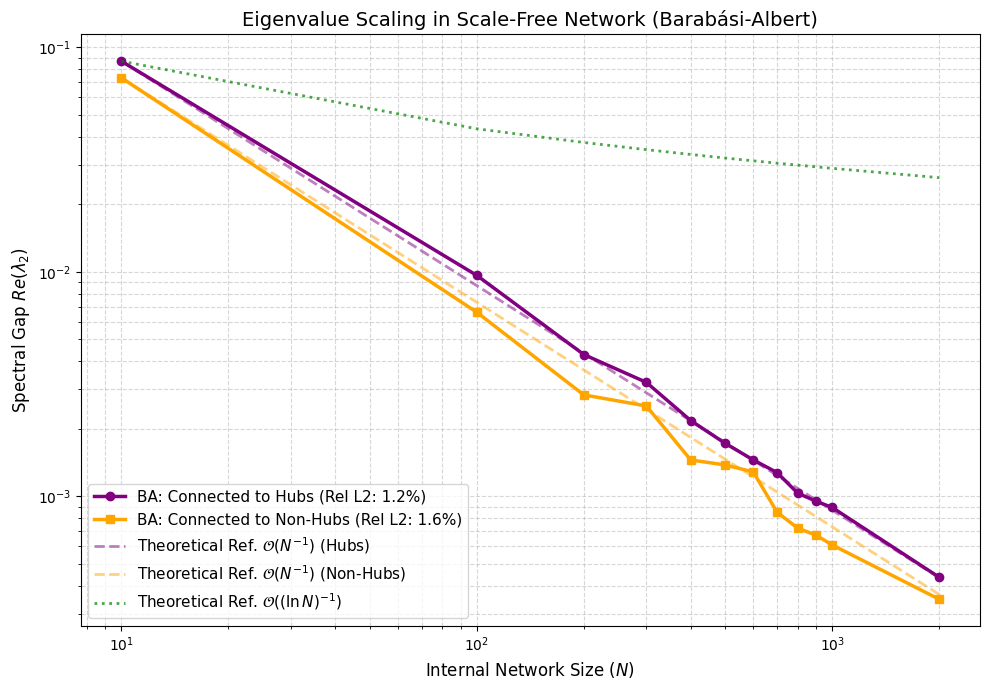}
        \caption{Scaling of the spectral gap $\text{Re}(\lambda_2)$ as a function of $N$ for the Barabási-Albert (BA) topology, showing the $\mathcal{O}(N^{-1})$ decay, both for a standard and a network with preferential attachment of Input/Output to hubs.}
        \label{fig:spectral_gap_ba}
    \end{minipage}\hfill
    \begin{minipage}[t]{0.49\textwidth}
        \vspace{0pt}
        \centering
        \includegraphics[width=\textwidth]{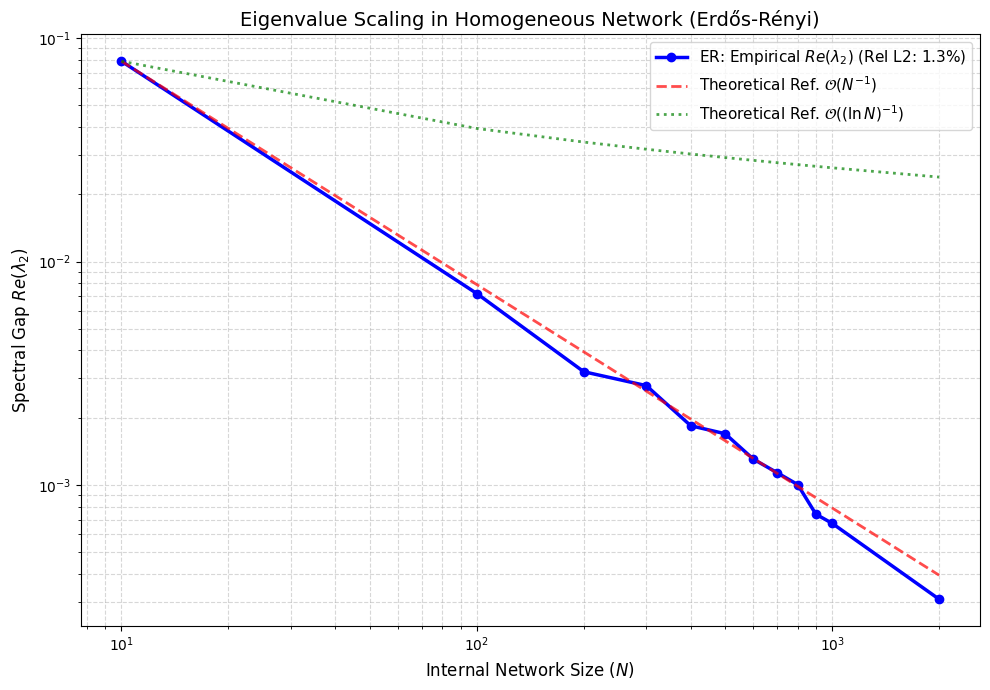}
        \caption{Scaling of the spectral gap $\text{Re}(\lambda_2)$ as a function of $N$ for the Erdős–Rényi (ER) topology, confirming the identical $\mathcal{O}(N^{-1})$ decay limit.}
        \label{fig:spectral_gap_er}
    \end{minipage}
\end{figure}

Although the internal spectral gap of isolated complex networks inherently depends on their topology---decaying and bounded above by $\mathcal{O}((\ln N)^{-1})$ in sparse Erd\H{o}s-R\'enyi (ER) graphs \cite{benjamini2014mixing} and remaining strictly lower-bounded by a constant $\mathcal{O}(1)$ in Barab\'asi-Albert (BA) scale-free networks \cite{mihail2006connectivity}---the strict structural bottleneck introduced by our specific boundary conditions profoundly alters the macroscopic transport dynamics. 
In fact, within the INPUT $\rightarrow$ network $\rightarrow$  OUTPUT setup, not all paths are of equal relevance, since the system will completely relax only when the signal will propagate to the OUTPUT node. This means that the shortest path from INPUT to OUTPUT becomes the most relevant for the behavior of spectral gap and this alters its behavior.
As a result, our numerical spectral analysis shows that the physical spectral gap $\text{Re}(\lambda_2)$ collapses, scaling approximately as $\mathcal{O}(N^{-1})$ for both the ER and BA setups, as it can be seen in Figure \ref{fig:spectral_gap_ba}-\ref{fig:spectral_gap_er}.

Therefore, to maintain a constant effective macroscopic coupling ($\sigma \text{Re}(\lambda_2) \approx \text{constant}$) and ensure that the operator learns a consistent continuous limit regardless of the spatial discretization size $N$, the physical diffusion coefficient must systematically scale as $\sigma \propto N$.


For our experiments we chose $\sigma=0.1*N$.

\clearpage

\section{Numerical Integration}
\label{app:B}
Here we analyze the numerical techniques used to efficiently integrate system \eqref{eq:coupled_FHN}.
First, we should preface that this system is stiff \cite{hairer1996solving}. In the context of numerical analysis, a system is considered stiff when it exhibits dynamics acting on widely different time scales (such as the fast action potential spikes and the slow recovery phases in electrophysiological models). This property causes standard explicit numerical methods to become unstable unless prohibitively small time steps are used. Therefore, finding an adequate numerical algorithm that scales well with the size of the network is a non-trivial task.

To overcome the stability constraints imposed by the stiff diffusion term while avoiding the computational burden of solving non-linear algebraic systems at each time step, we adopted an Implicit-Explicit (IMEX) integration scheme \cite{ascher1995implicit, hundsdorfer2003numerical}. Specifically, we employed a first-order IMEX Euler method, which treats the local non-linear reaction kinetics explicitly and the linear spatial coupling implicitly. 

Let us rewrite the networked system defined in Equation \eqref{eq:coupled_FHN} in a compact vector notation. Let $\mathbf{V}(t) = [V_1(t), \dots, V_{|\mathcal{V}|}(t)]^T$ and $\mathbf{w}(t) = [w_1(t), \dots, w_{|\mathcal{V}|}(t)]^T$ be the state vectors of the network at time $t$. The dynamics can be separated as follows:
\begin{equation}
\label{eq:vector_system}
\begin{aligned}
\frac{d\mathbf{V}}{dt} &= \mathbf{F}(\mathbf{V}, \mathbf{w}, t) - \sigma \mathbf{L} \mathbf{V} \\
\frac{d\mathbf{w}}{dt} &= \mathbf{G}(\mathbf{V}, \mathbf{w})
\end{aligned}
\end{equation}
where $\mathbf{F}(\mathbf{V}, \mathbf{w}, t)$ incorporates the non-linear local FitzHugh-Nagumo dynamics and the external applied current, while $\mathbf{G}(\mathbf{V}, \mathbf{w})$ represents the linear dynamics of the recovery variables.

In our IMEX formulation, given a time step $\Delta t = t_{n+1} - t_n$, the temporal update from step $n$ to $n+1$ is performed in two stages. First, we evaluate the explicit forward Euler step \footnote{The use of a Runge-Kutta method would be a more rigorous approach to the problem; however since our main aim was to test the performance of FNO and not to analyze this particular system, we preferred this faster integration technique.} for the local reaction terms to compute an intermediate predictor state $\mathbf{V}^*$, alongside the full update for the recovery variables $\mathbf{w}^{n+1}$:
\begin{equation}
\label{eq:imex_explicit}
\begin{aligned}
\mathbf{V}^* &= \mathbf{V}^n + \Delta t \mathbf{F}(\mathbf{V}^n, \mathbf{w}^n, t_n) \\
\mathbf{w}^{n+1} &= \mathbf{w}^n + \Delta t \mathbf{G}(\mathbf{V}^n, \mathbf{w}^n)
\end{aligned}
\end{equation}

Subsequently, the implicit backward Euler step is applied exclusively to the linear diffusion term. The final potential $\mathbf{V}^{n+1}$ is obtained by solving:
\begin{equation}
\label{eq:imex_implicit_1}
\mathbf{V}^{n+1} = \mathbf{V}^* - \Delta t \sigma \mathbf{L} \mathbf{V}^{n+1}
\end{equation}

By rearranging Equation \eqref{eq:imex_implicit_1}, we isolate the unknown vector $\mathbf{V}^{n+1}$:
\begin{equation}
\label{eq:imex_implicit_2}
\left( \mathbf{I} + \Delta t \sigma \mathbf{L} \right) \mathbf{V}^{n+1} = \mathbf{V}^*
\end{equation}
where $\mathbf{I}$ is the $|\mathcal{V}| \times |\mathcal{V}|$ identity matrix. 

A standard fully implicit method would require assembling and inverting a Jacobian matrix at every time step due to the non-linearities. Conversely, in our IMEX splitting, the matrix $\mathbf{M} = \left( \mathbf{I} + \Delta t \sigma \mathbf{L} \right)$ governing the implicit update is completely linear and time-invariant. Consequently, its inverse $\mathbf{M}^{-1}$ \footnote{Since, unlike $\mathbf{M}$, $\mathbf{M}^{-1}$ is a dense matrix, solving system \eqref{eq:imex_implicit_2} in this way is not recommendable for very large graphs (i.e.: $N > 50000$.)} can be pre-computed exactly once before the start of the temporal integration loop \footnote{Another, maybe more efficient, method for solving system \eqref{eq:imex_implicit_2} is an LU method.}.

The implicit update then reduces to a simple matrix-vector multiplication:
\begin{equation}
\label{eq:imex_final}
\mathbf{V}^{n+1} = \mathbf{M}^{-1} \mathbf{V}^*
\end{equation}

Instead of simulating single trajectories sequentially, this formulation allowed us to process large batches simultaneously, directly leveraging the parallelization capabilities of a Graphics Processing Unit (GPU)\footnote{All simulations were run on an NVIDIA Quadro RTX 5000 with 16 GB of VRAM.}.

A fundamental advantage of this IMEX formulation resides in its stability properties. If a standard fully explicit forward Euler scheme had been applied to the entire system \eqref{eq:coupled_FHN}, the integration time step $\Delta t$ would have been strictly constrained by the absolute stability region of the diffusion term. Specifically, for a diffusion process governed by a graph Laplacian, the stability condition requires $\Delta t \le \frac{2}{\sigma \lambda_{\text{max}}(\mathbf{L})}$, where $\lambda_{\text{max}}(\mathbf{L})$ is the largest eigenvalue of the Laplacian matrix  \cite{obrien1950study}. Since $\lambda_{\text{max}}(\mathbf{L})$ grows with the maximum degree of the network, an explicit method implies that $\Delta t = \mathcal{O}\left( \frac{1}{\sigma d_{\text{max}}} \right)$.

Consequently, without the IMEX approach, analyzing networks with denser connectivity or increasing the global diffusion intensity $\sigma$ would have forced the integration step size to decrease drastically, leading to prohibitively long simulation times. 

In contrast, the implicit backward Euler step applied to the linear diffusion operator is unconditionally stable. By decoupling the spatial coupling from the explicit update, the stability of the diffusion term is no longer a limiting factor. This mathematical property allowed us to safely maintain a fixed, relatively large integration time step of $\Delta t = 0.04$ \footnote{For our configuration of $t \in [0,100]$ and a number of samples per function $= 2500$.} across all simulations, completely independent of $\sigma$ or the underlying graph topology. The step size becomes constrained solely by the accuracy requirements of the fast local reaction kinetics of the FitzHugh-Nagumo model, guaranteeing both robust convergence and highly efficient dataset generation.


\clearpage

\section{Supplementary Material}
\label{app:C}
Here we report all the additional figures related to the analysis of Section \ref{Chp:4}.

\subsection{ER Network}
\begin{figure}[htbp]
    \centering
    \begin{subfigure}[b]{0.32\textwidth}
        \centering
        \includegraphics[width=\textwidth]{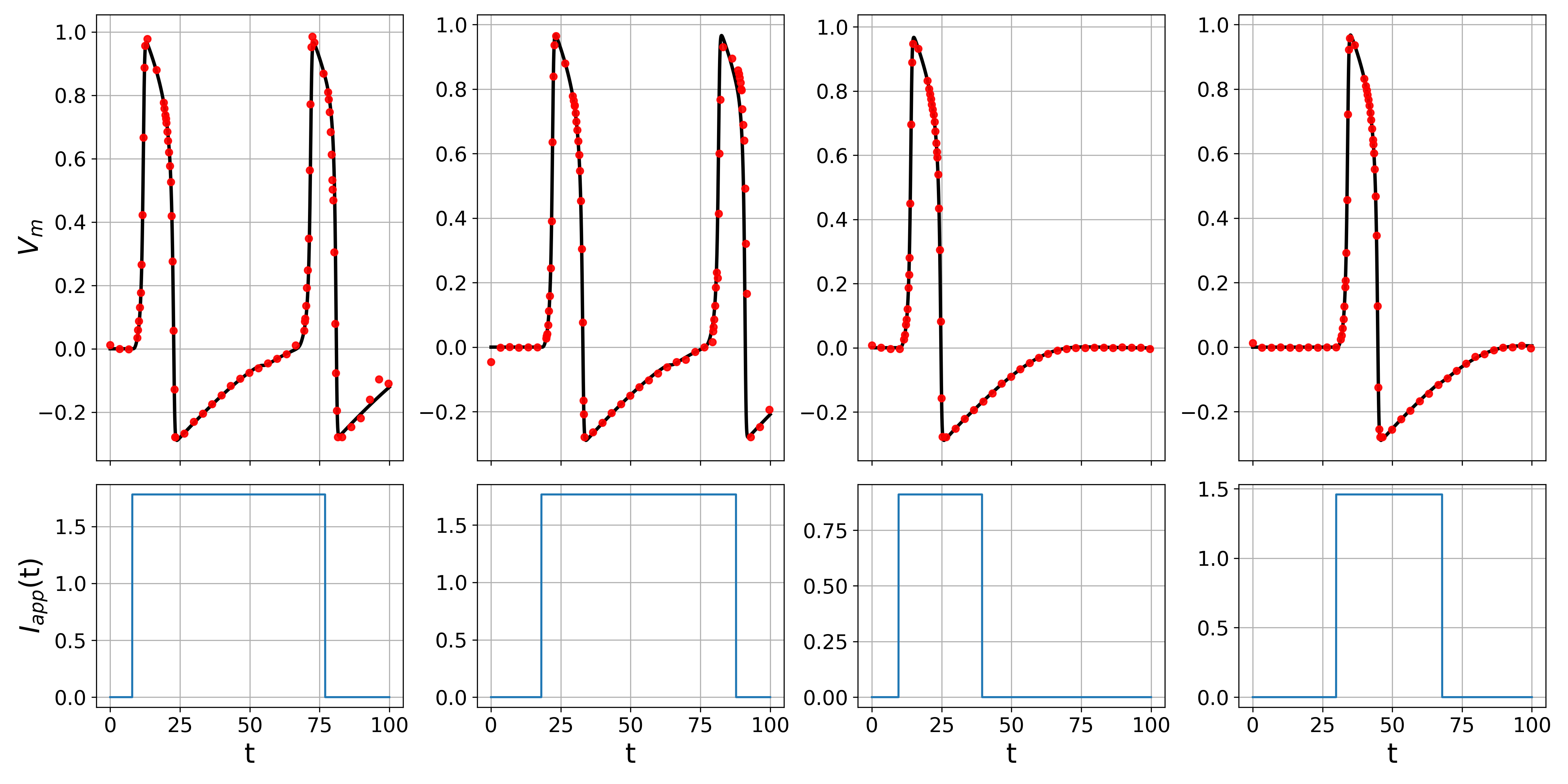}
        \caption{$N=10$}
    \end{subfigure}
    \hfill
    \begin{subfigure}[b]{0.32\textwidth}
        \centering
        \includegraphics[width=\textwidth]{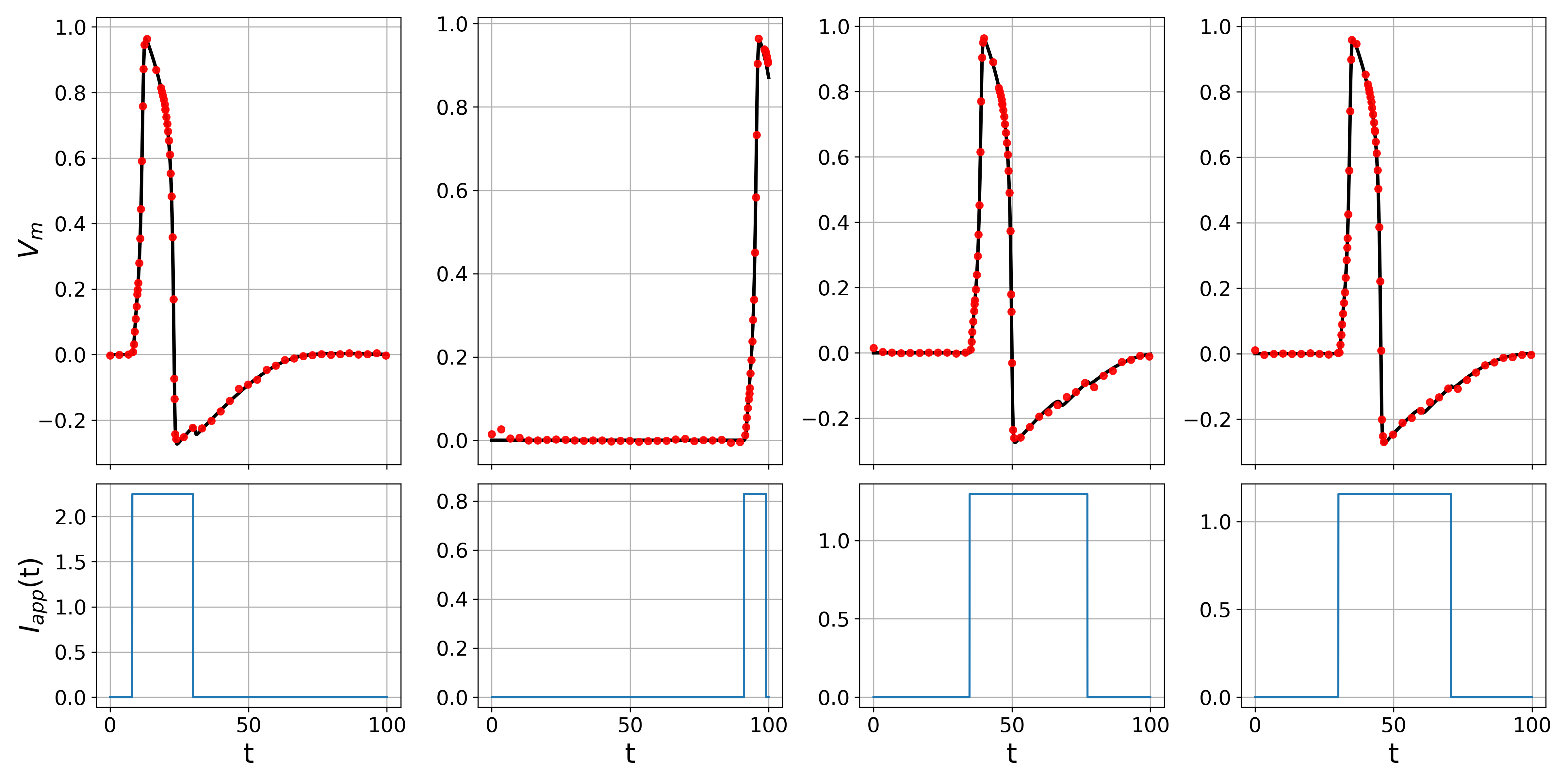}
        \caption{$N=100$}
    \end{subfigure}
    \hfill
    \begin{subfigure}[b]{0.32\textwidth}
        \centering
        \includegraphics[width=\textwidth]{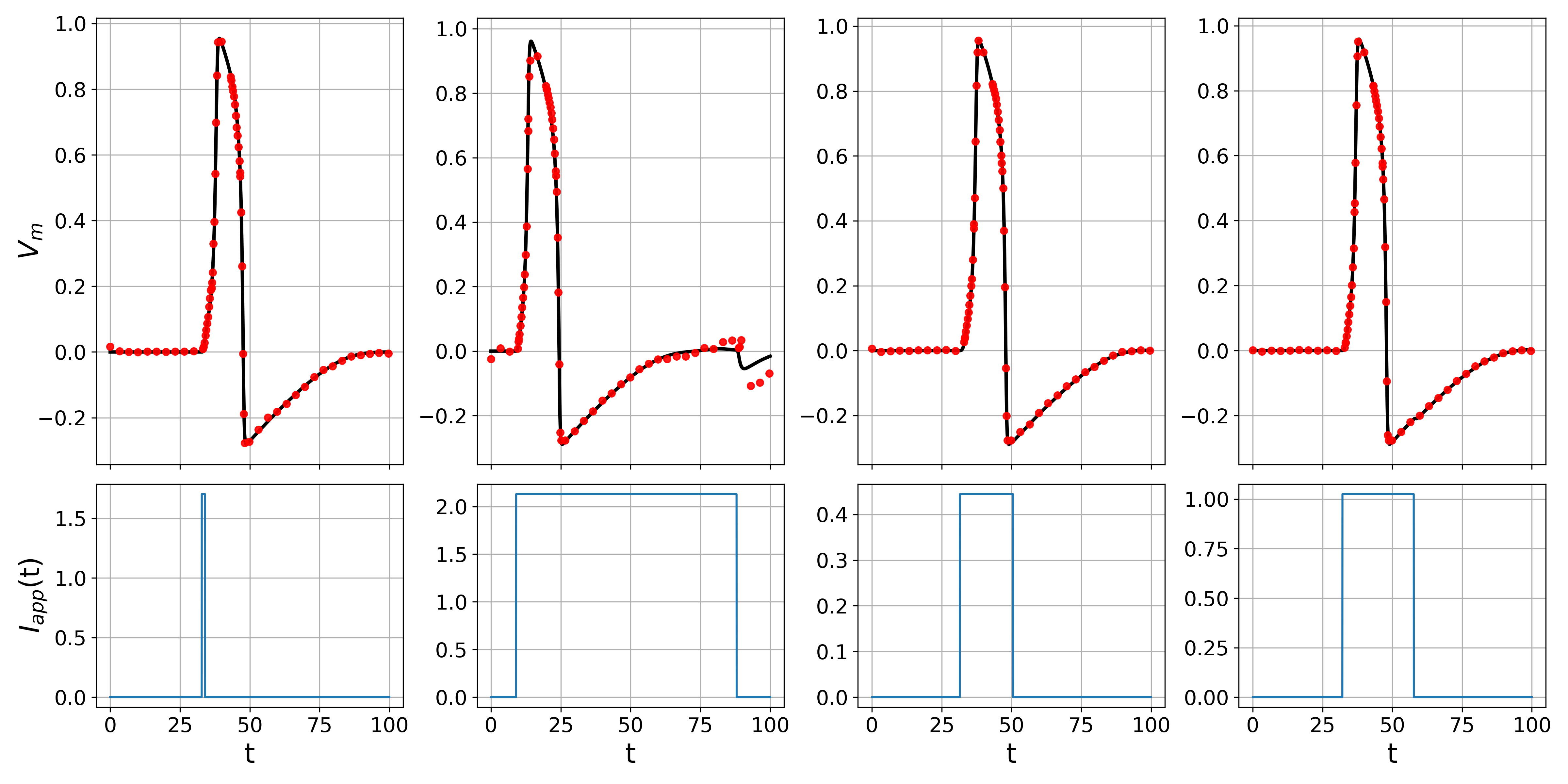}
        \caption{$N=200$}
    \end{subfigure}
    
    \vspace{0.4cm} 
    
    \begin{subfigure}[b]{0.32\textwidth}
        \centering
        \includegraphics[width=\textwidth]{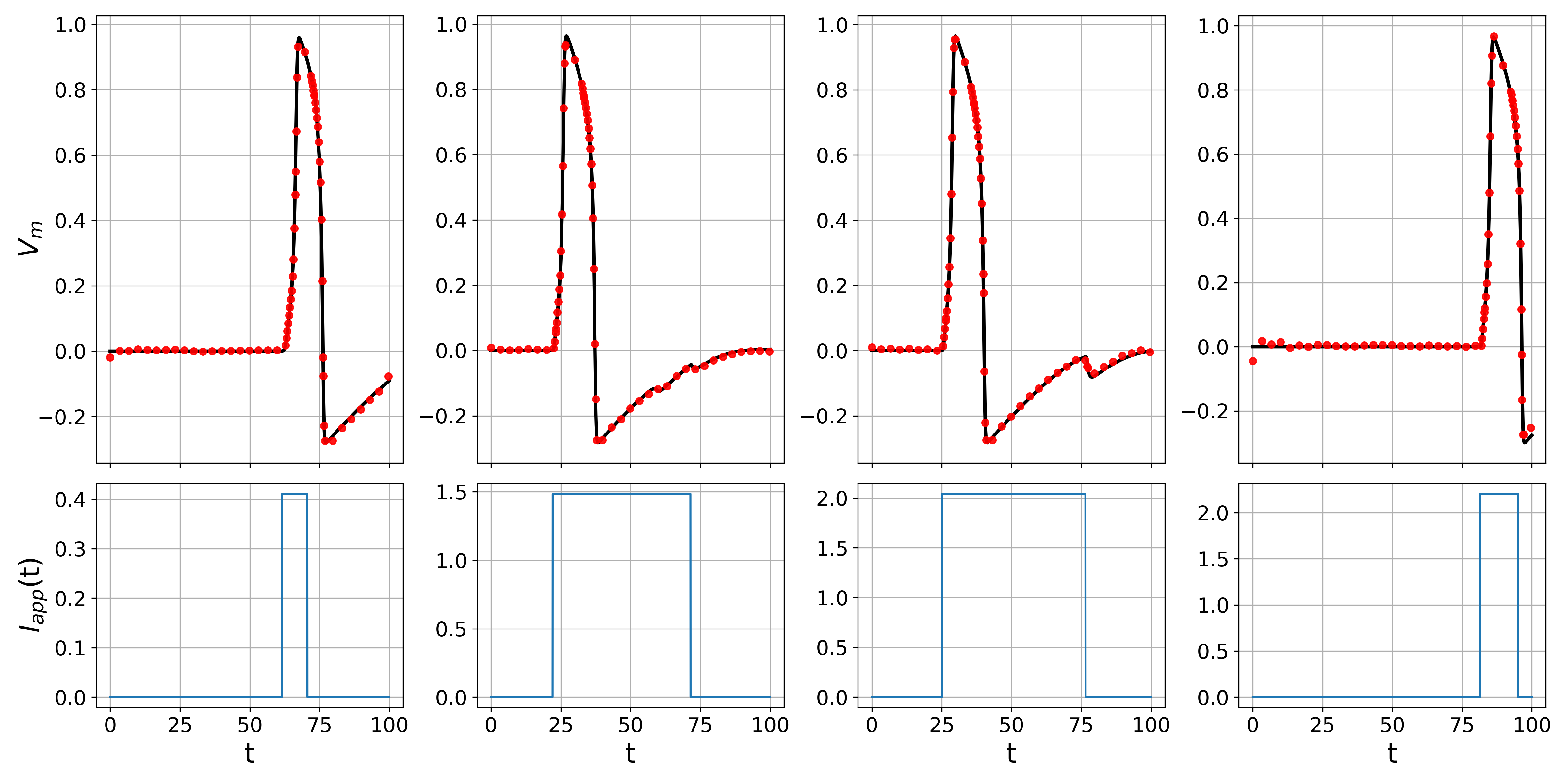}
        \caption{$N=300$}
    \end{subfigure}
    \hfill
    \begin{subfigure}[b]{0.32\textwidth}
        \centering
        \includegraphics[width=\textwidth]{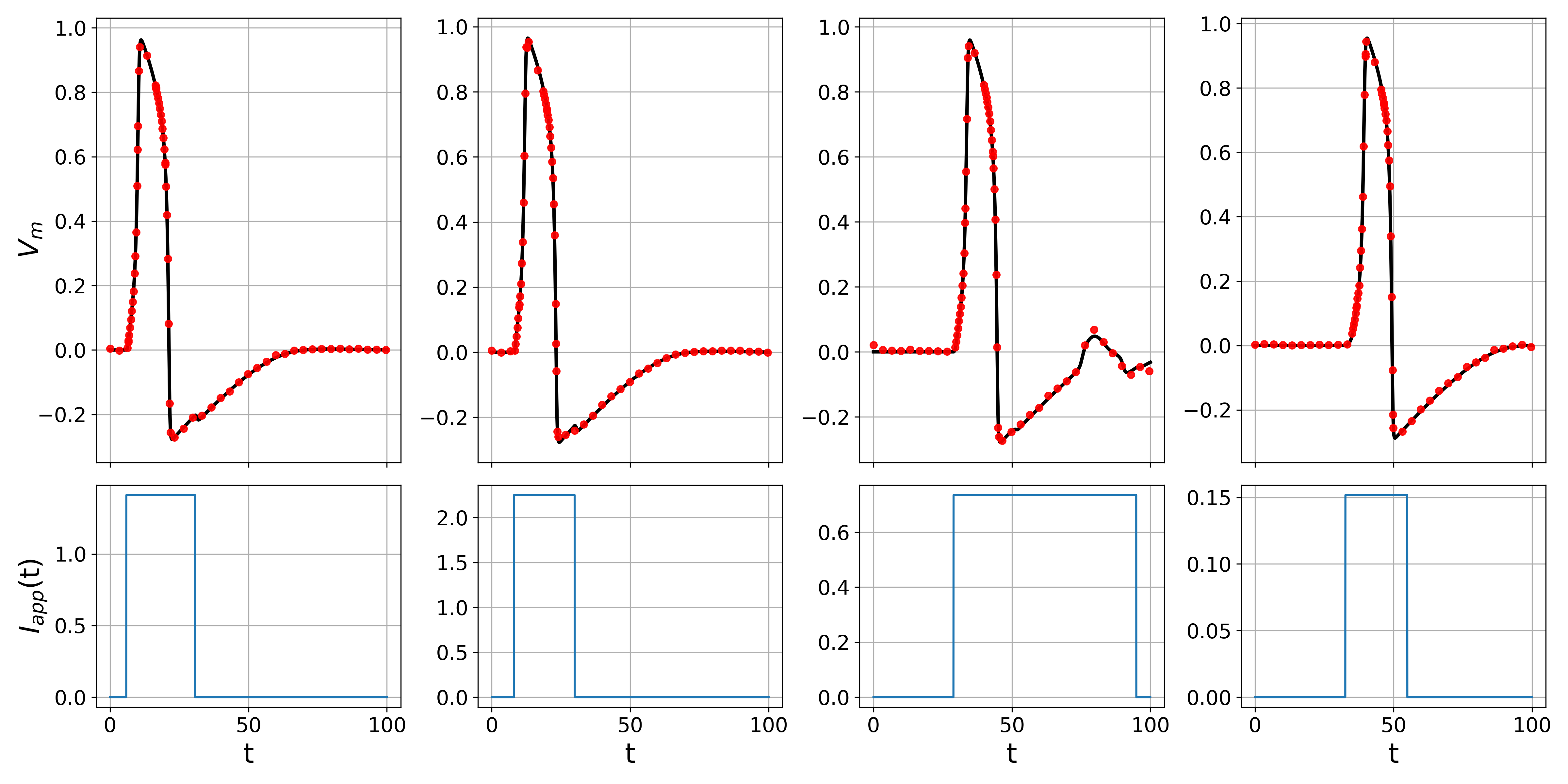}
        \caption{$N=400$}
    \end{subfigure}
    \hfill
    \begin{subfigure}[b]{0.32\textwidth}
        \centering
        \includegraphics[width=\textwidth]{ER_Network_FNO_HHpeaks_N=500.png}
        \caption{$N=500$}
    \end{subfigure}
    
    \vspace{0.4cm}
    
    \begin{subfigure}[b]{0.32\textwidth}
        \centering
        \includegraphics[width=\textwidth]{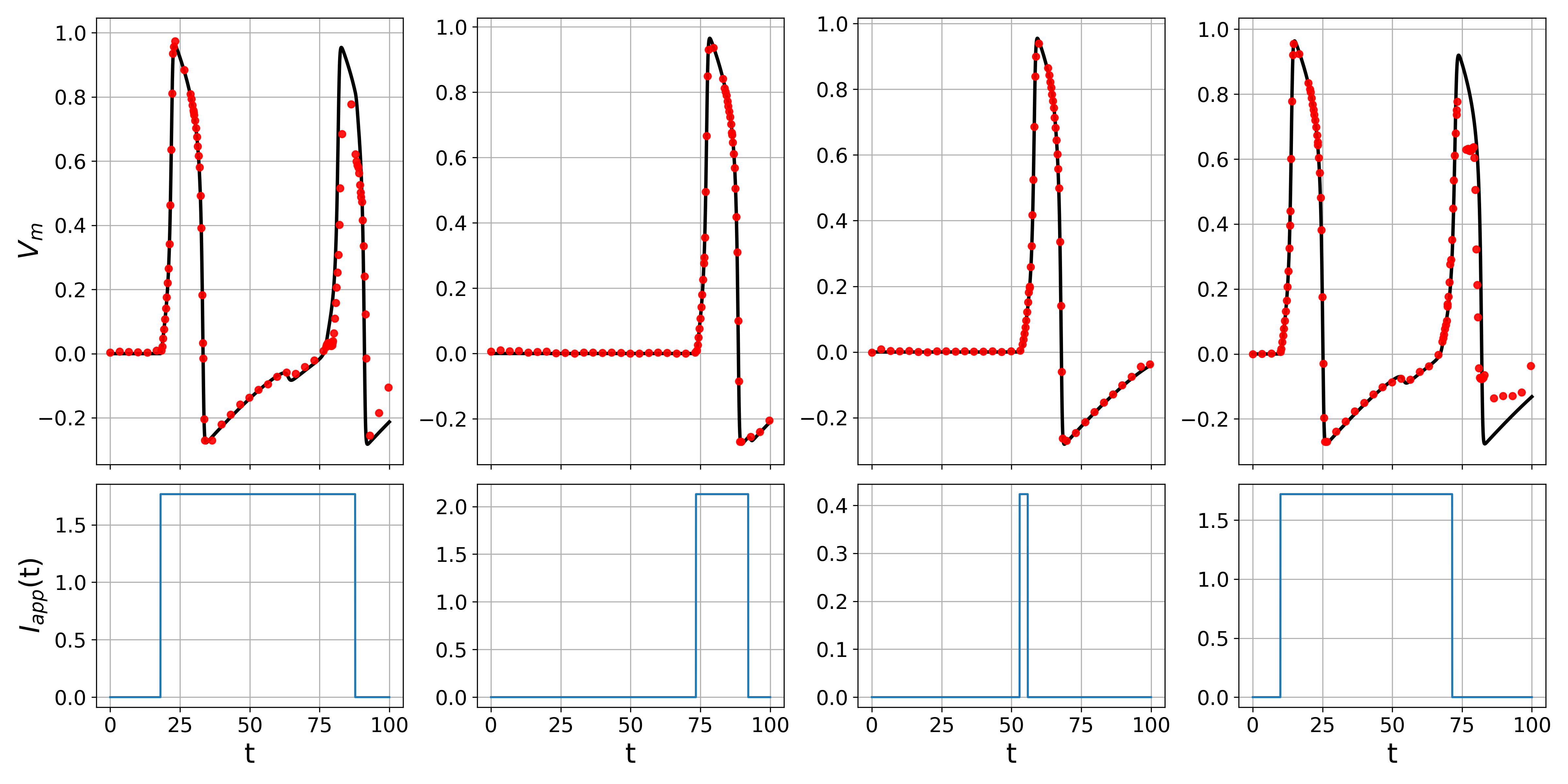}
        \caption{$N=600$}
    \end{subfigure}
    \hfill
    \begin{subfigure}[b]{0.32\textwidth}
        \centering
        \includegraphics[width=\textwidth]{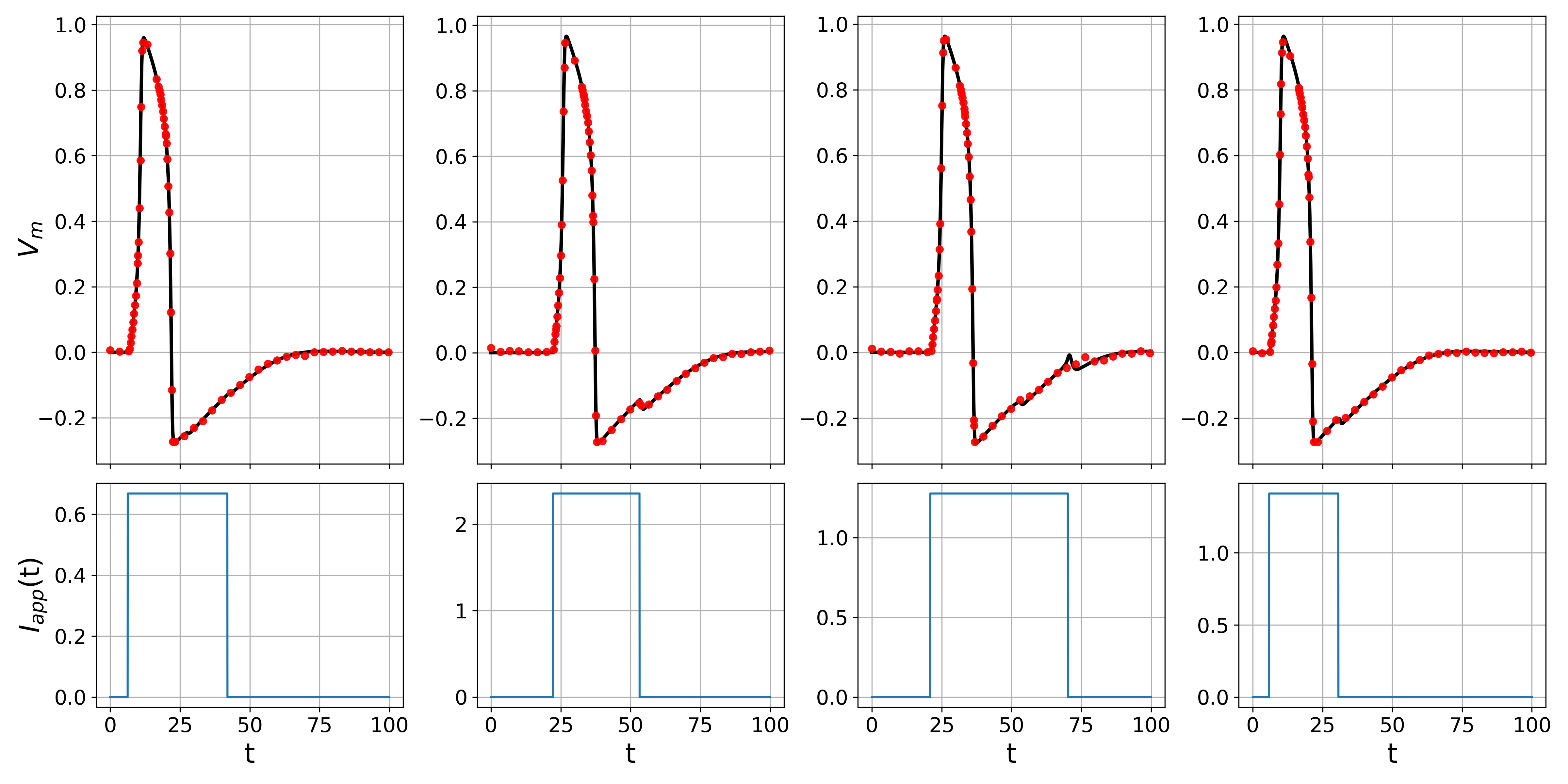}
        \caption{$N=700$}
    \end{subfigure}
    \hfill
    \begin{subfigure}[b]{0.32\textwidth}
        \centering
        \includegraphics[width=\textwidth]{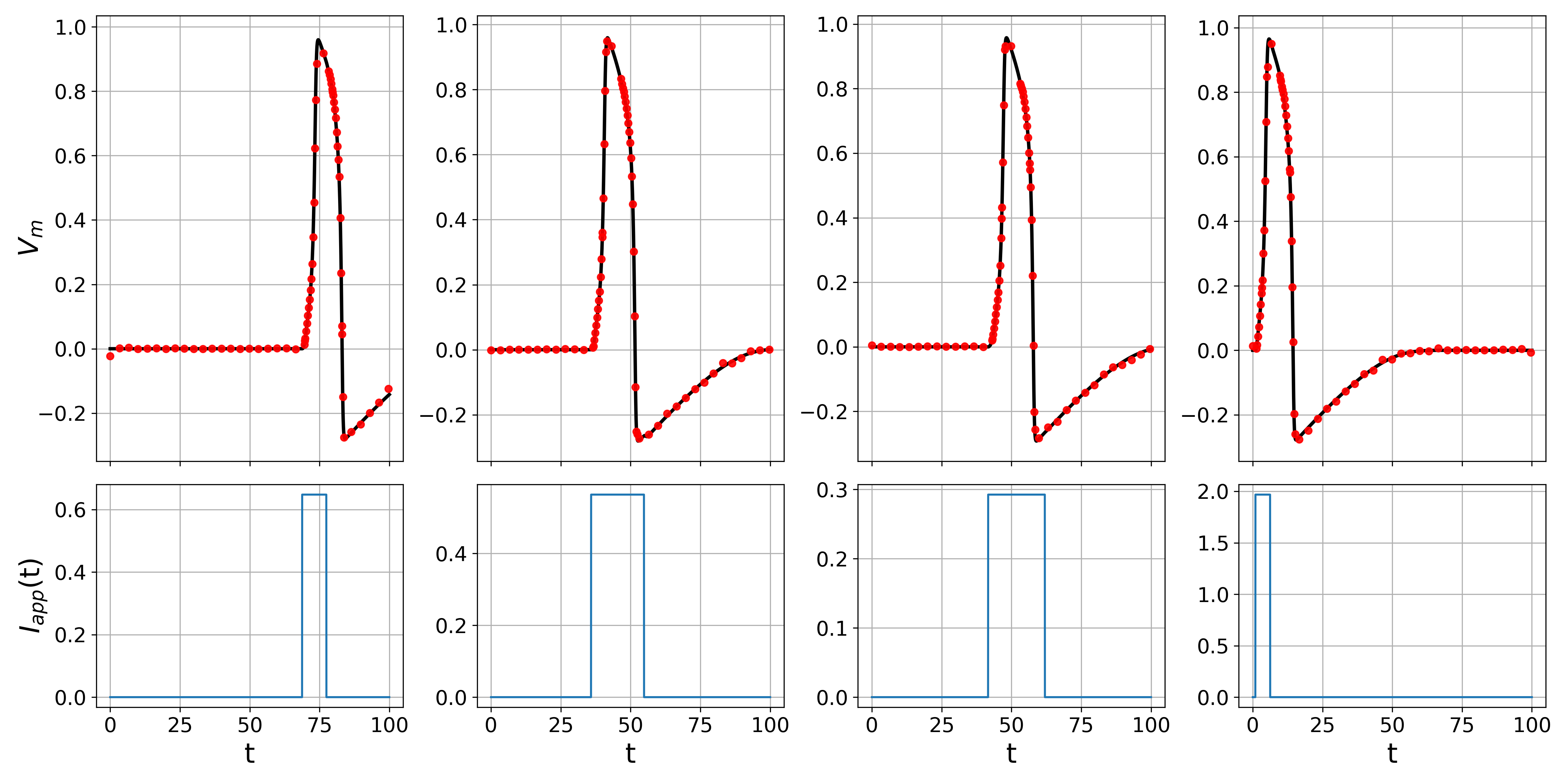}
        \caption{$N=800$}
    \end{subfigure}
    
    \vspace{0.4cm}
    
    \begin{subfigure}[b]{0.32\textwidth}
        \centering
        \includegraphics[width=\textwidth]{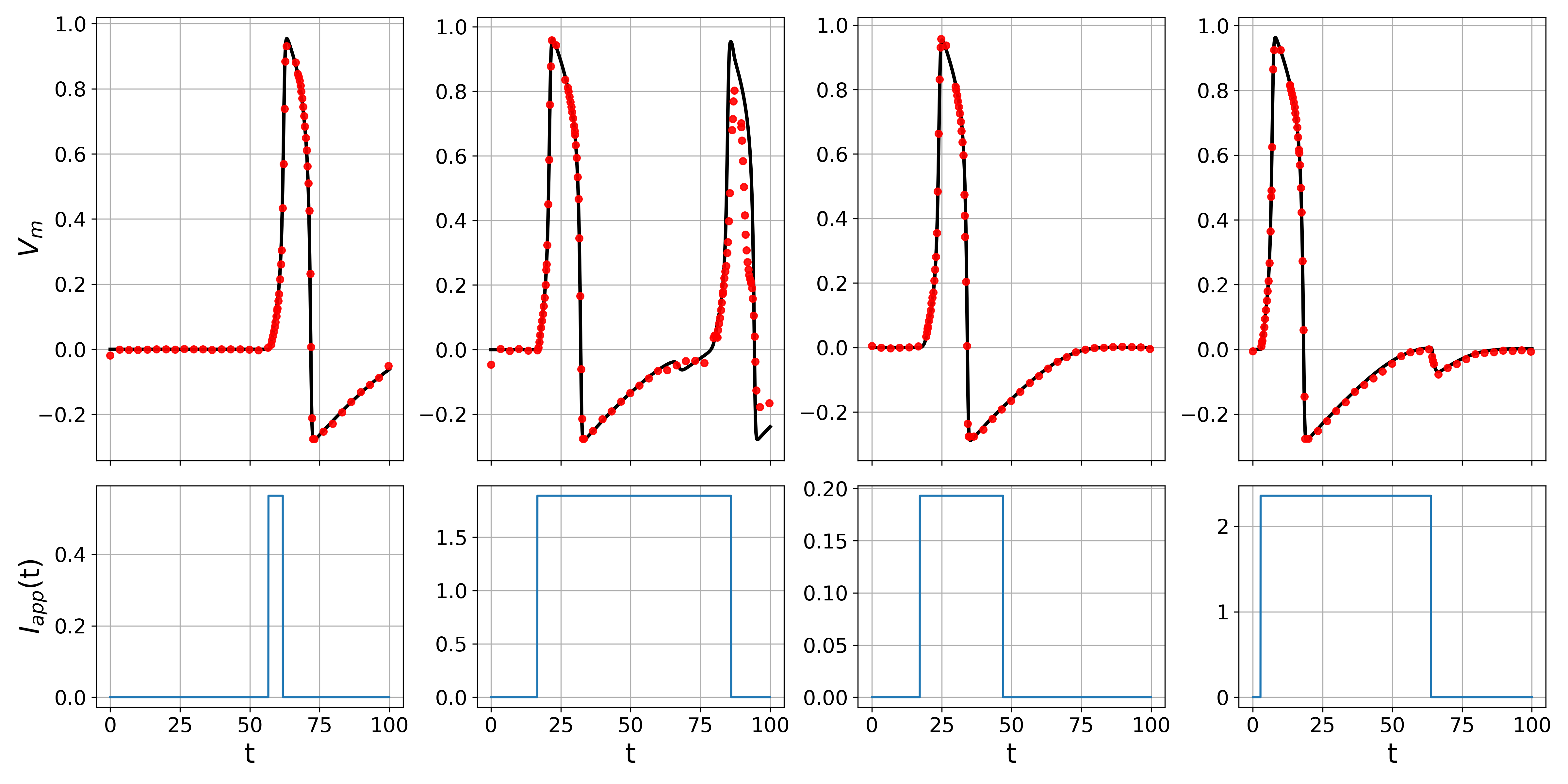}
        \caption{$N=900$}
    \end{subfigure}
    \hfill
    \begin{subfigure}[b]{0.32\textwidth}
        \centering
        \includegraphics[width=\textwidth]{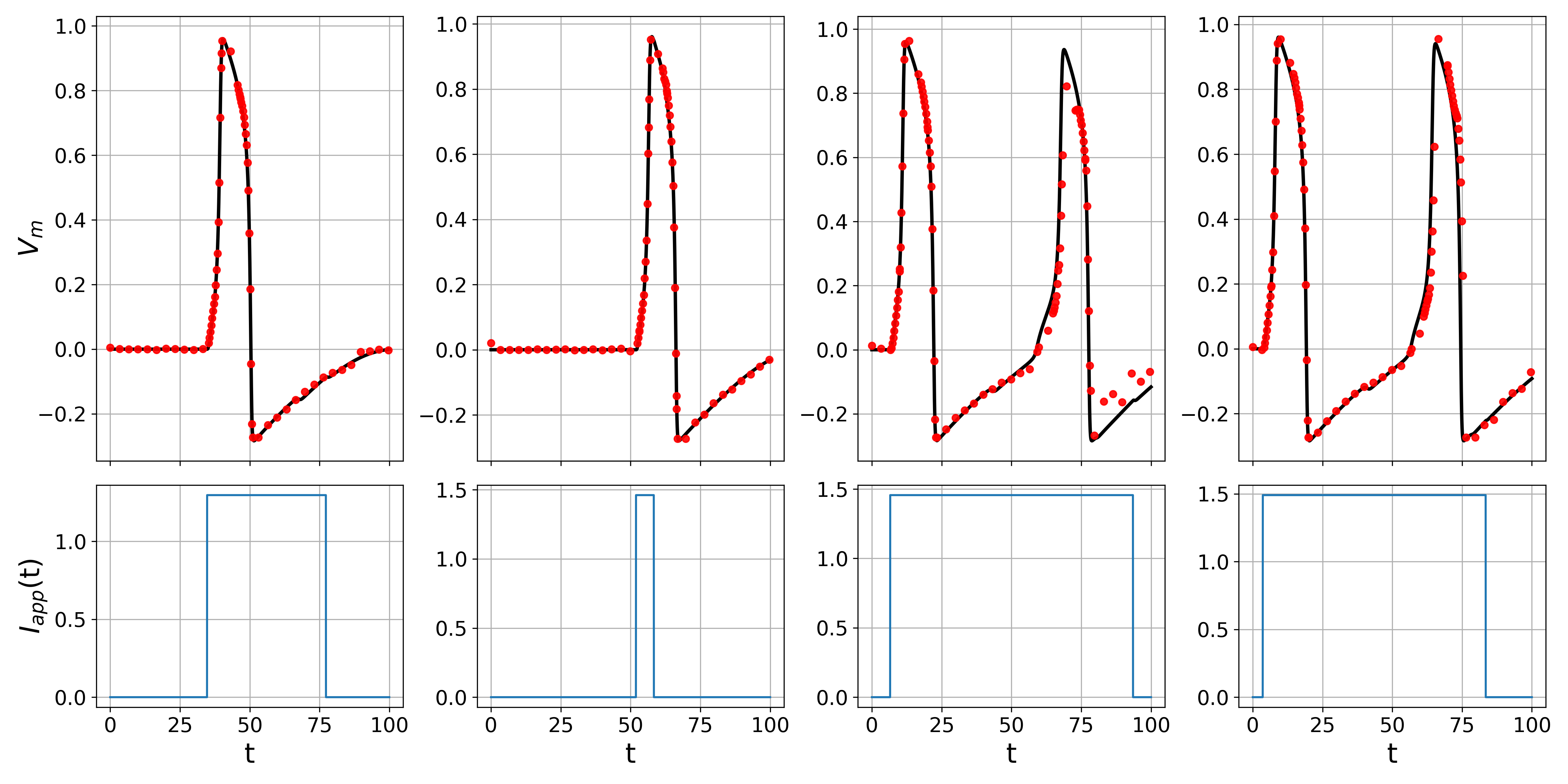}
        \caption{$N=1000$}
    \end{subfigure}
    \hfill
    \begin{subfigure}[b]{0.32\textwidth}
        \centering
        \includegraphics[width=\textwidth]{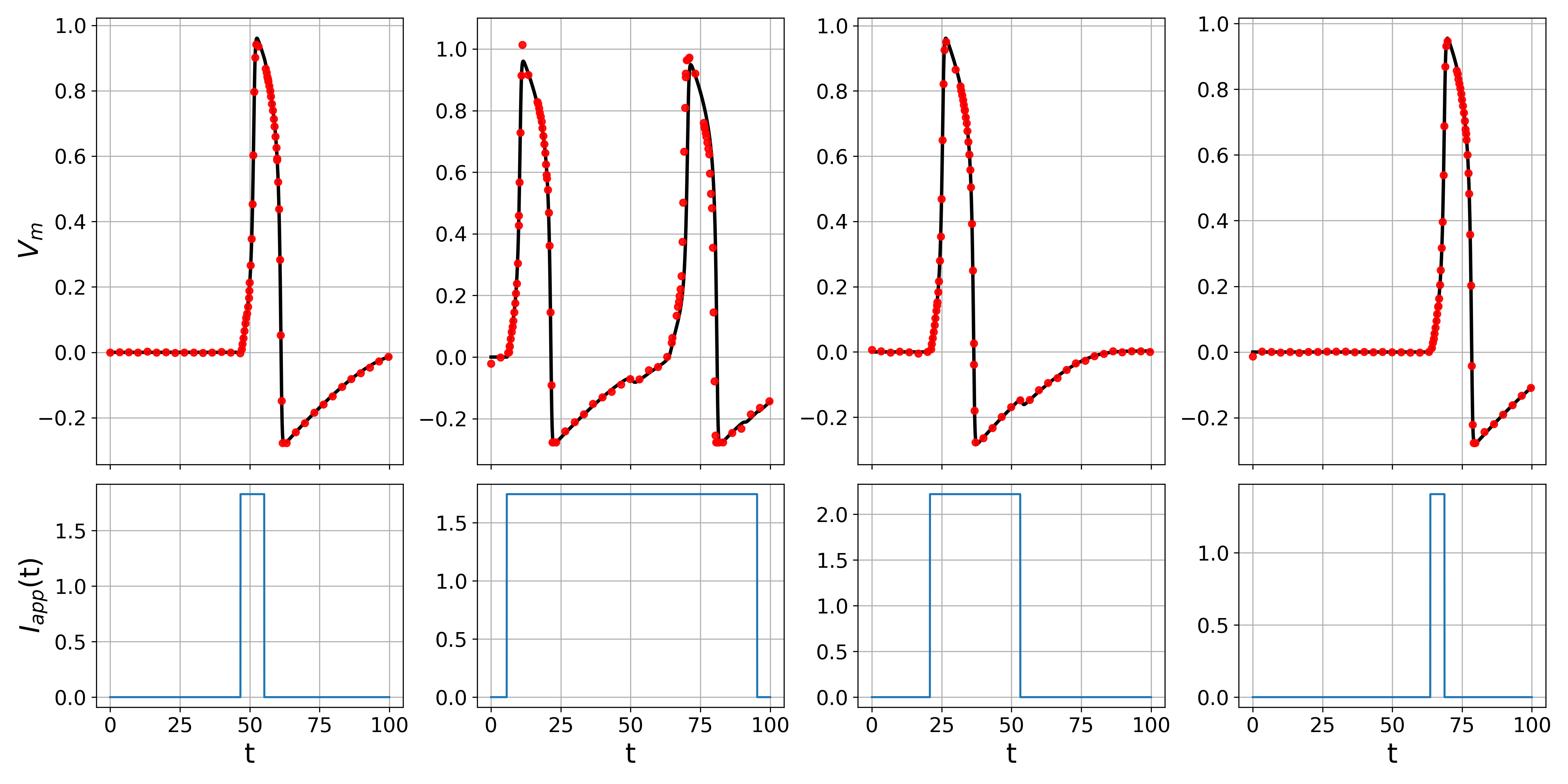}
        \caption{$N=2000$}
    \end{subfigure}
    
    \caption{Some instances of the test set for different Erd\H{o}s-R\'enyi network sizes ($N$).  For each column we can see the applied input current (bottom) with its corresponding potential response (top): the continuous line represents the result of the simulations, whereas the red dots indicate the output of the NN.}
    \label{fig:er_peaks_grid}
\end{figure}


\begin{figure}[htbp]
    \centering
    \begin{subfigure}[b]{0.29\textwidth}
        \centering
        \includegraphics[width=\textwidth]{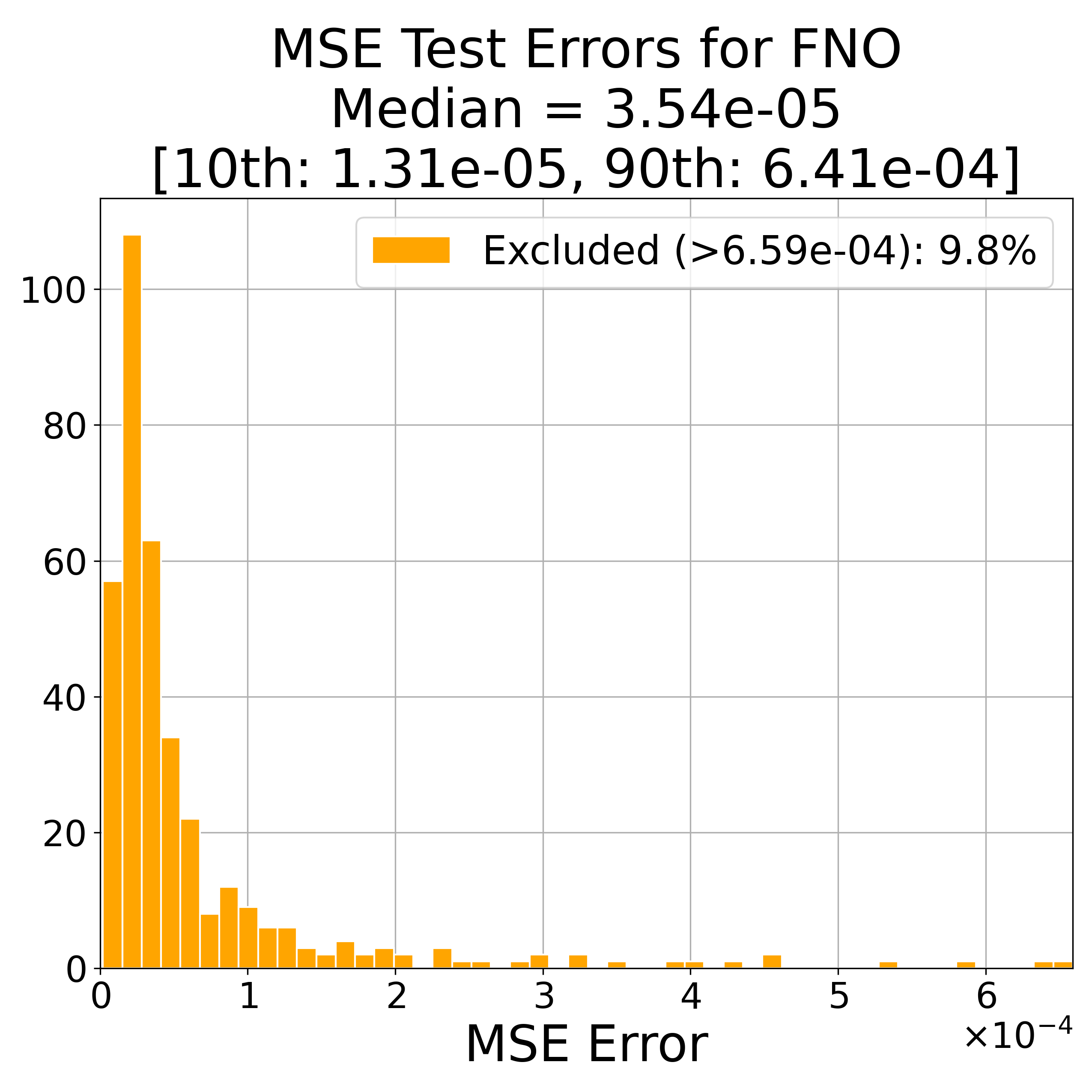}
        \caption{$N=10$}
    \end{subfigure}
    \hfill
    \begin{subfigure}[b]{0.29\textwidth}
        \centering
        \includegraphics[width=\textwidth]{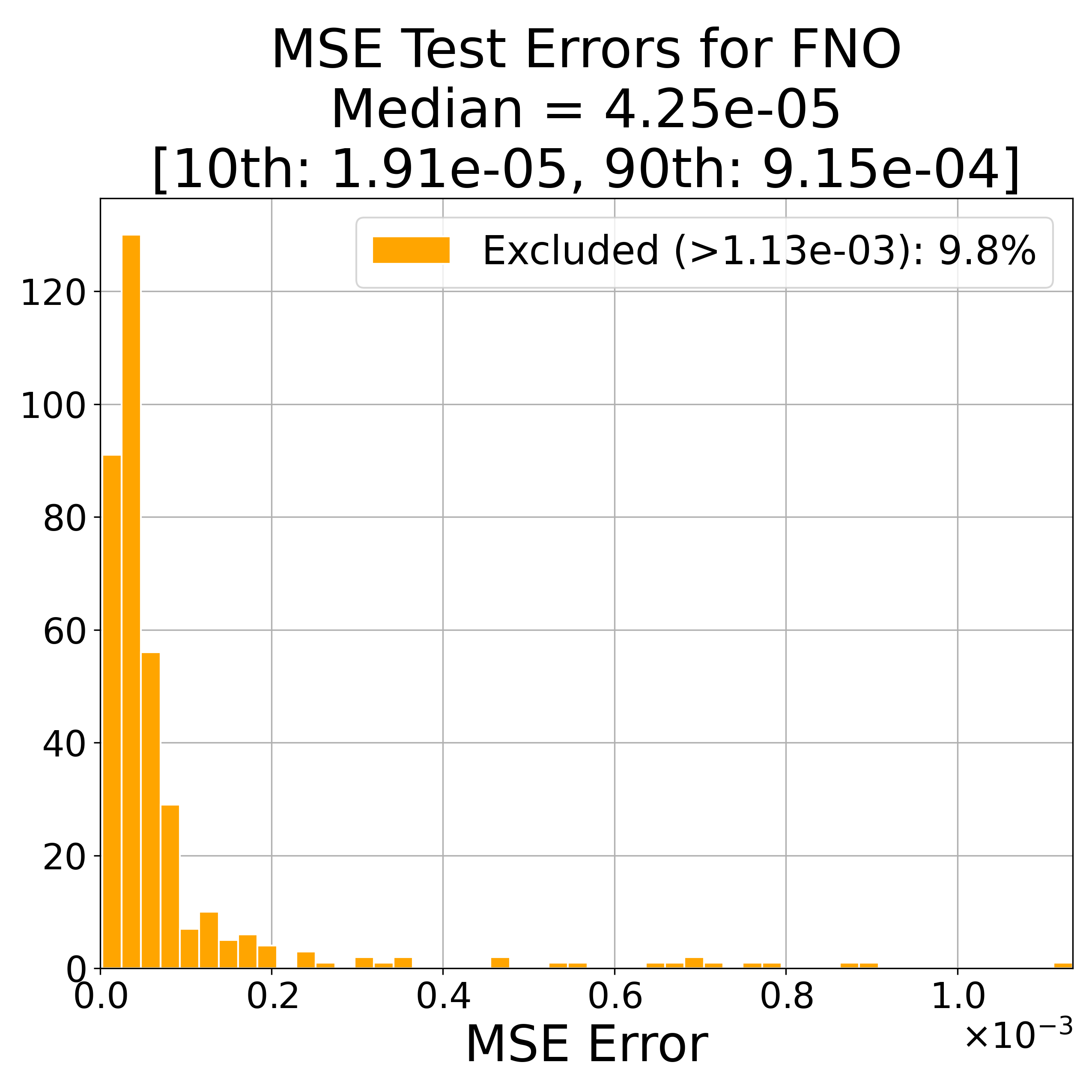}
        \caption{$N=100$}
    \end{subfigure}
    \hfill
    \begin{subfigure}[b]{0.29\textwidth}
        \centering
        \includegraphics[width=\textwidth]{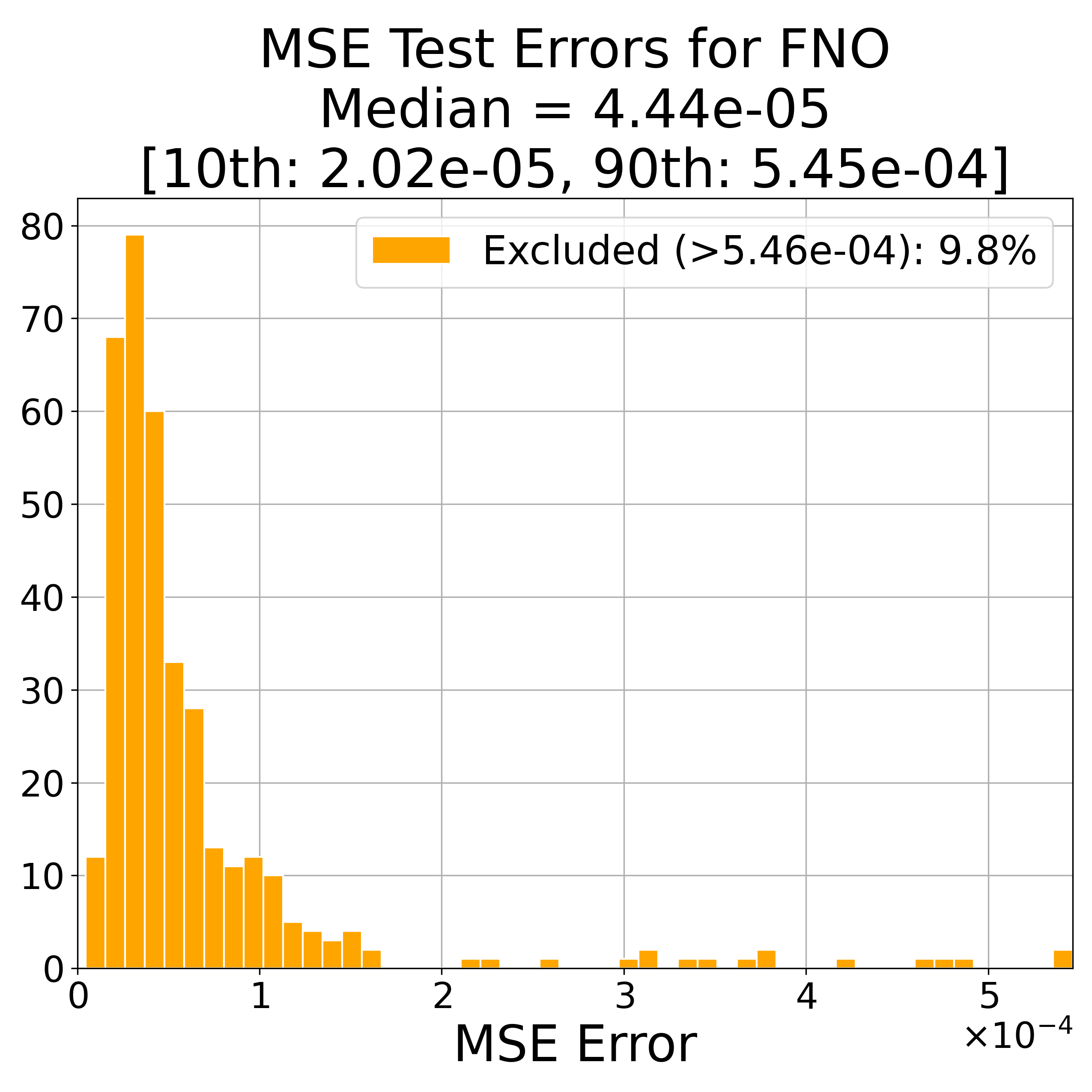}
        \caption{$N=200$}
    \end{subfigure}
    
    \vspace{0.1cm} 
    
    \begin{subfigure}[b]{0.29\textwidth}
        \centering
        \includegraphics[width=\textwidth]{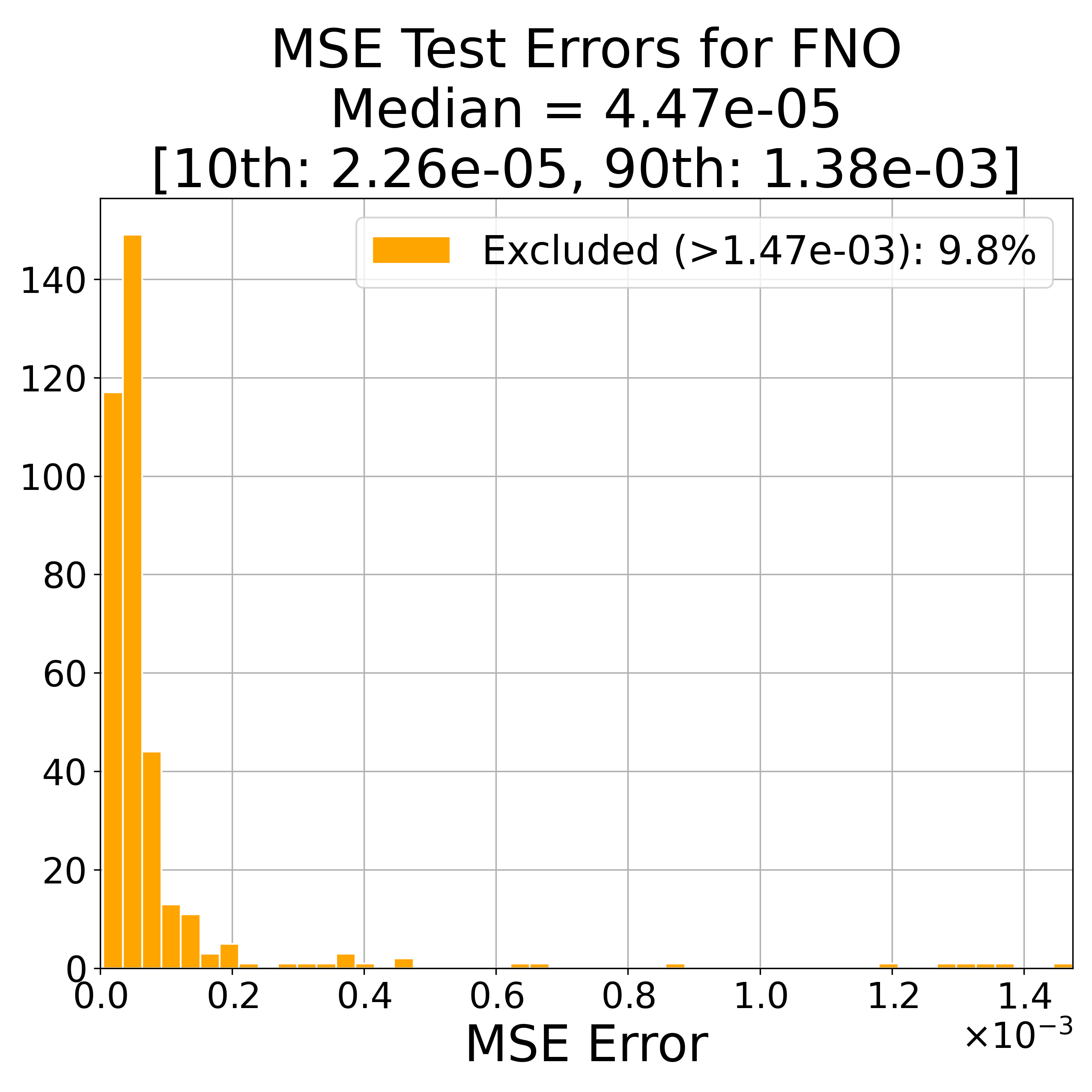}
        \caption{$N=300$}
    \end{subfigure}
    \hfill
    \begin{subfigure}[b]{0.29\textwidth}
        \centering
        \includegraphics[width=\textwidth]{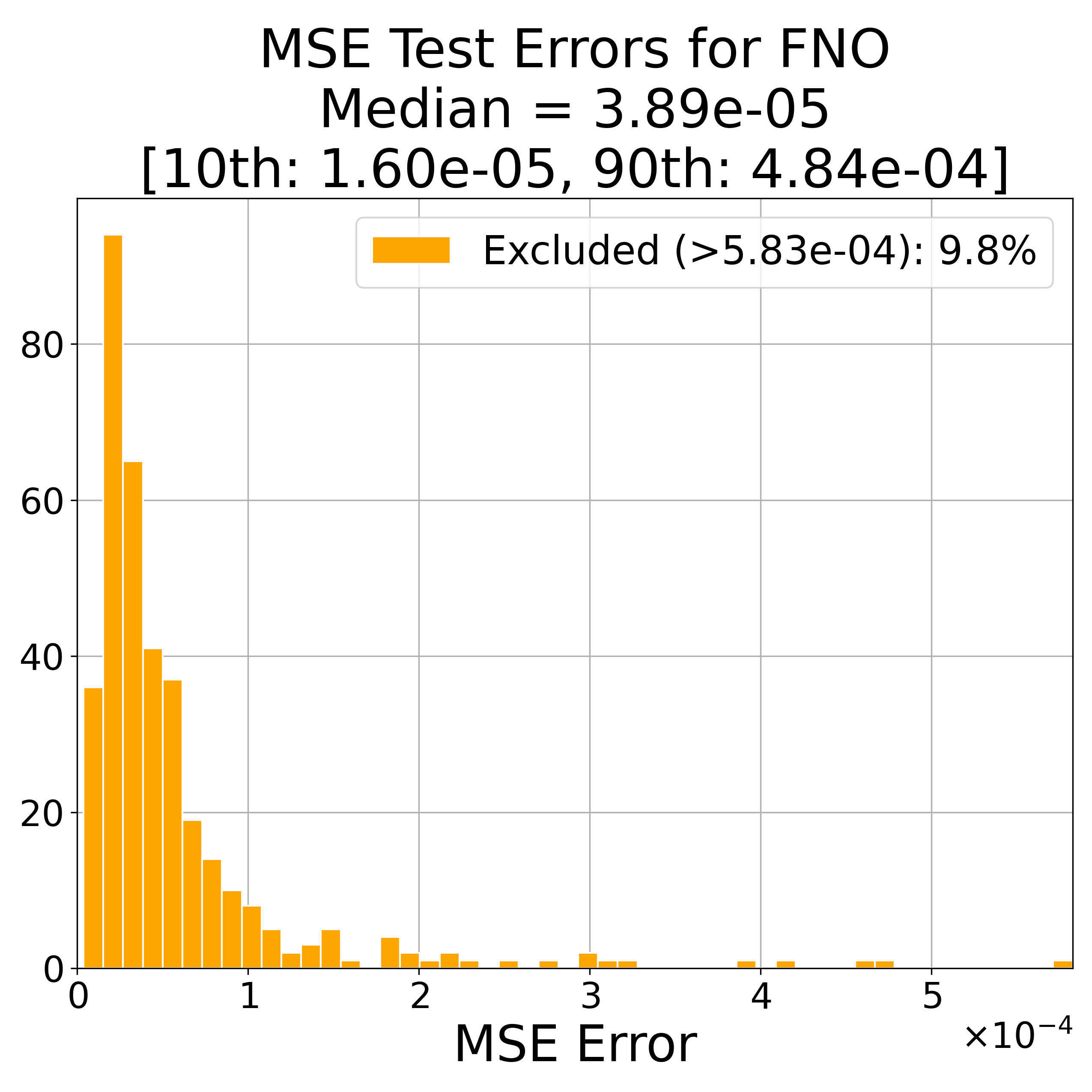}
        \caption{$N=400$}
    \end{subfigure}
    \hfill
    \begin{subfigure}[b]{0.29\textwidth}
        \centering
        \includegraphics[width=\textwidth]{ER_Network_FNO_L2histoFNO_N=500.png}
        \caption{$N=500$}
    \end{subfigure}
    
    \vspace{0.1cm} 
    
    \begin{subfigure}[b]{0.29\textwidth}
        \centering
        \includegraphics[width=\textwidth]{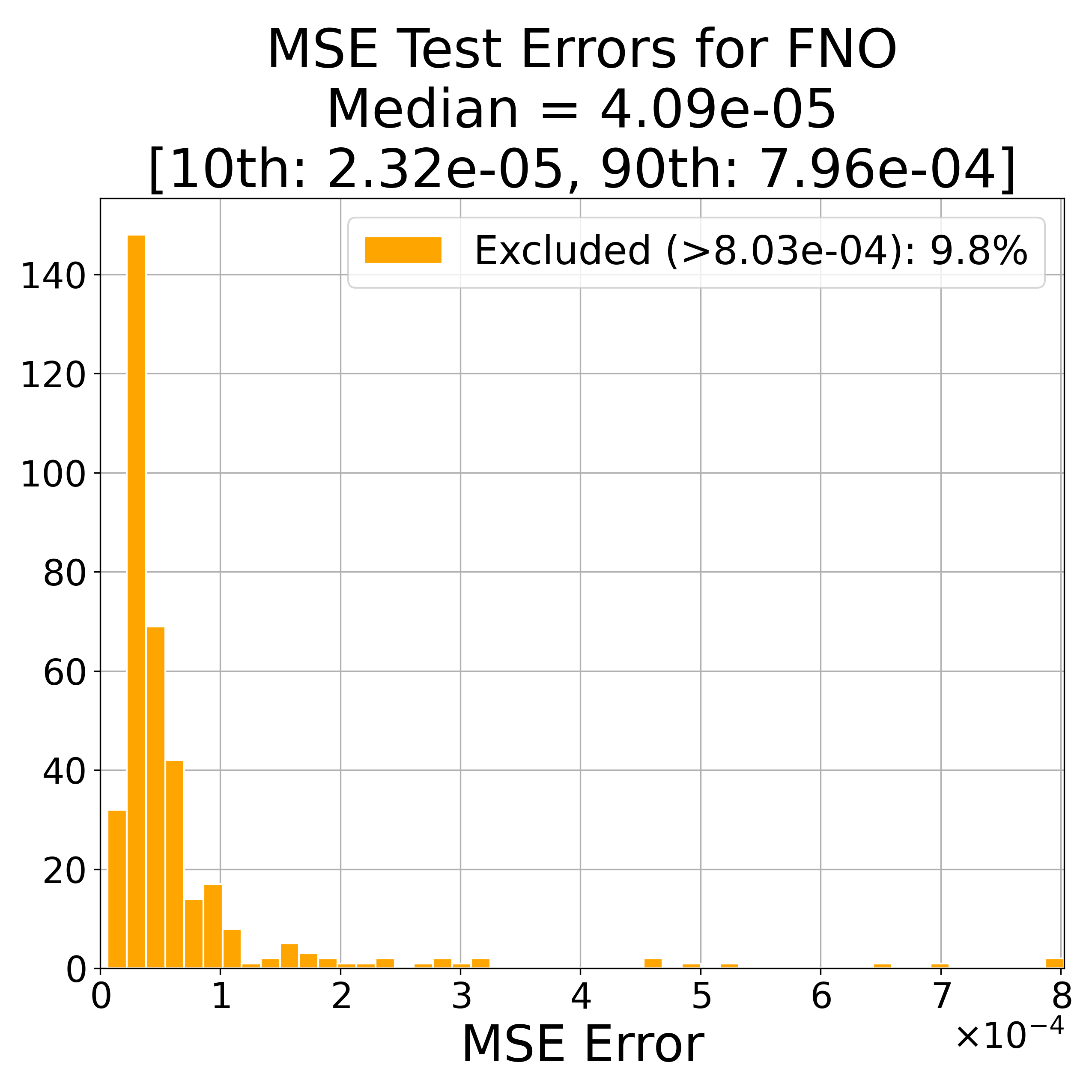}
        \caption{$N=600$}
    \end{subfigure}
    \hfill
    \begin{subfigure}[b]{0.29\textwidth}
        \centering
        \includegraphics[width=\textwidth]{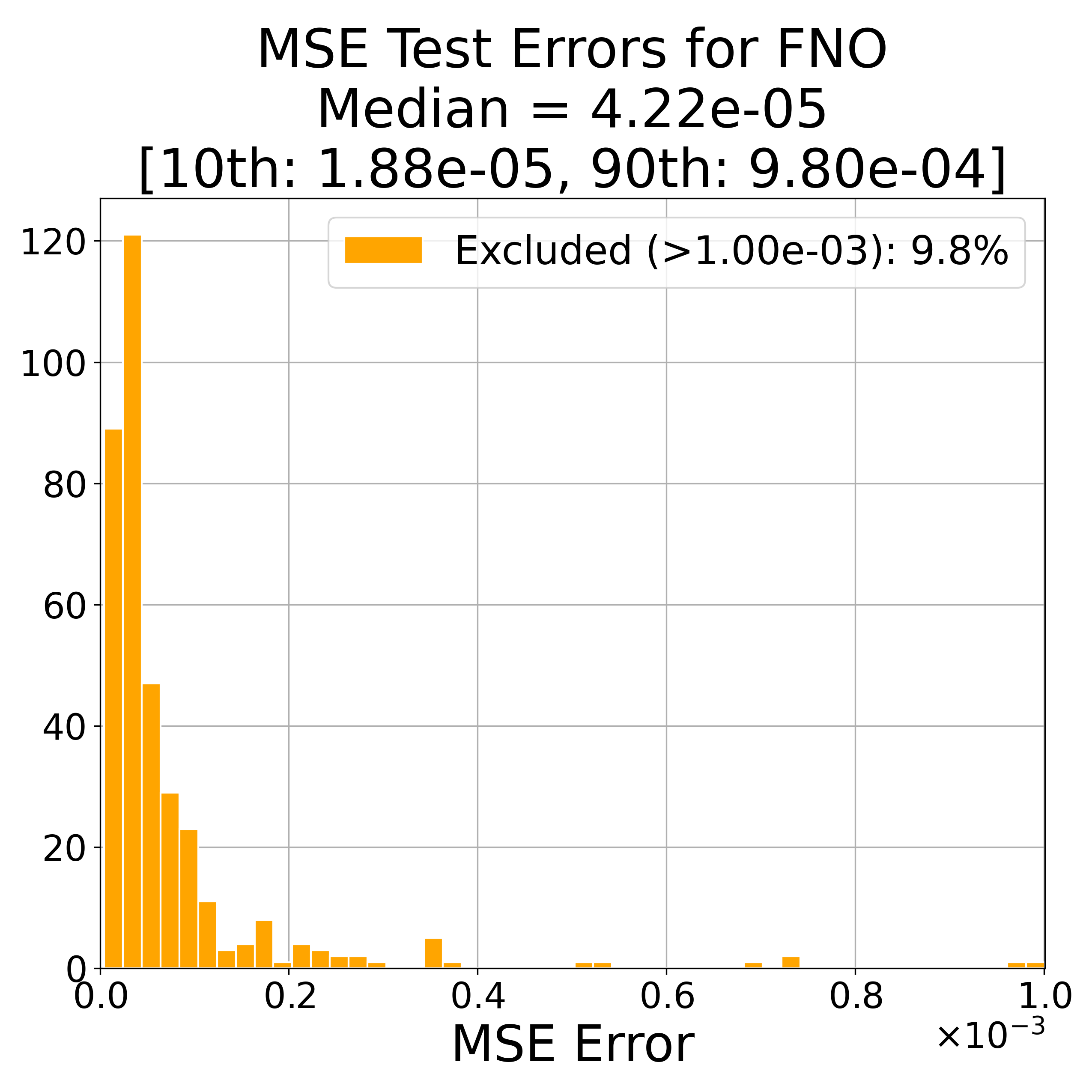}
        \caption{$N=700$}
    \end{subfigure}
    \hfill
    \begin{subfigure}[b]{0.29\textwidth}
        \centering
        \includegraphics[width=\textwidth]{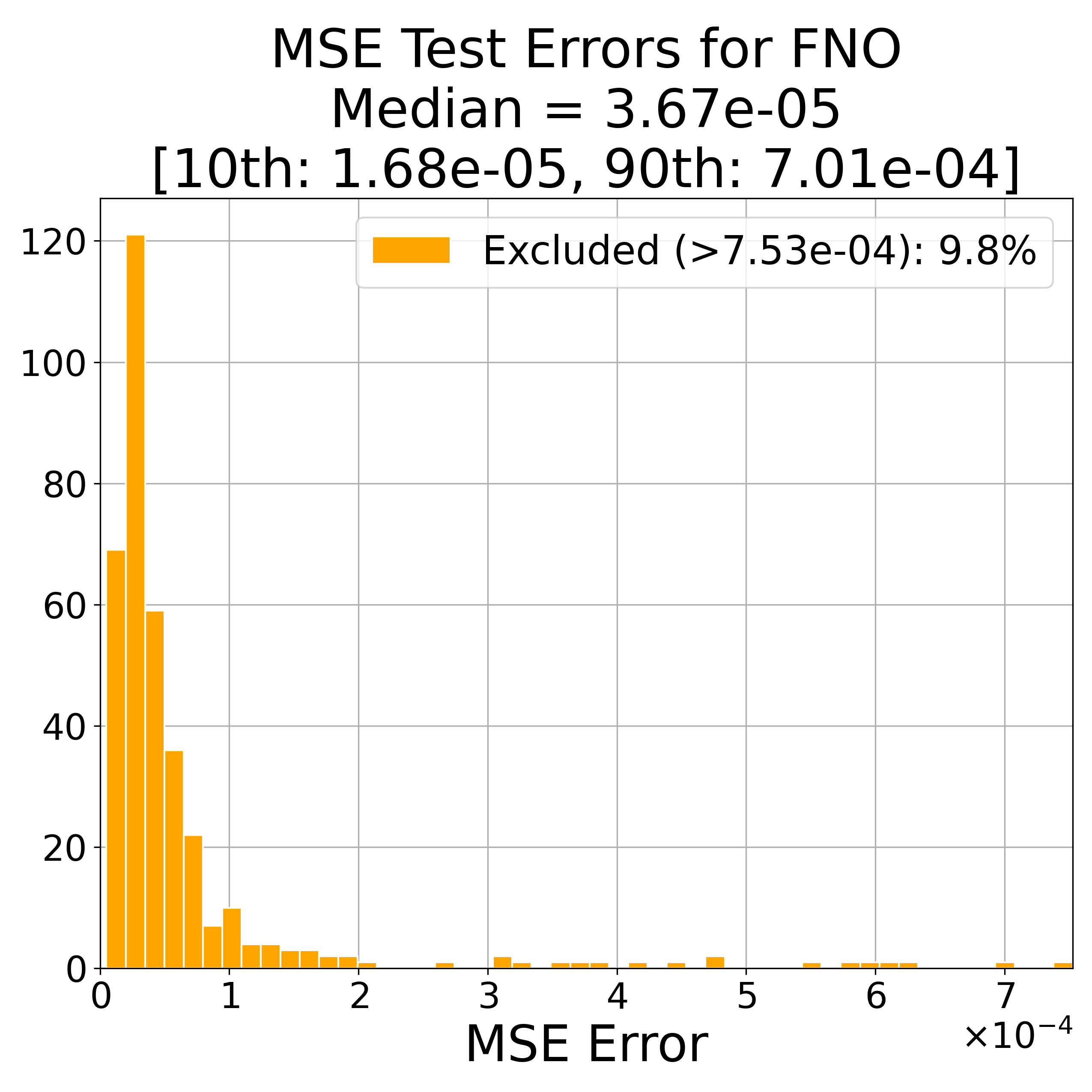}
        \caption{$N=800$}
    \end{subfigure}
    
    \vspace{0.1cm} 
    
    \begin{subfigure}[b]{0.29\textwidth}
        \centering
        \includegraphics[width=\textwidth]{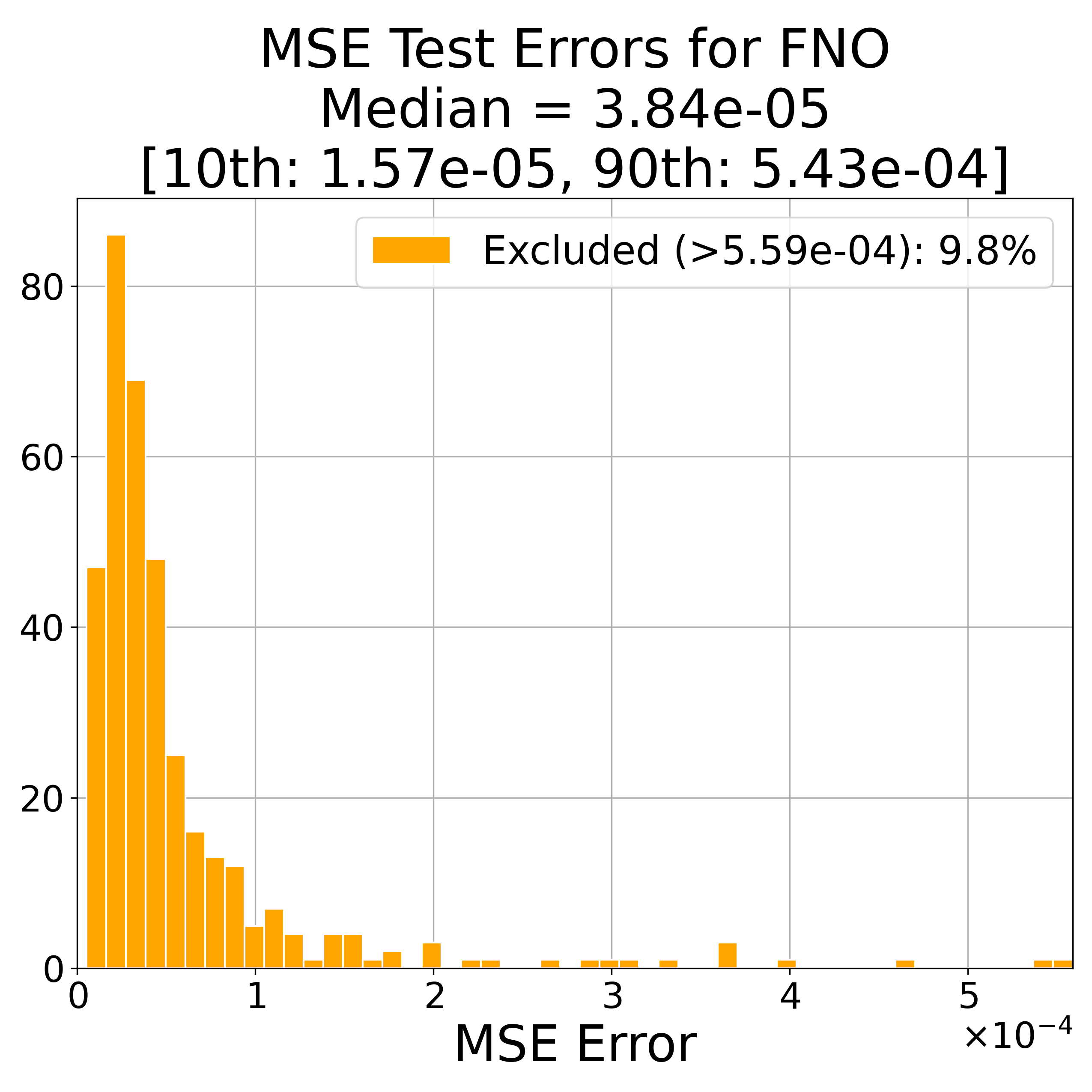}
        \caption{$N=900$}
    \end{subfigure}
    \hfill
    \begin{subfigure}[b]{0.29\textwidth}
        \centering
        \includegraphics[width=\textwidth]{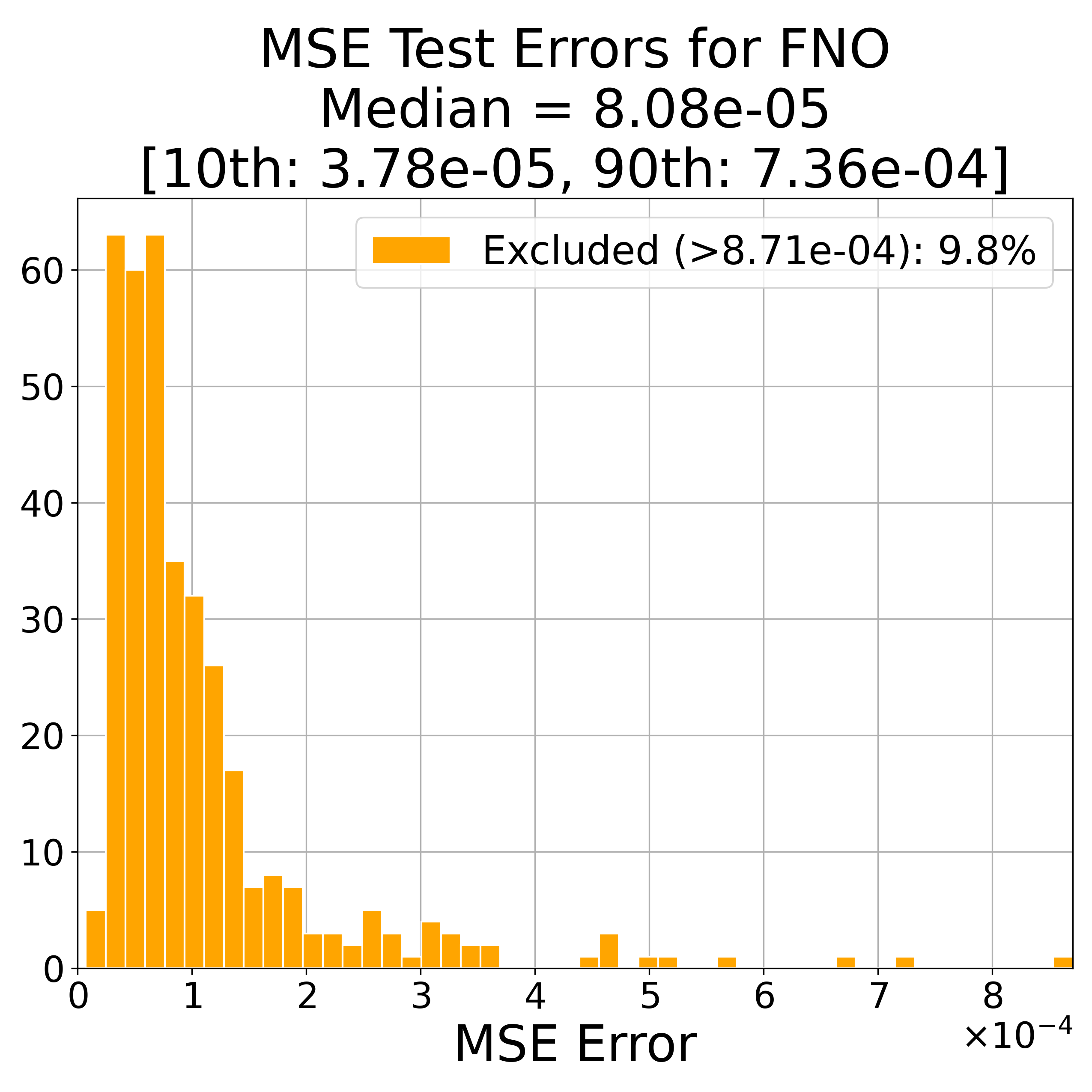}
        \caption{$N=1000$}
    \end{subfigure}
    \hfill
    \begin{subfigure}[b]{0.29\textwidth}
        \centering
        \includegraphics[width=\textwidth]{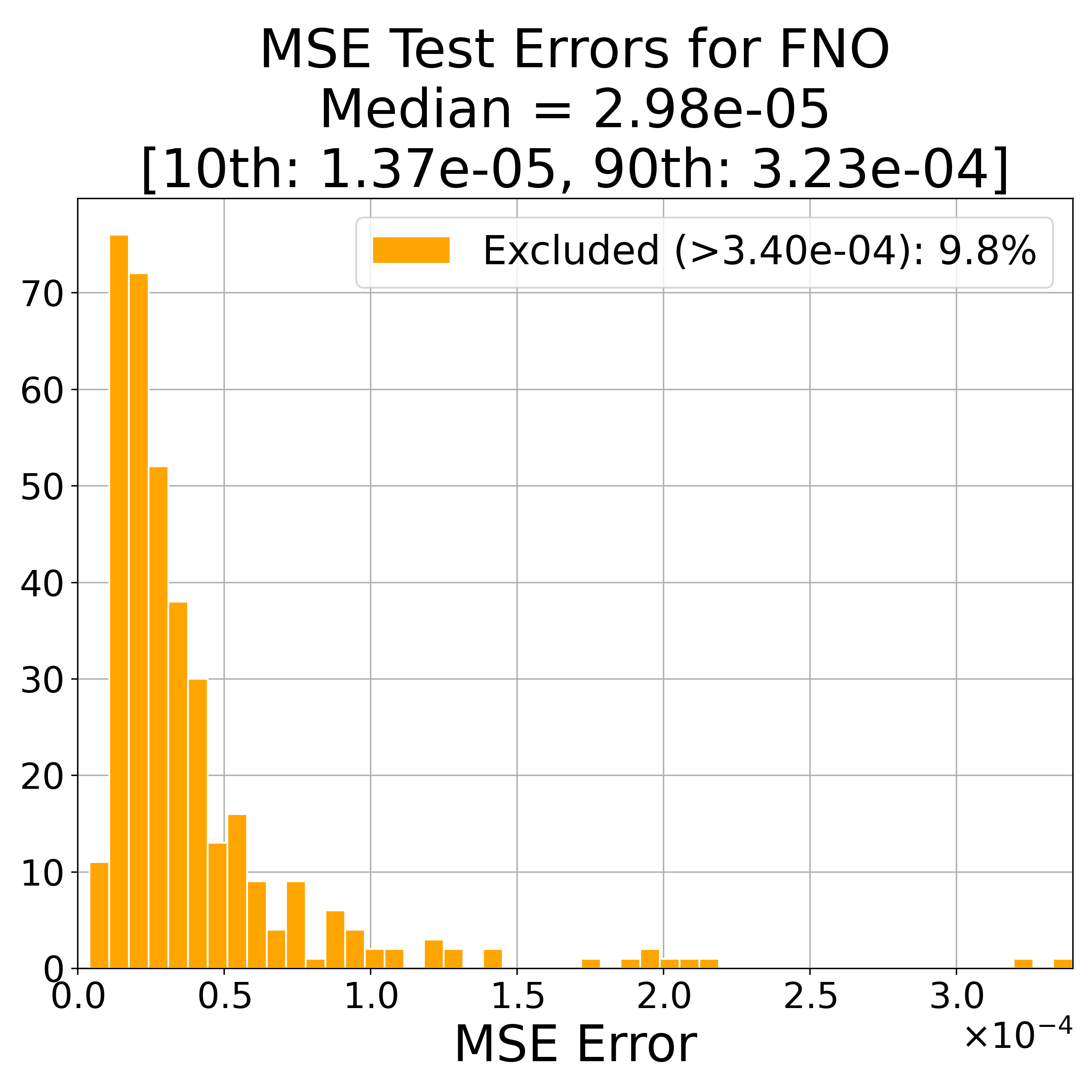}
        \caption{$N=2000$}
    \end{subfigure}
    
    \caption{MSE distributions across the test set for different Erd\H{o}s-R\'enyi network sizes ($N$).}
    \label{fig:er_histo_grid}
\end{figure}

\begin{figure}[htbp]
    \centering
    \begin{subfigure}[b]{0.32\textwidth}
        \centering
        \includegraphics[width=\textwidth]{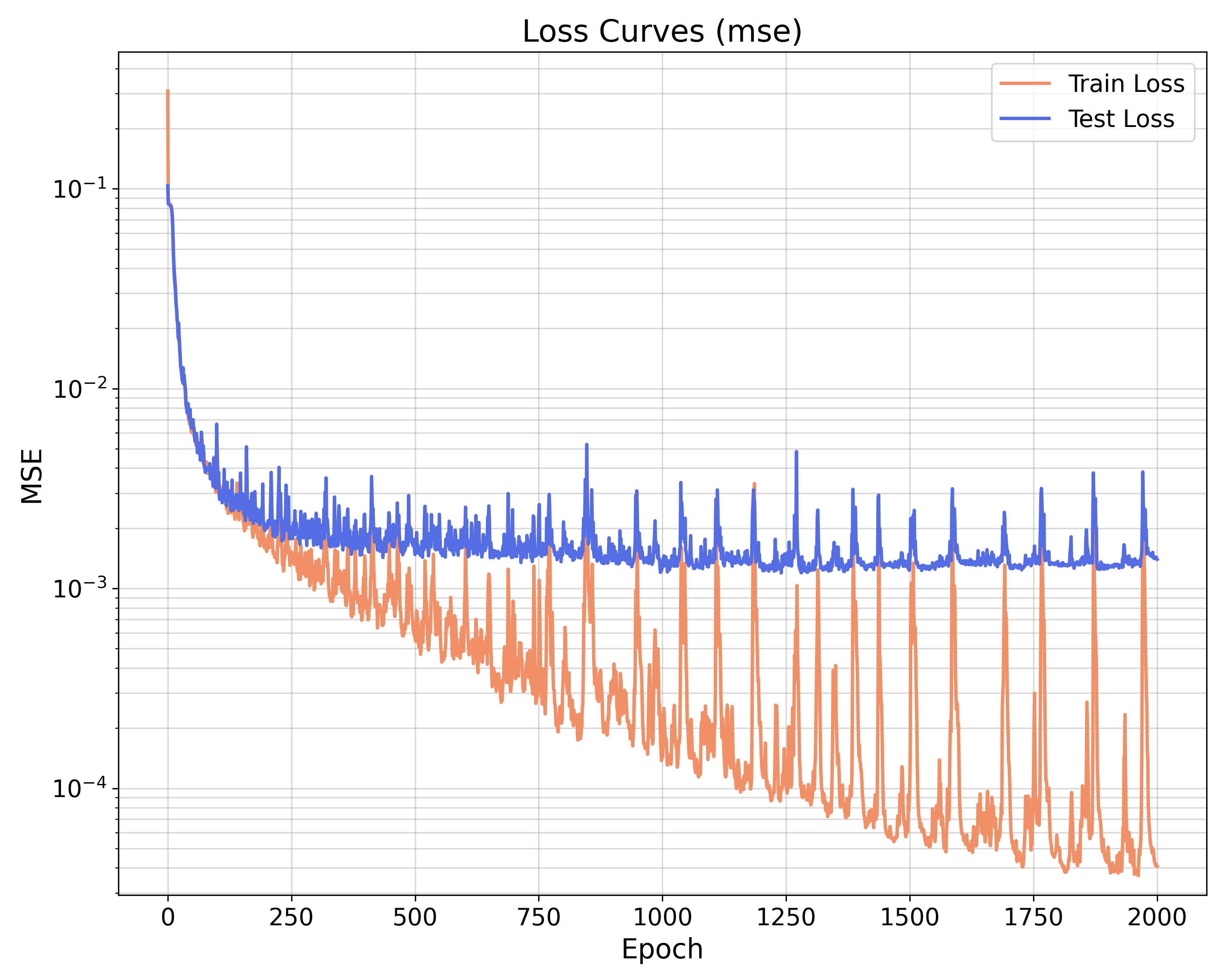}
        \caption{$N=10$}
    \end{subfigure}
    \hfill
    \begin{subfigure}[b]{0.32\textwidth}
        \centering
        \includegraphics[width=\textwidth]{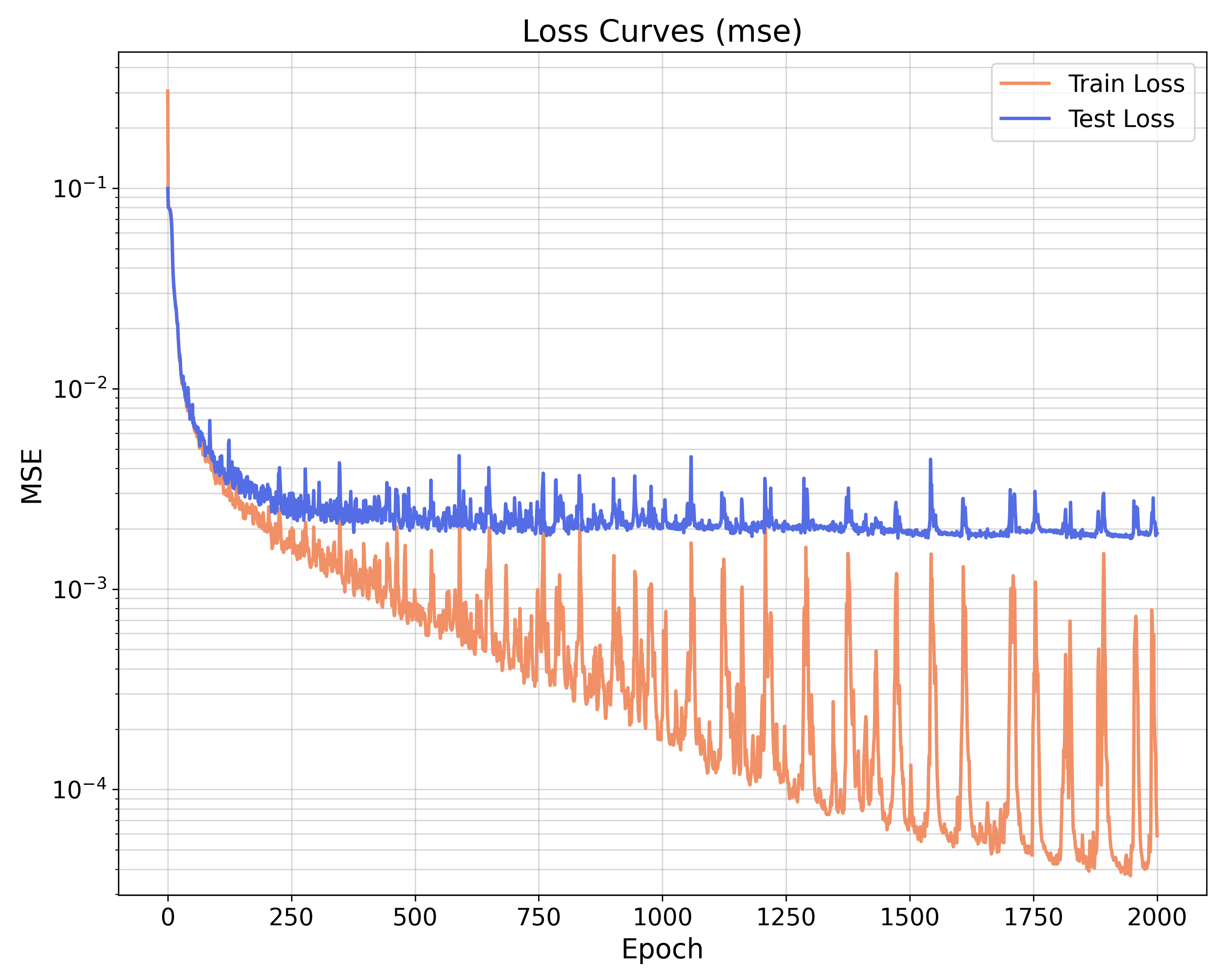}
        \caption{$N=100$}
    \end{subfigure}
    \hfill
    \begin{subfigure}[b]{0.32\textwidth}
        \centering
        \includegraphics[width=\textwidth]{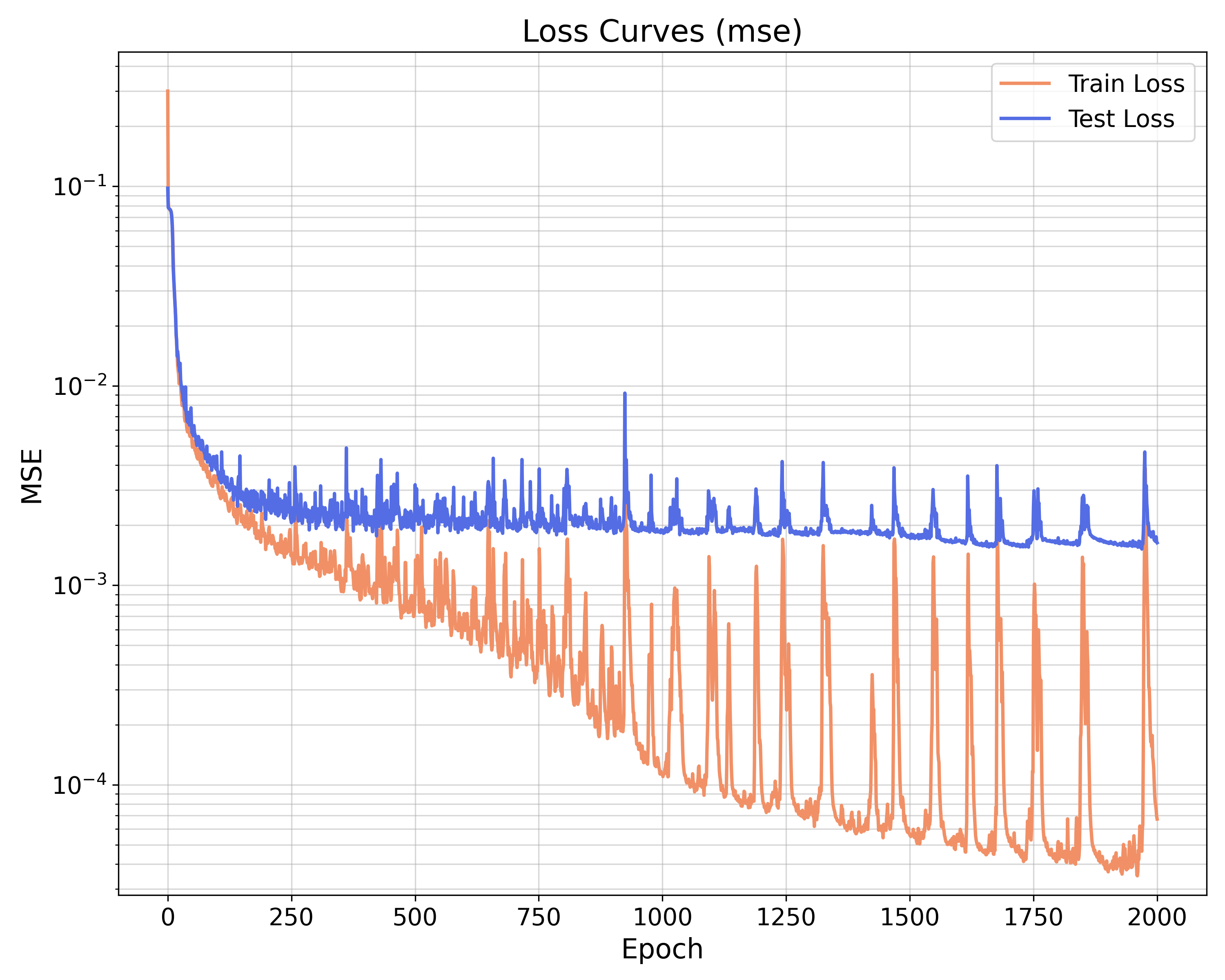}
        \caption{$N=200$}
    \end{subfigure}
    
    \vspace{0.4cm}
    
    \begin{subfigure}[b]{0.32\textwidth}
        \centering
        \includegraphics[width=\textwidth]{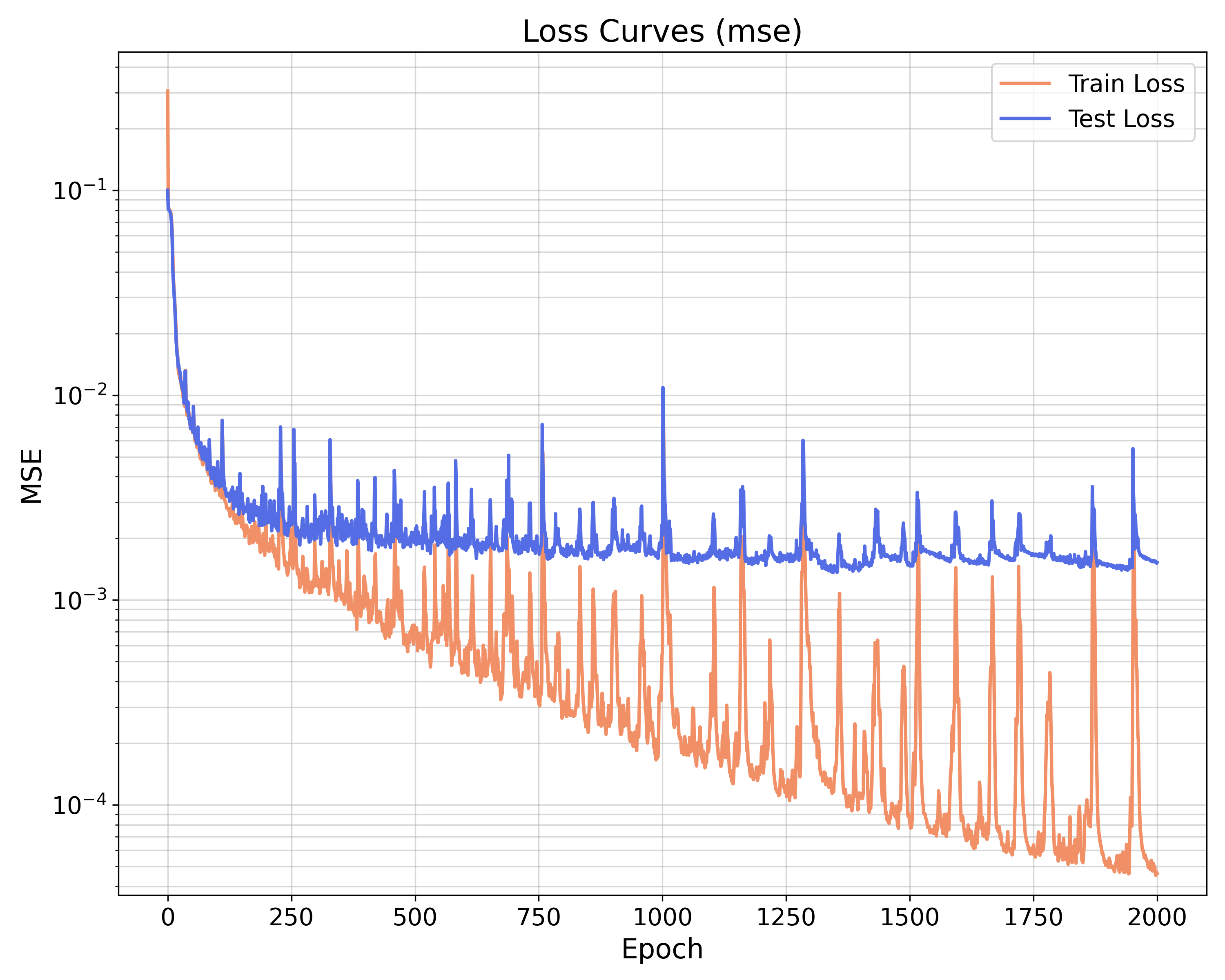}
        \caption{$N=300$}
    \end{subfigure}
    \hfill
    \begin{subfigure}[b]{0.32\textwidth}
        \centering
        \includegraphics[width=\textwidth]{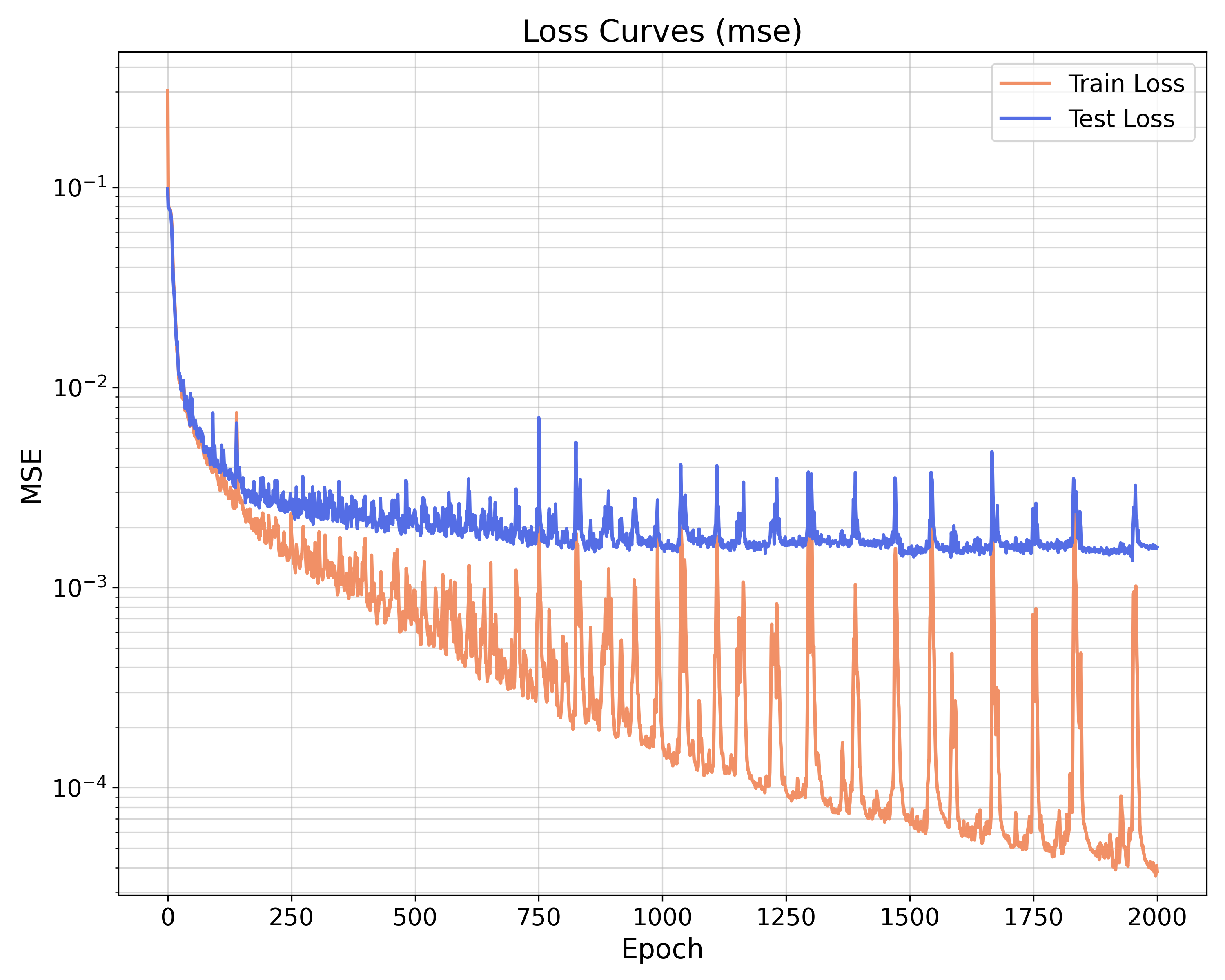}
        \caption{$N=400$}
    \end{subfigure}
    \hfill
    \begin{subfigure}[b]{0.32\textwidth}
        \centering
        \includegraphics[width=\textwidth]{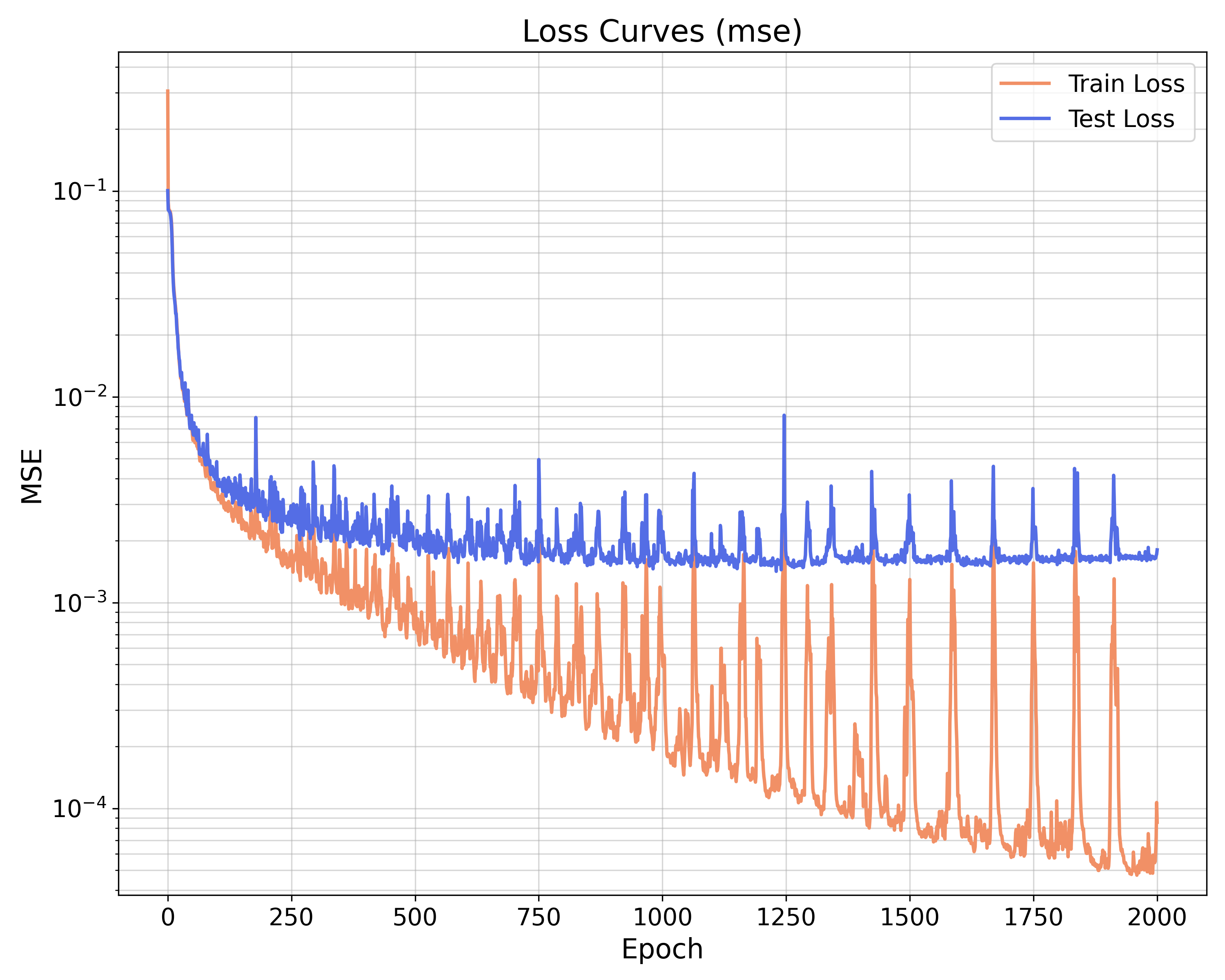}
        \caption{$N=500$}
    \end{subfigure}
    
    \vspace{0.4cm}
    
    \begin{subfigure}[b]{0.32\textwidth}
        \centering
        \includegraphics[width=\textwidth]{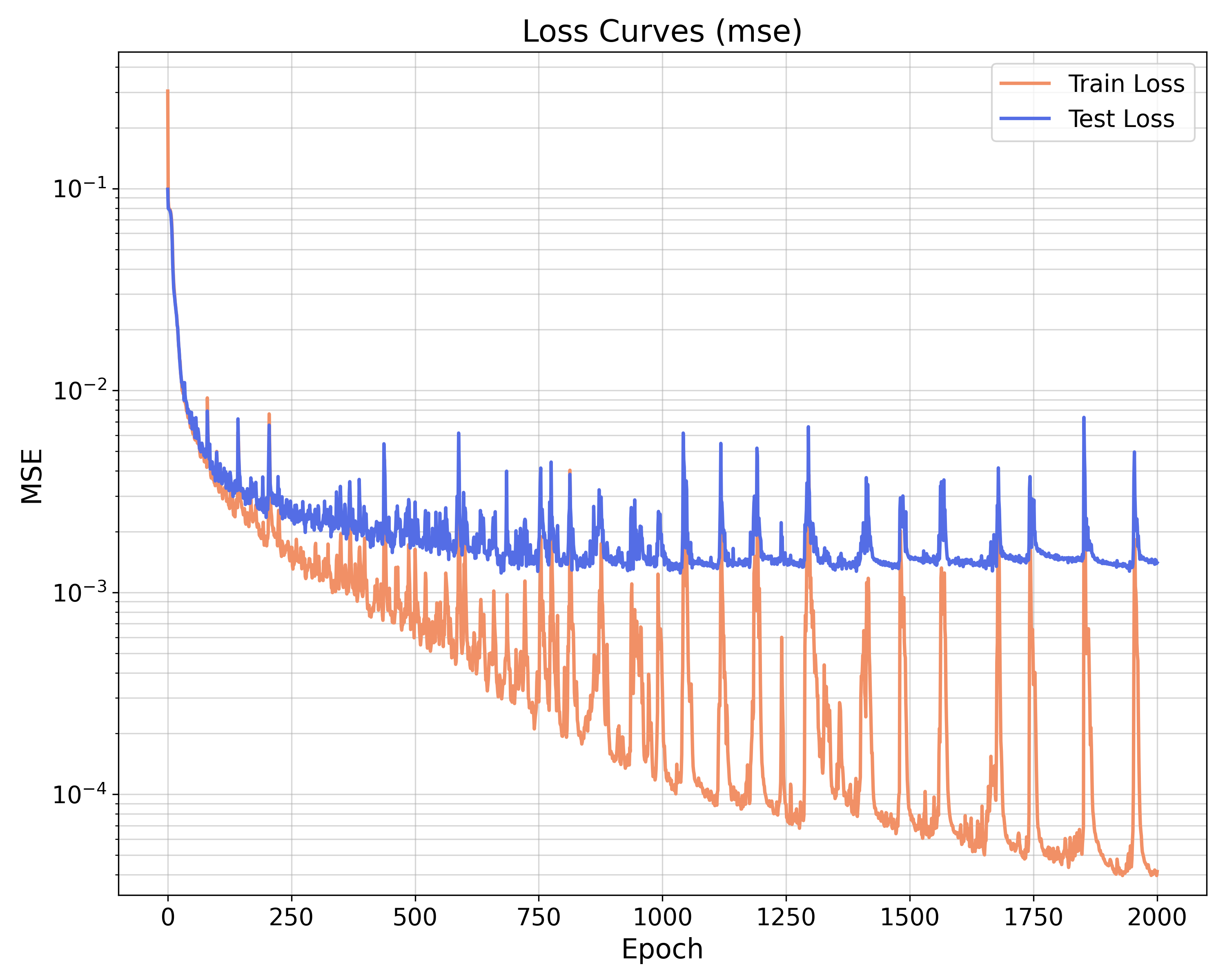}
        \caption{$N=600$}
    \end{subfigure}
    \hfill
    \begin{subfigure}[b]{0.32\textwidth}
        \centering
        \includegraphics[width=\textwidth]{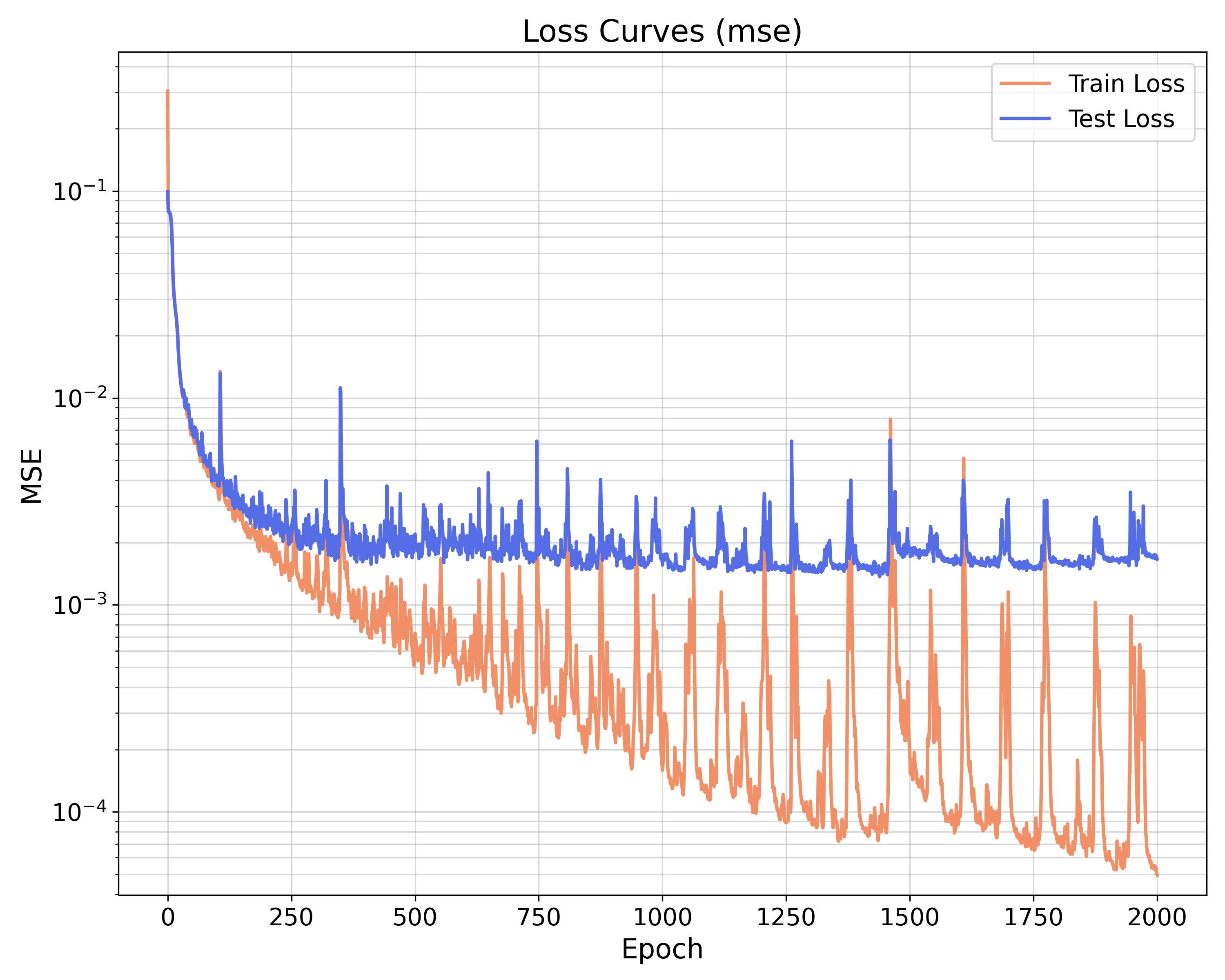}
        \caption{$N=700$}
    \end{subfigure}
    \hfill
    \begin{subfigure}[b]{0.32\textwidth}
        \centering
        \includegraphics[width=\textwidth]{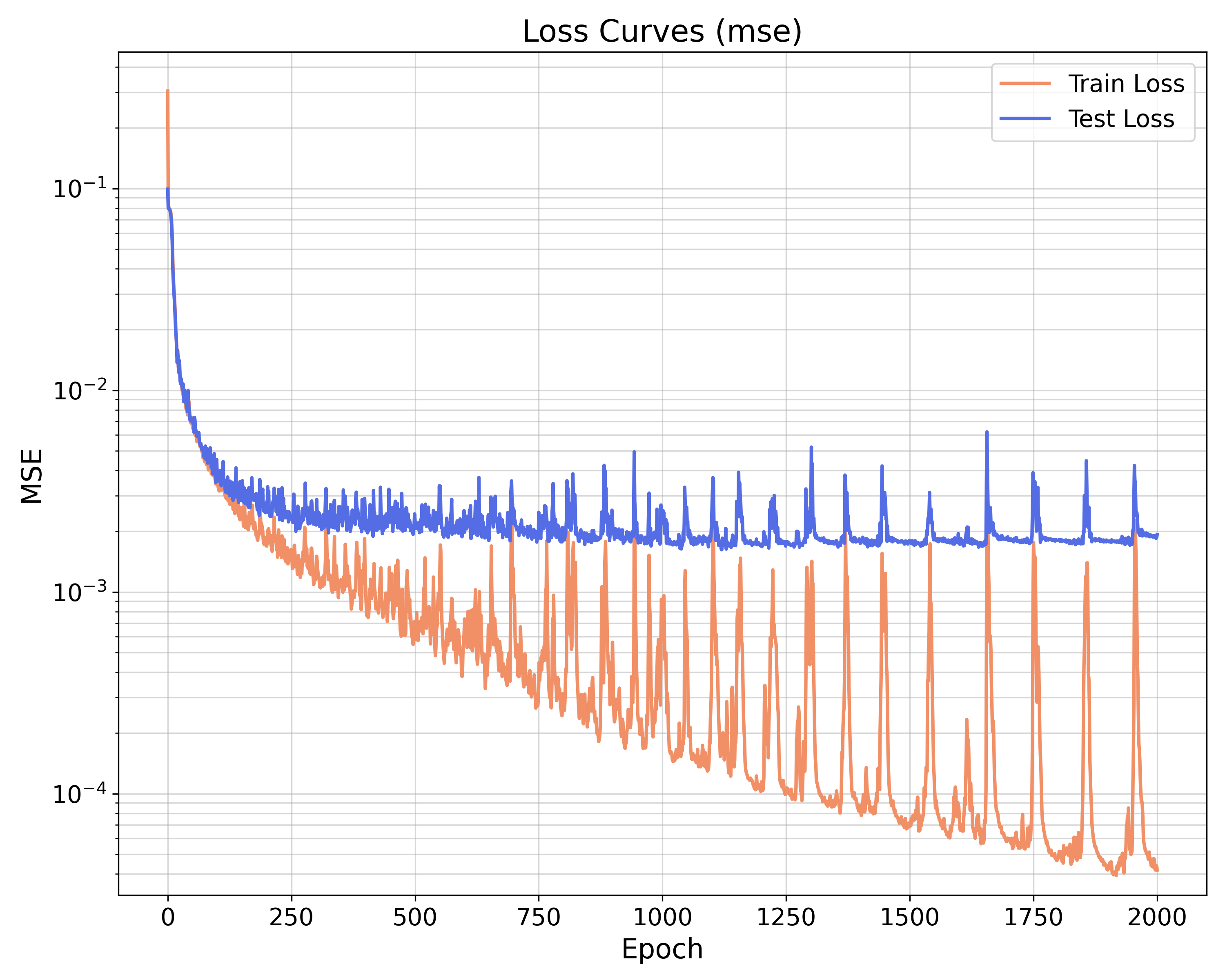}
        \caption{$N=800$}
    \end{subfigure}
    
    \vspace{0.4cm}
    
    \begin{subfigure}[b]{0.32\textwidth}
        \centering
        \includegraphics[width=\textwidth]{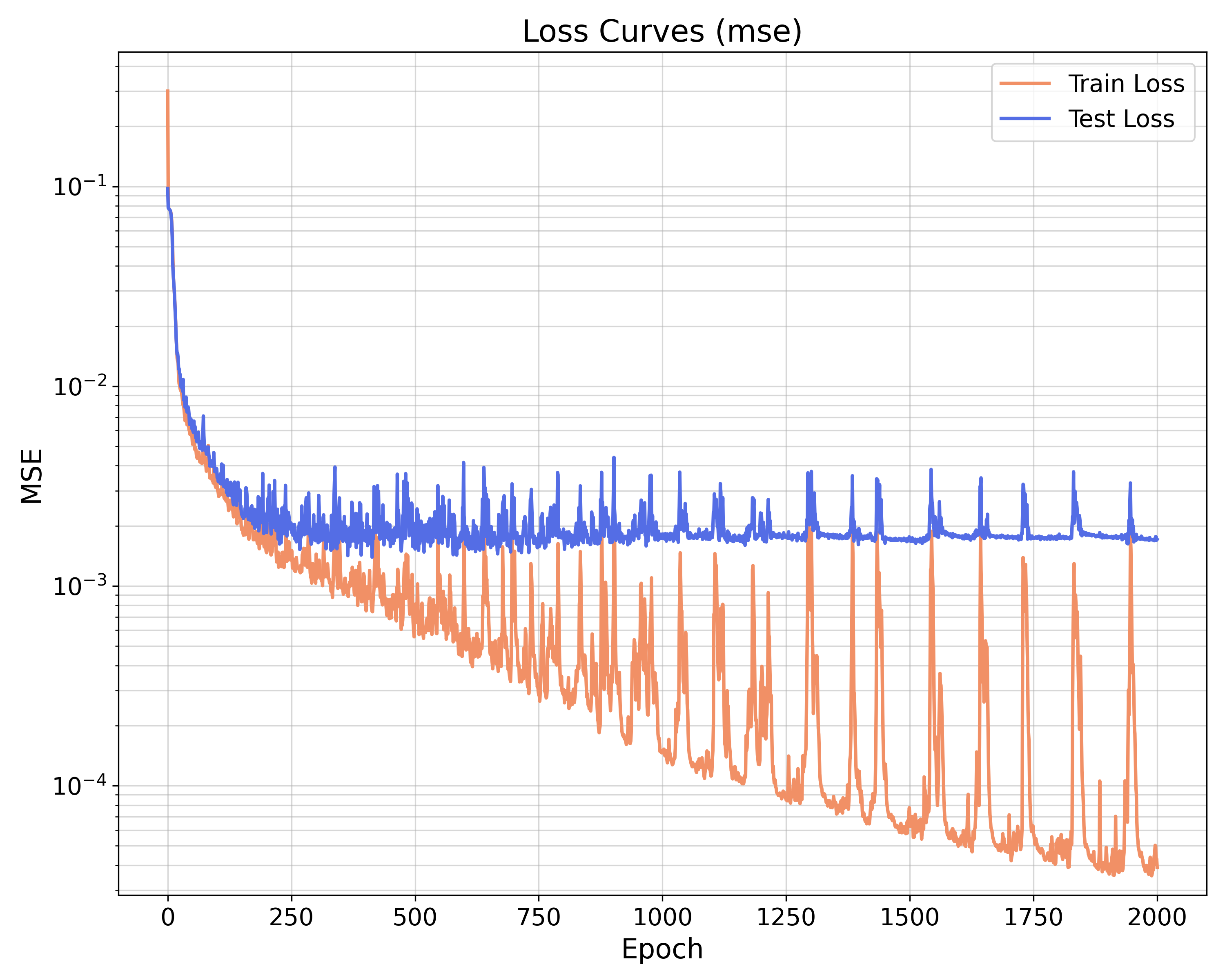}
        \caption{$N=900$}
    \end{subfigure}
    \hfill
    \begin{subfigure}[b]{0.32\textwidth}
        \centering
        \includegraphics[width=\textwidth]{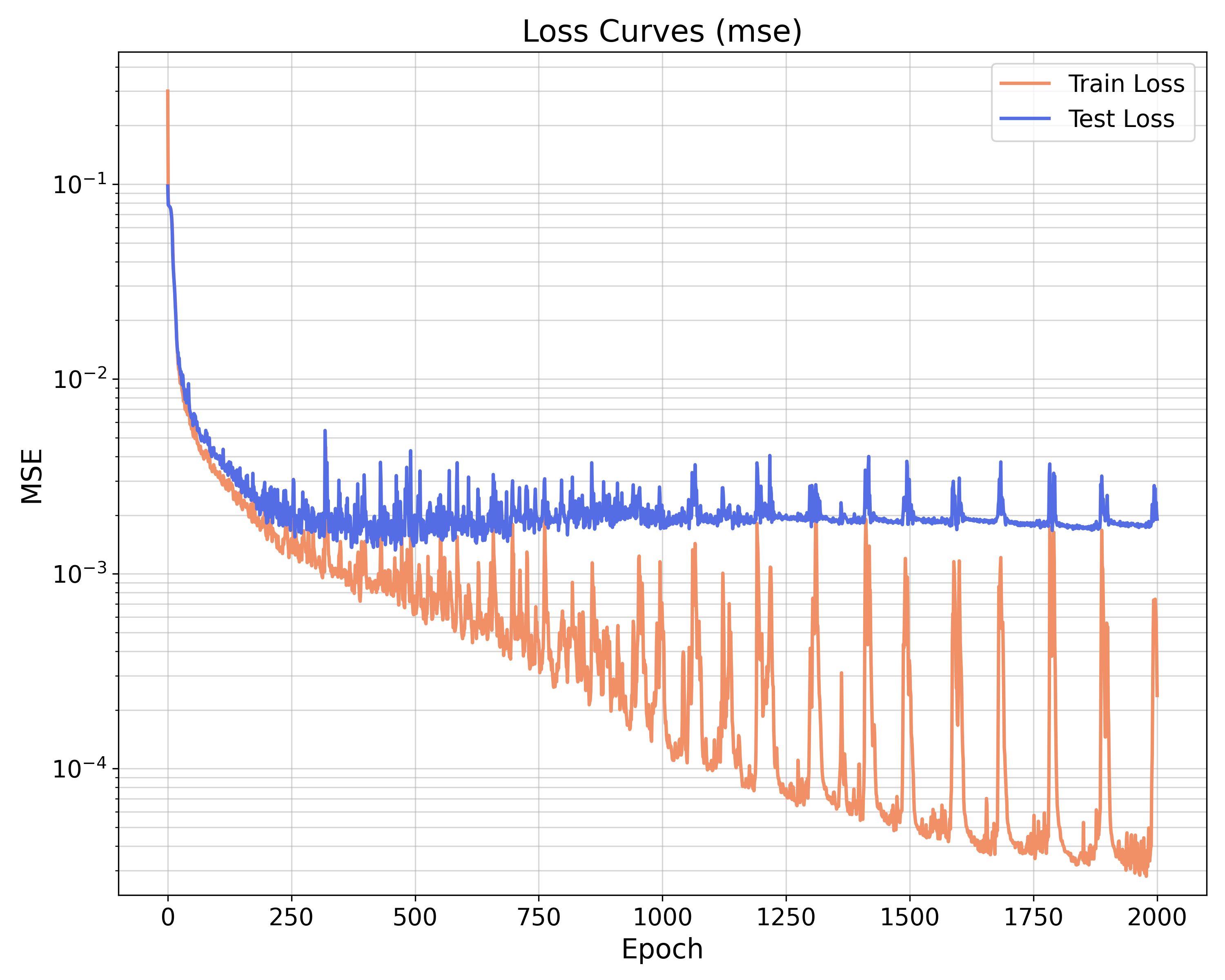}
        \caption{$N=1000$}
    \end{subfigure}
    \hfill
    \begin{subfigure}[b]{0.32\textwidth}
        \centering
        \includegraphics[width=\textwidth]{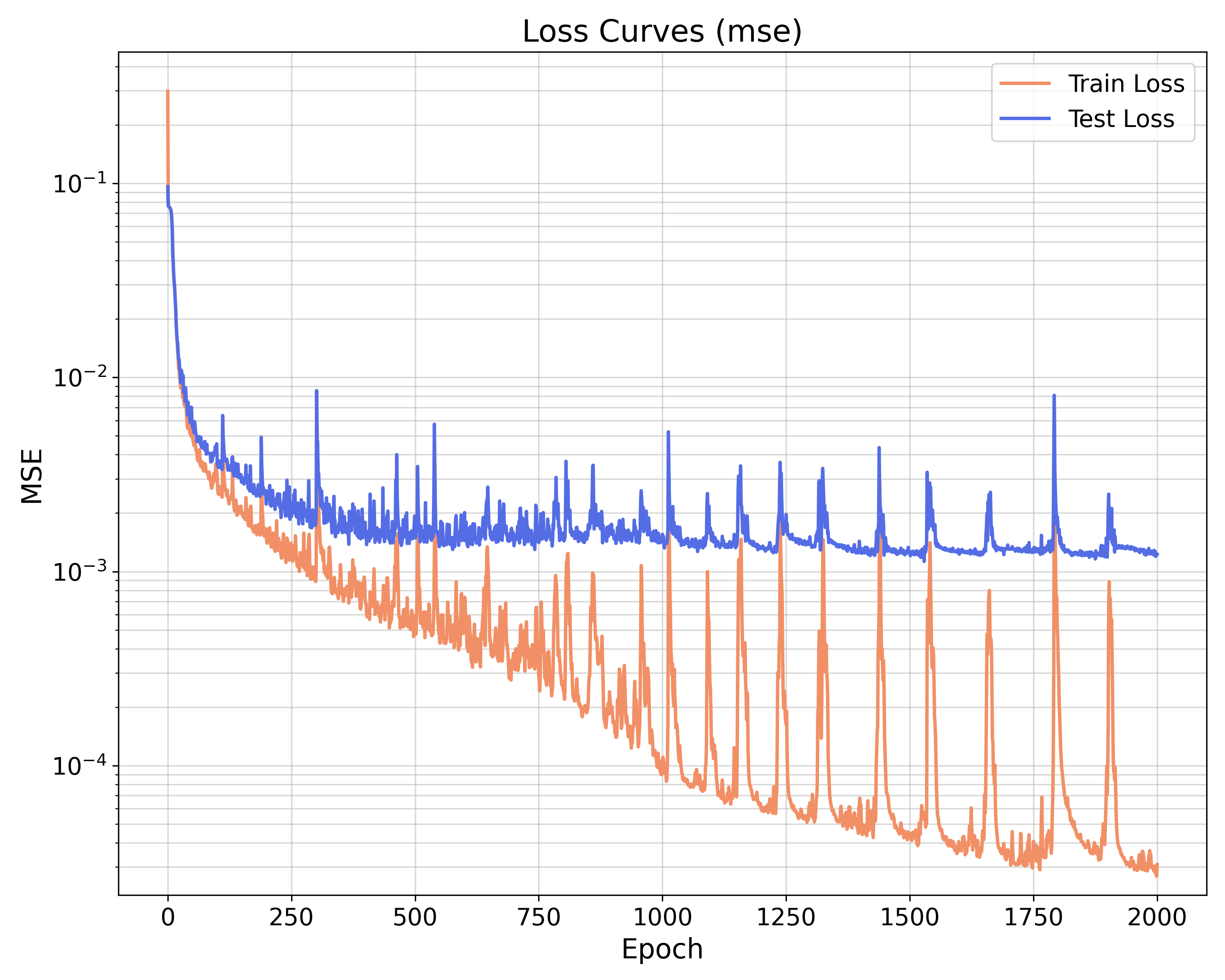}
        \caption{$N=2000$}
    \end{subfigure}
    
    \caption{Training and test loss convergence for different Erd\H{o}s-R\'enyi network sizes ($N$).}
    \label{fig:er_loss_grid}
\end{figure}

\clearpage

\subsection{BA Network}
\begin{figure}[htbp]
    \centering
    \begin{subfigure}[b]{0.32\textwidth}
        \centering
        \includegraphics[width=\textwidth]{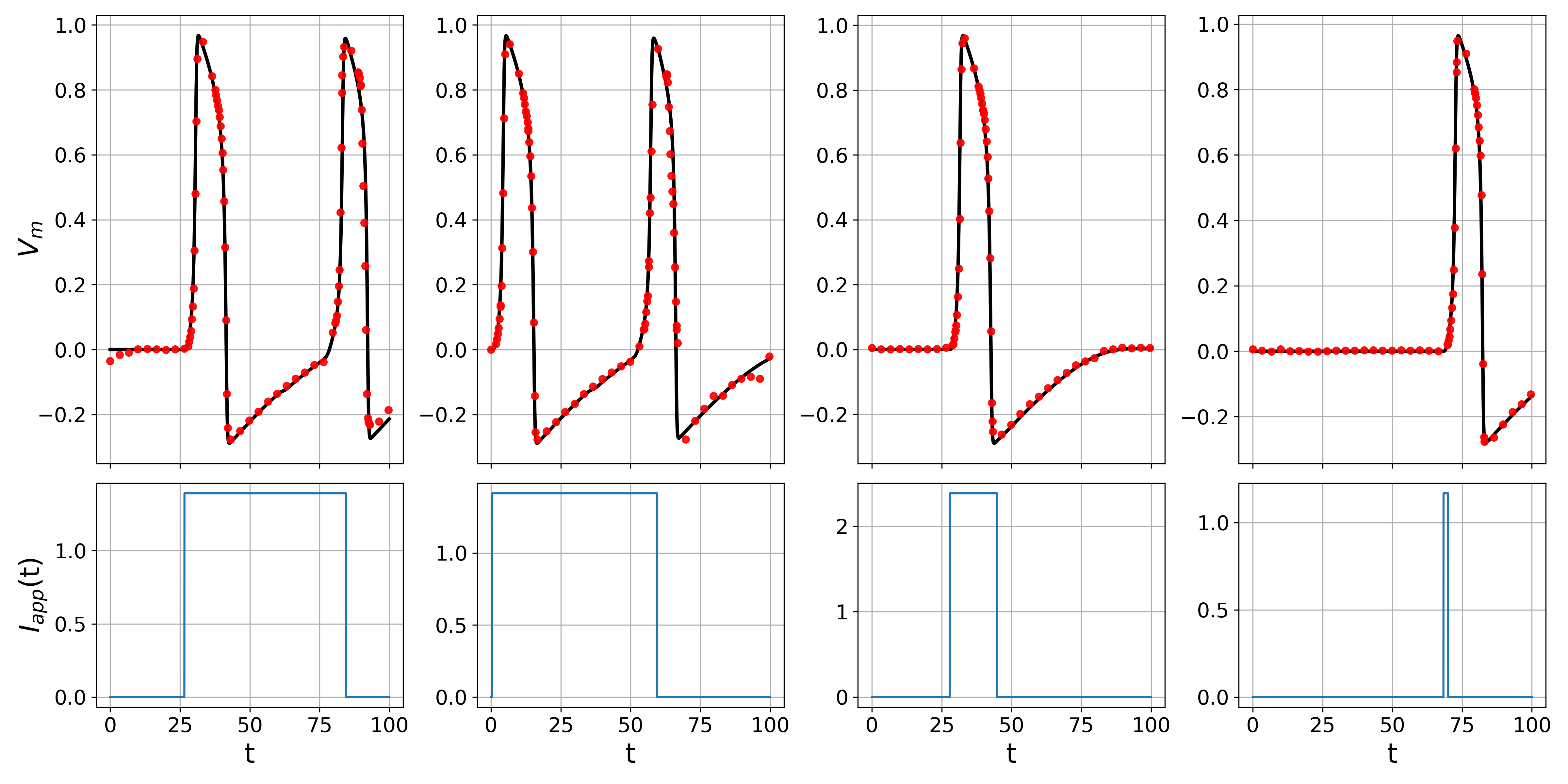}
        \caption{$N=10$}
    \end{subfigure}
    \hfill
    \begin{subfigure}[b]{0.32\textwidth}
        \centering
        \includegraphics[width=\textwidth]{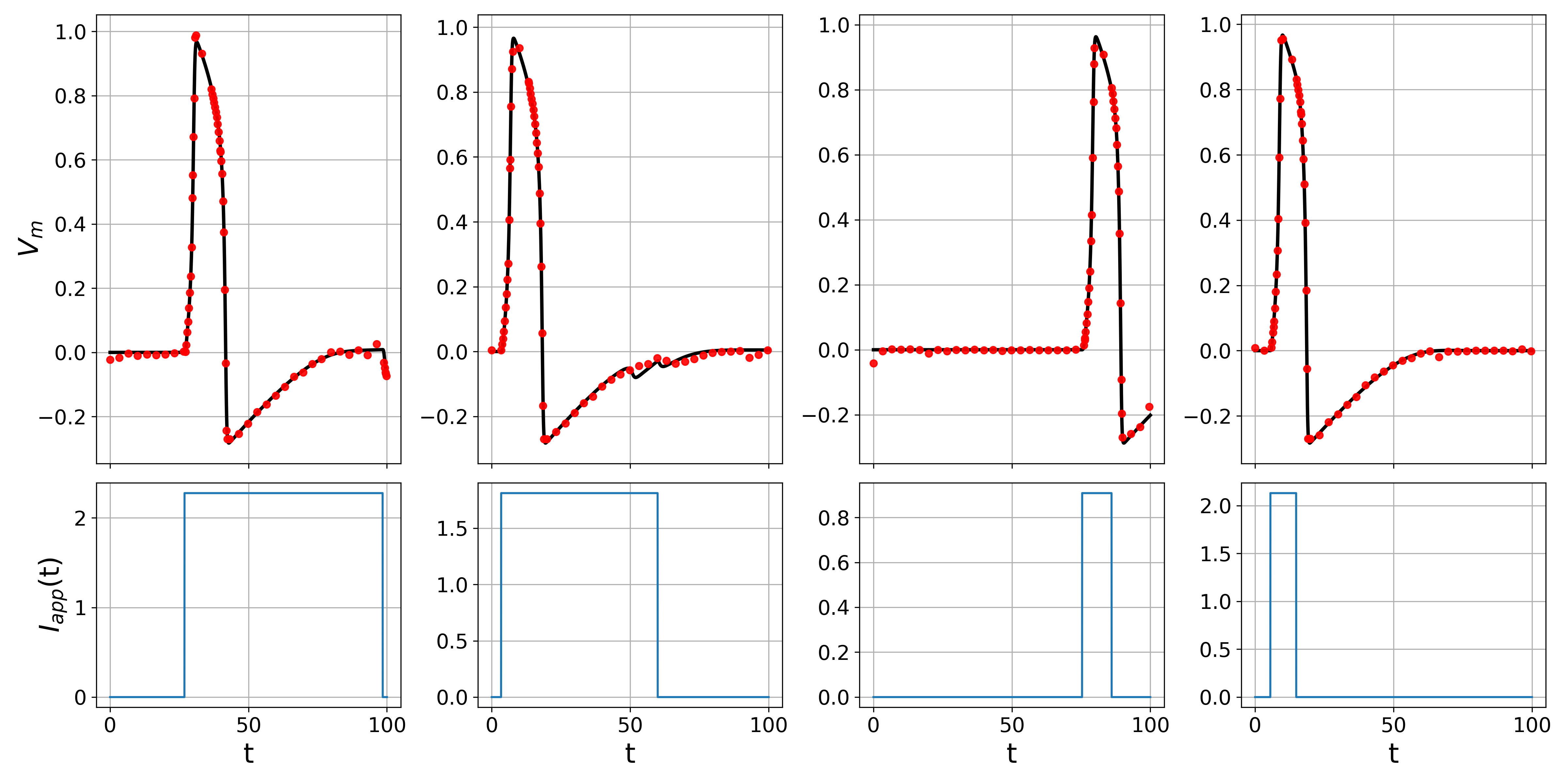}
        \caption{$N=100$}
    \end{subfigure}
    \hfill
    \begin{subfigure}[b]{0.32\textwidth}
        \centering
        \includegraphics[width=\textwidth]{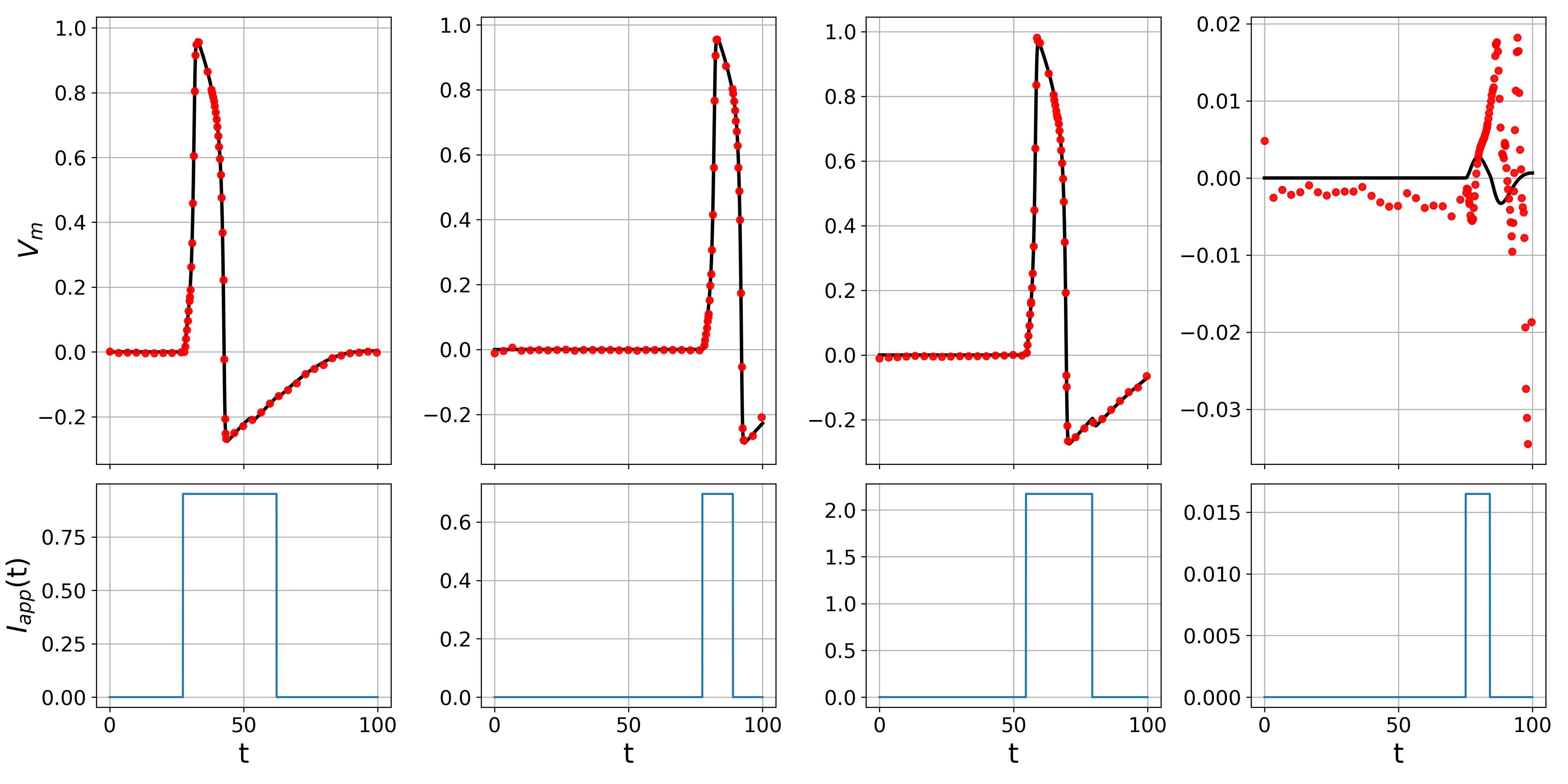}
        \caption{$N=200$}
    \end{subfigure}
    
    \vspace{0.4cm}
    
    \begin{subfigure}[b]{0.32\textwidth}
        \centering
        \includegraphics[width=\textwidth]{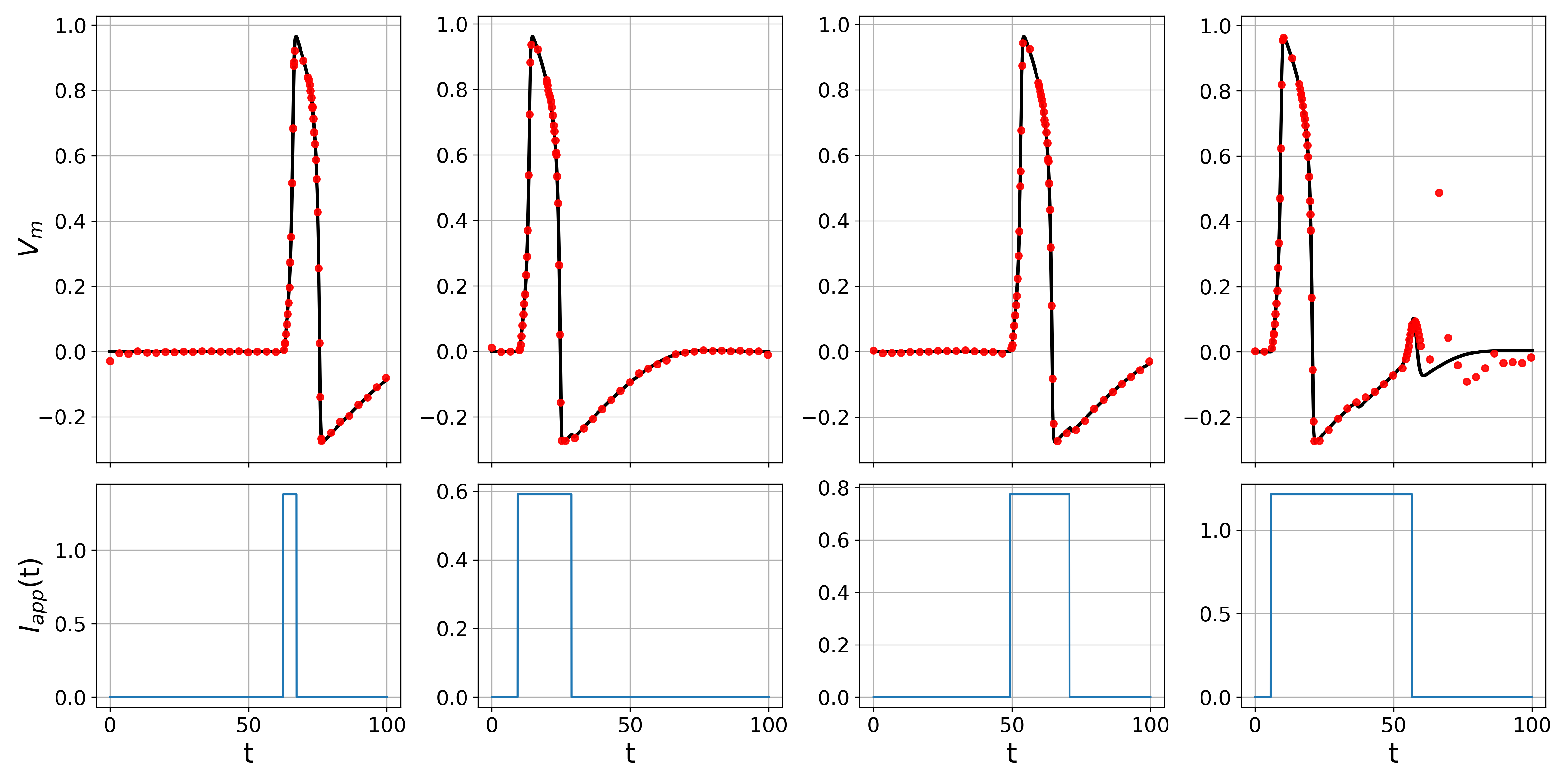}
        \caption{$N=300$}
    \end{subfigure}
    \hfill
    \begin{subfigure}[b]{0.32\textwidth}
        \centering
        \includegraphics[width=\textwidth]{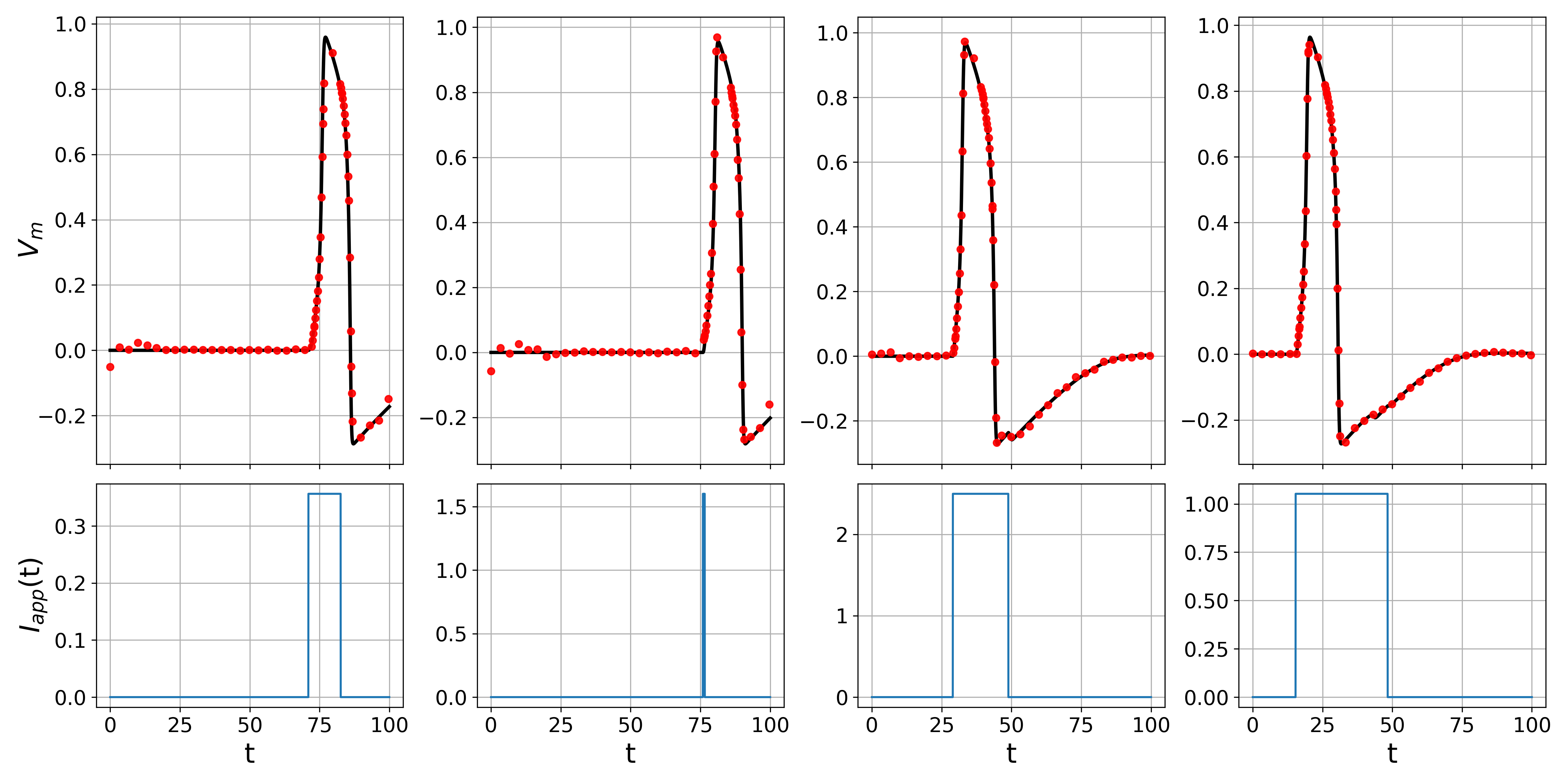}
        \caption{$N=400$}
    \end{subfigure}
    \hfill
    \begin{subfigure}[b]{0.32\textwidth}
        \centering
        \includegraphics[width=\textwidth]{BA_Network_FNO_HHpeaks_S=42_N=500.png}
        \caption{$N=500$}
    \end{subfigure}
    
    \vspace{0.4cm}
    
    \begin{subfigure}[b]{0.32\textwidth}
        \centering
        \includegraphics[width=\textwidth]{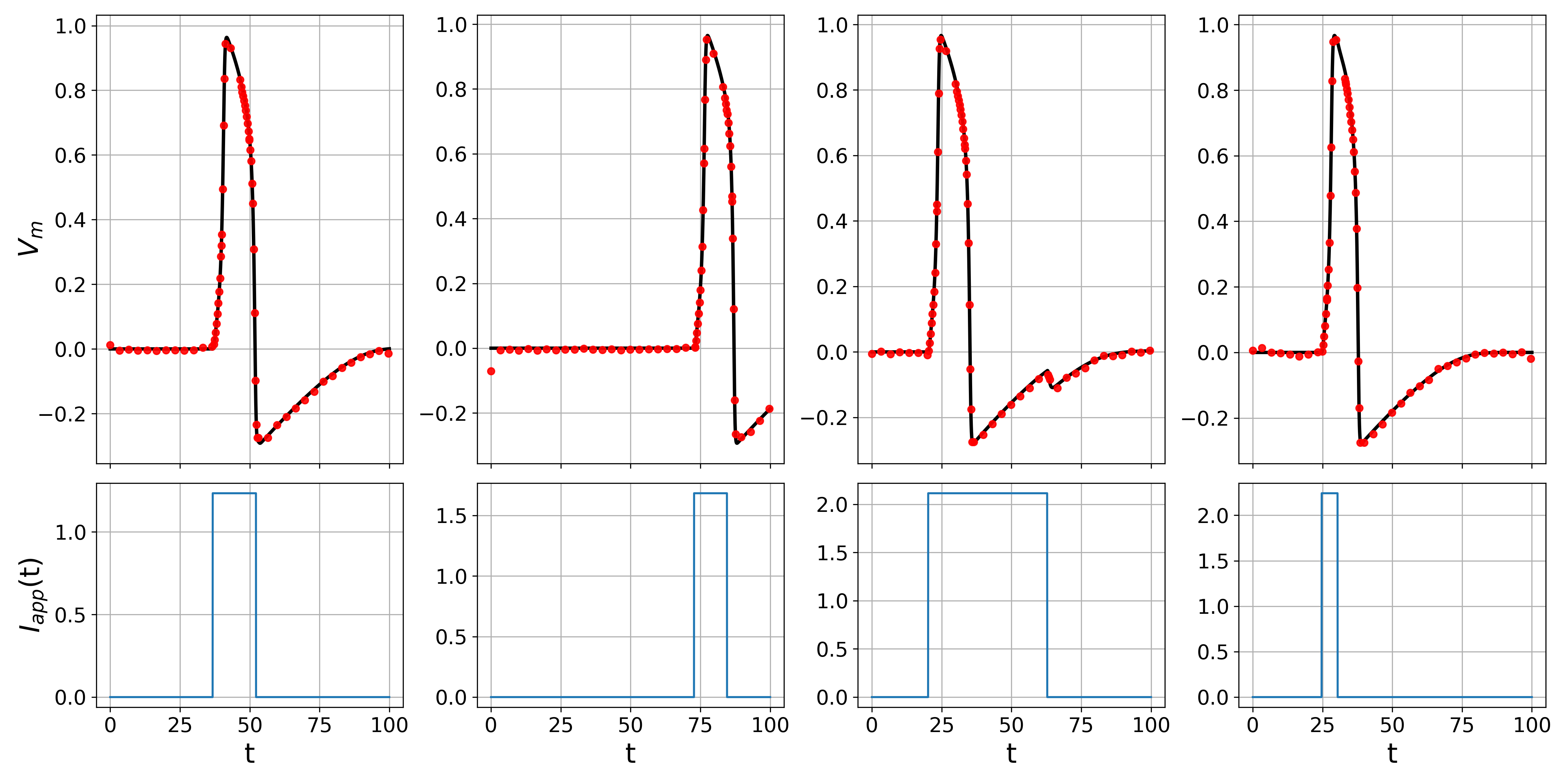}
        \caption{$N=600$}
    \end{subfigure}
    \hfill
    \begin{subfigure}[b]{0.32\textwidth}
        \centering
        \includegraphics[width=\textwidth]{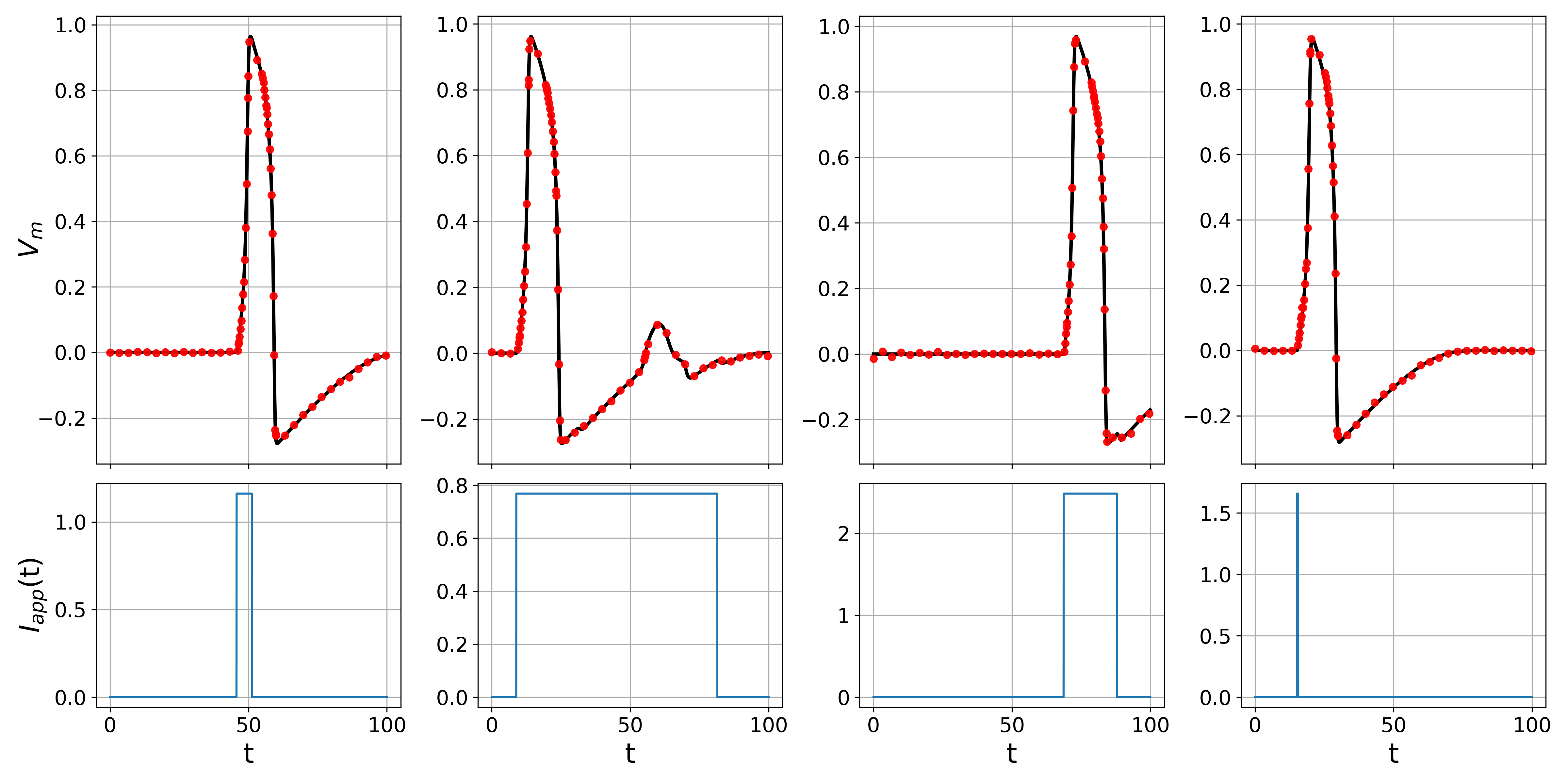}
        \caption{$N=700$}
    \end{subfigure}
    \hfill
    \begin{subfigure}[b]{0.32\textwidth}
        \centering
        \includegraphics[width=\textwidth]{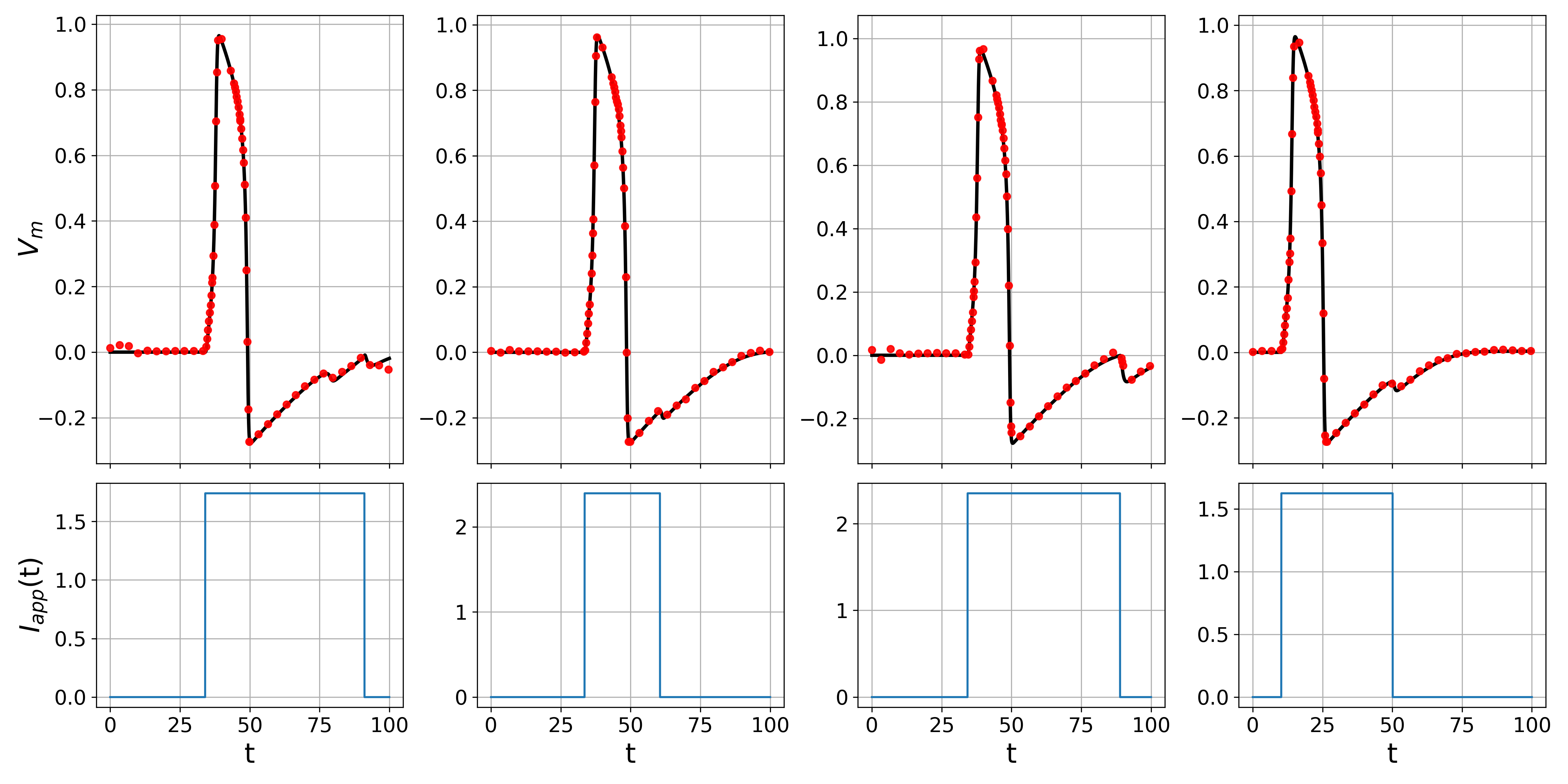}
        \caption{$N=800$}
    \end{subfigure}
    
    \vspace{0.4cm}
    
    \begin{subfigure}[b]{0.32\textwidth}
        \centering
        \includegraphics[width=\textwidth]{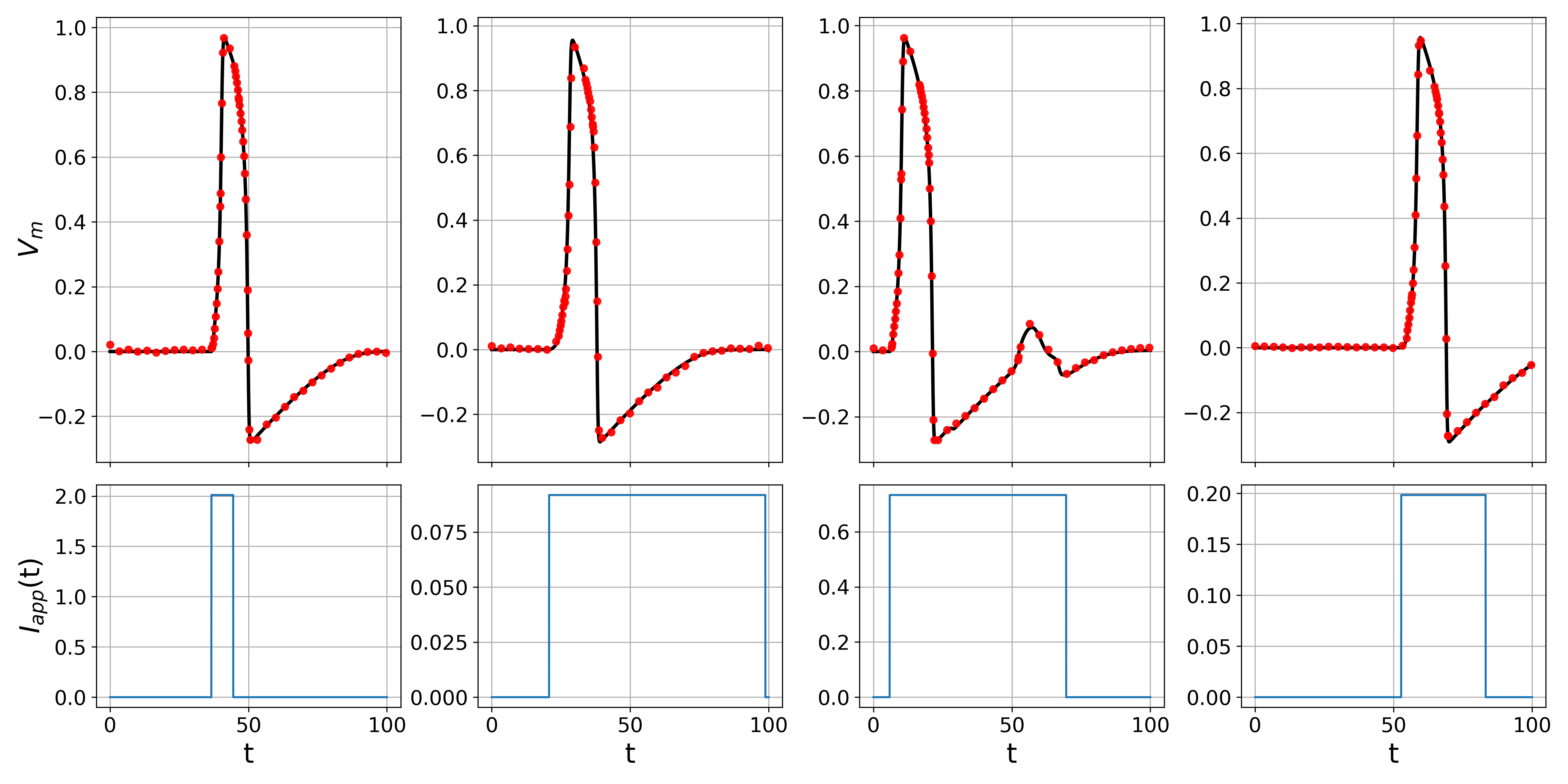}
        \caption{$N=900$}
    \end{subfigure}
    \hfill
    \begin{subfigure}[b]{0.32\textwidth}
        \centering
        \includegraphics[width=\textwidth]{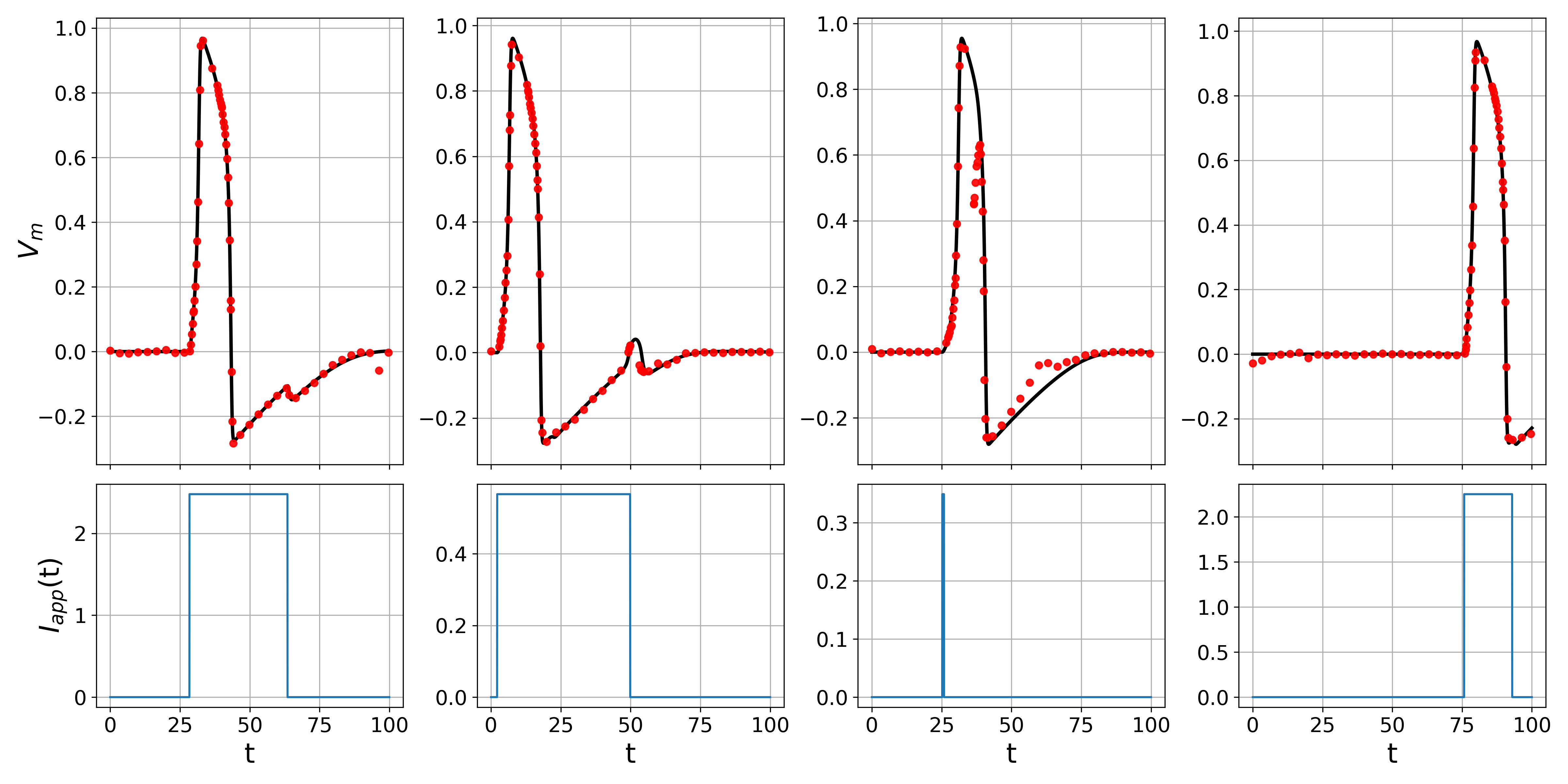}
        \caption{$N=1000$}
    \end{subfigure}
    \hfill
    \begin{subfigure}[b]{0.32\textwidth}
        \centering
        \includegraphics[width=\textwidth]{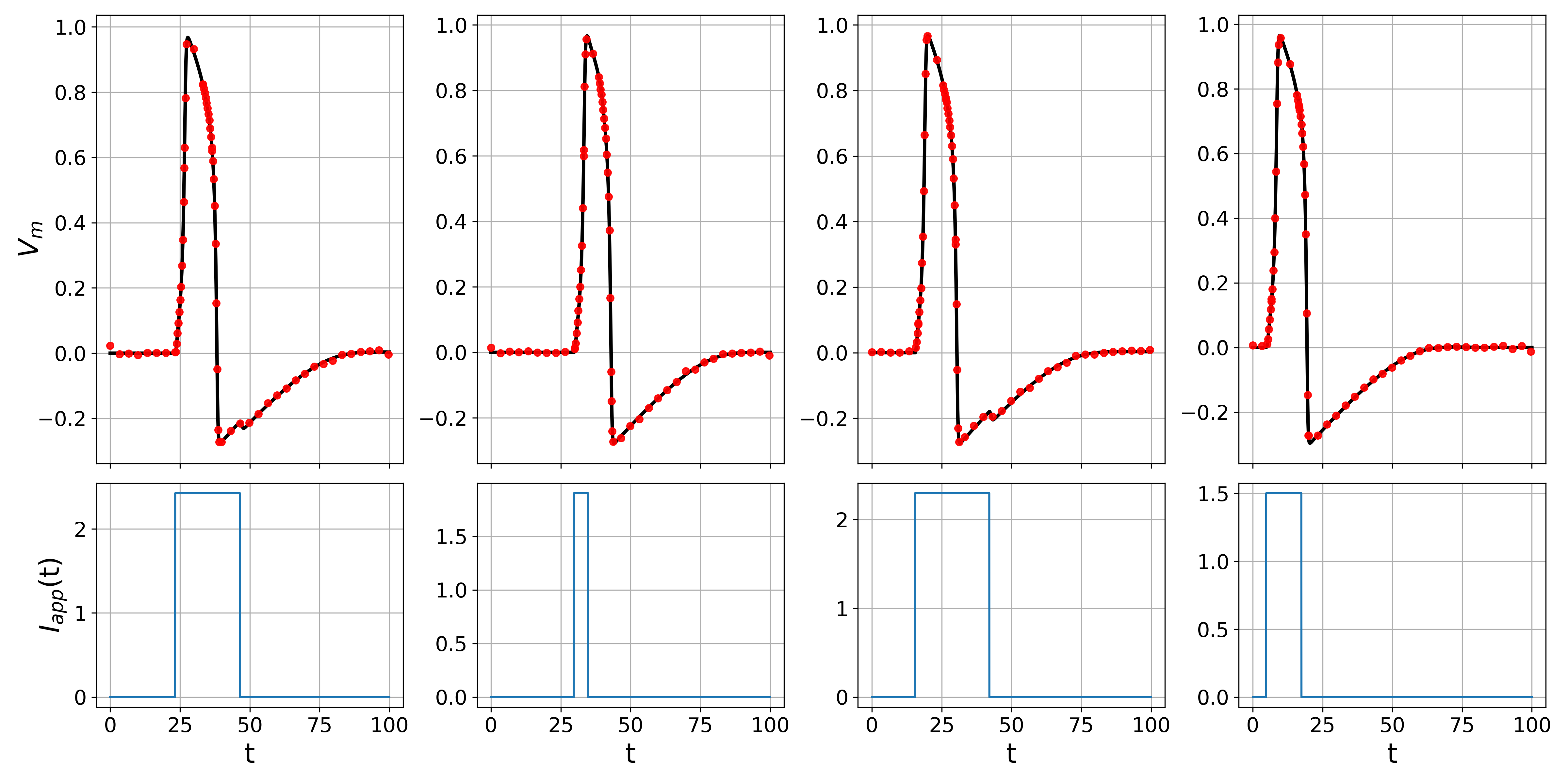}
        \caption{$N=2000$}
    \end{subfigure}
    
    \caption{Some instances of the test set for different Barab\'asi-Albert network sizes ($N$). For each column we can see the applied input current (bottom) with its corresponding potential response (top): the continuous line represents the result of the simulations, whereas the red dots indicate the output of the NN.}
    \label{fig:ba_peaks_grid}
\end{figure}


\begin{figure}[htbp]
    \centering
    \begin{subfigure}[b]{0.29\textwidth}
        \centering
        \includegraphics[width=\textwidth]{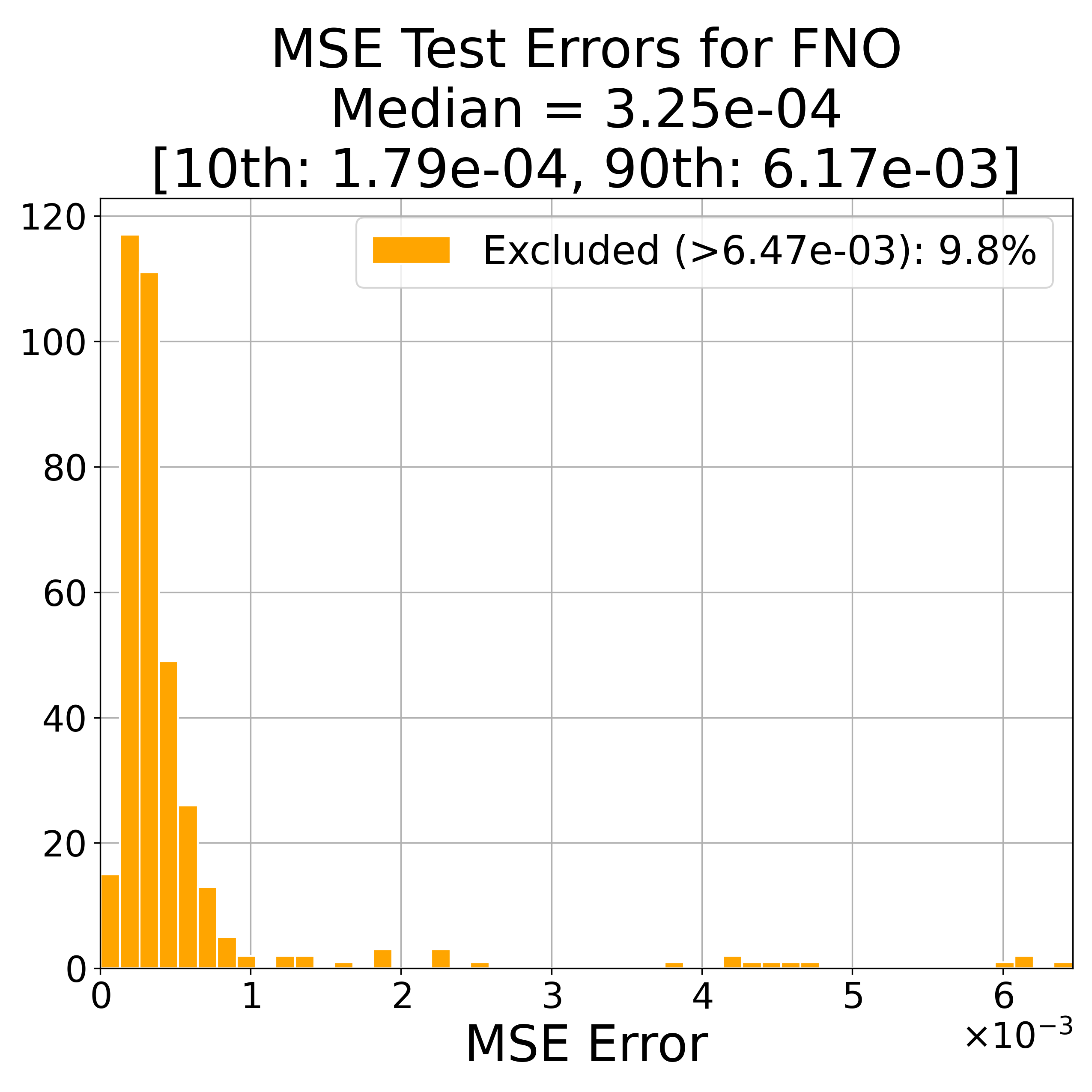}
        \caption{$N=10$}
    \end{subfigure}
    \hfill
    \begin{subfigure}[b]{0.29\textwidth}
        \centering
        \includegraphics[width=\textwidth]{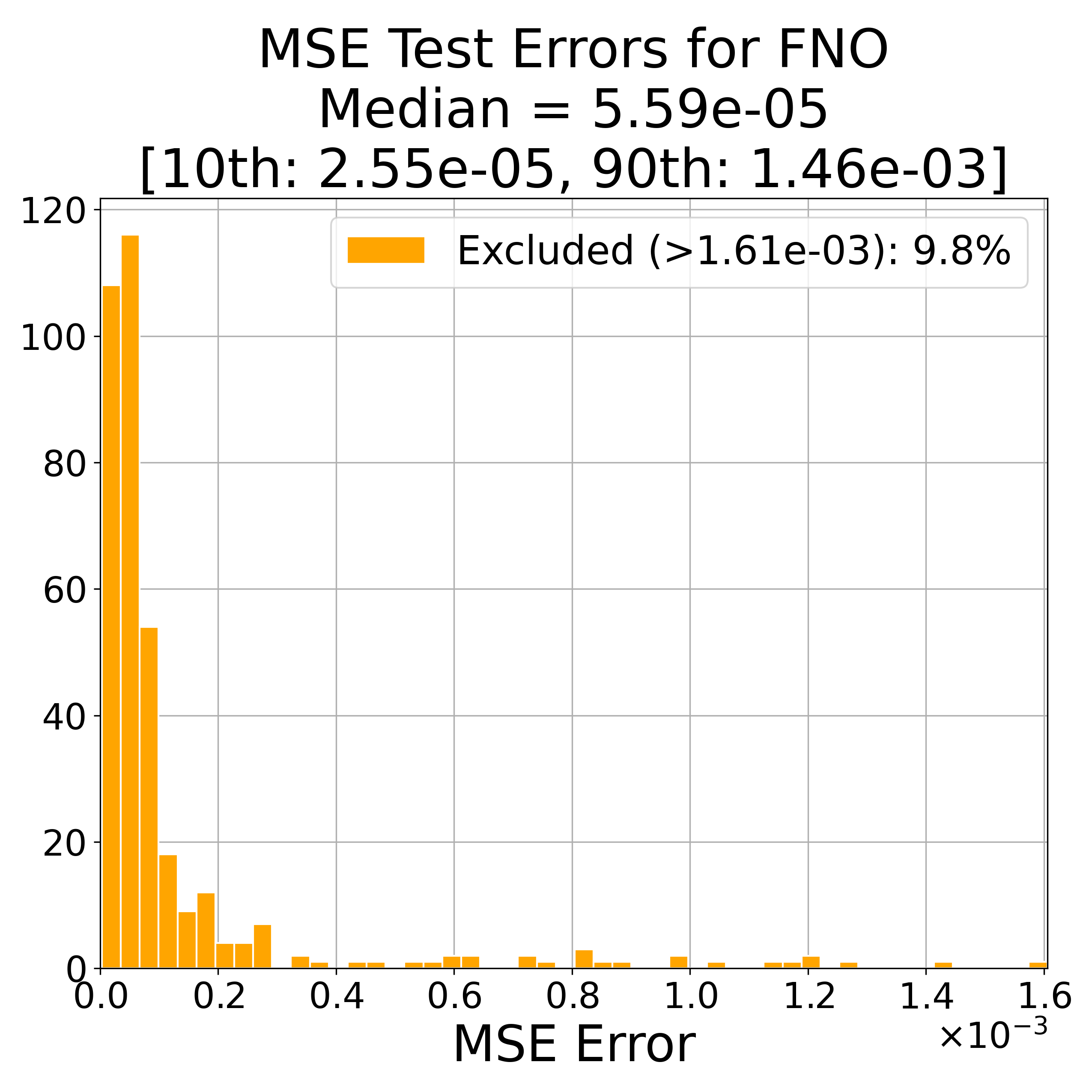}
        \caption{$N=100$}
    \end{subfigure}
    \hfill
    \begin{subfigure}[b]{0.29\textwidth}
        \centering
        \includegraphics[width=\textwidth]{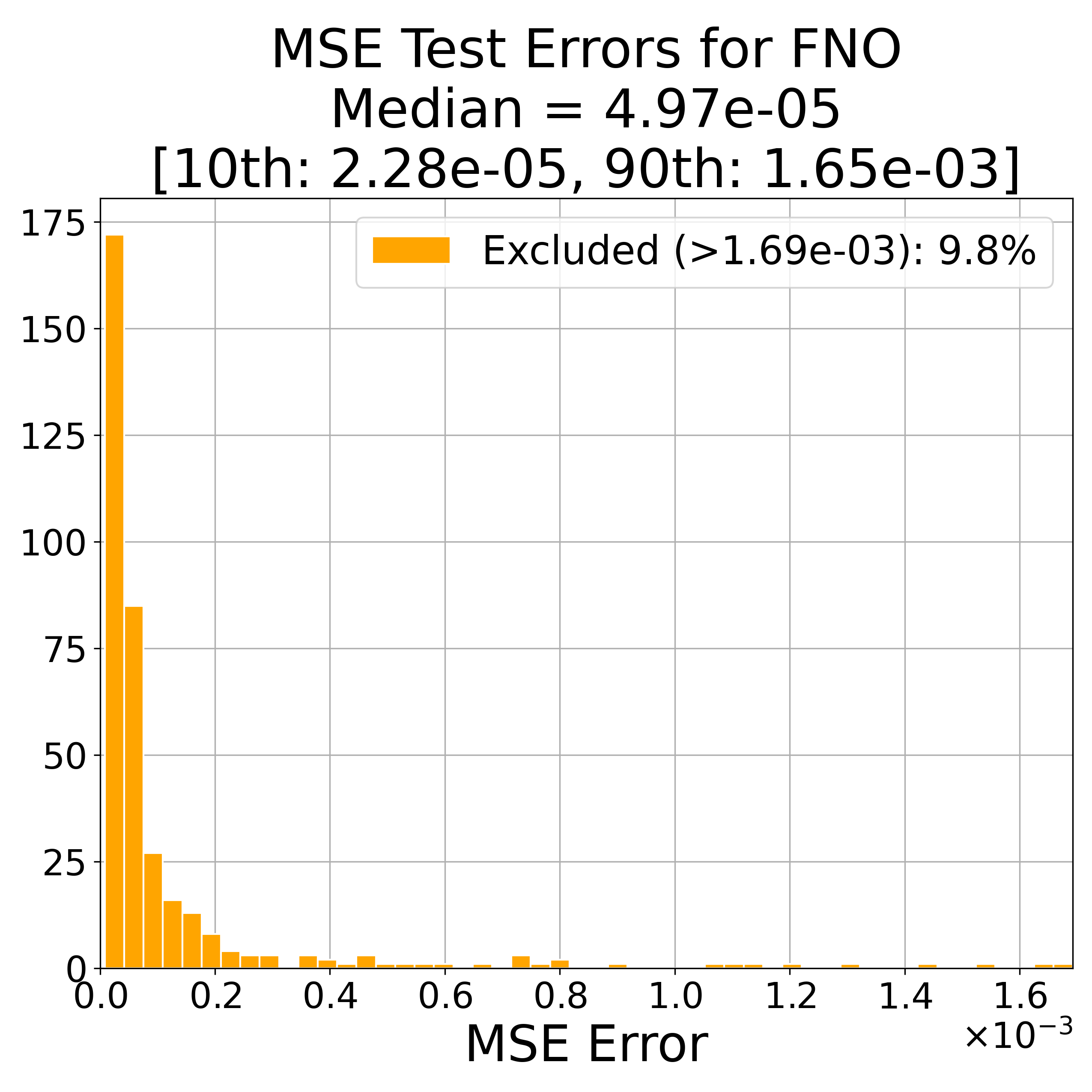}
        \caption{$N=200$}
    \end{subfigure}
    
    \vspace{0.1cm}
    
    \begin{subfigure}[b]{0.29\textwidth}
        \centering
        \includegraphics[width=\textwidth]{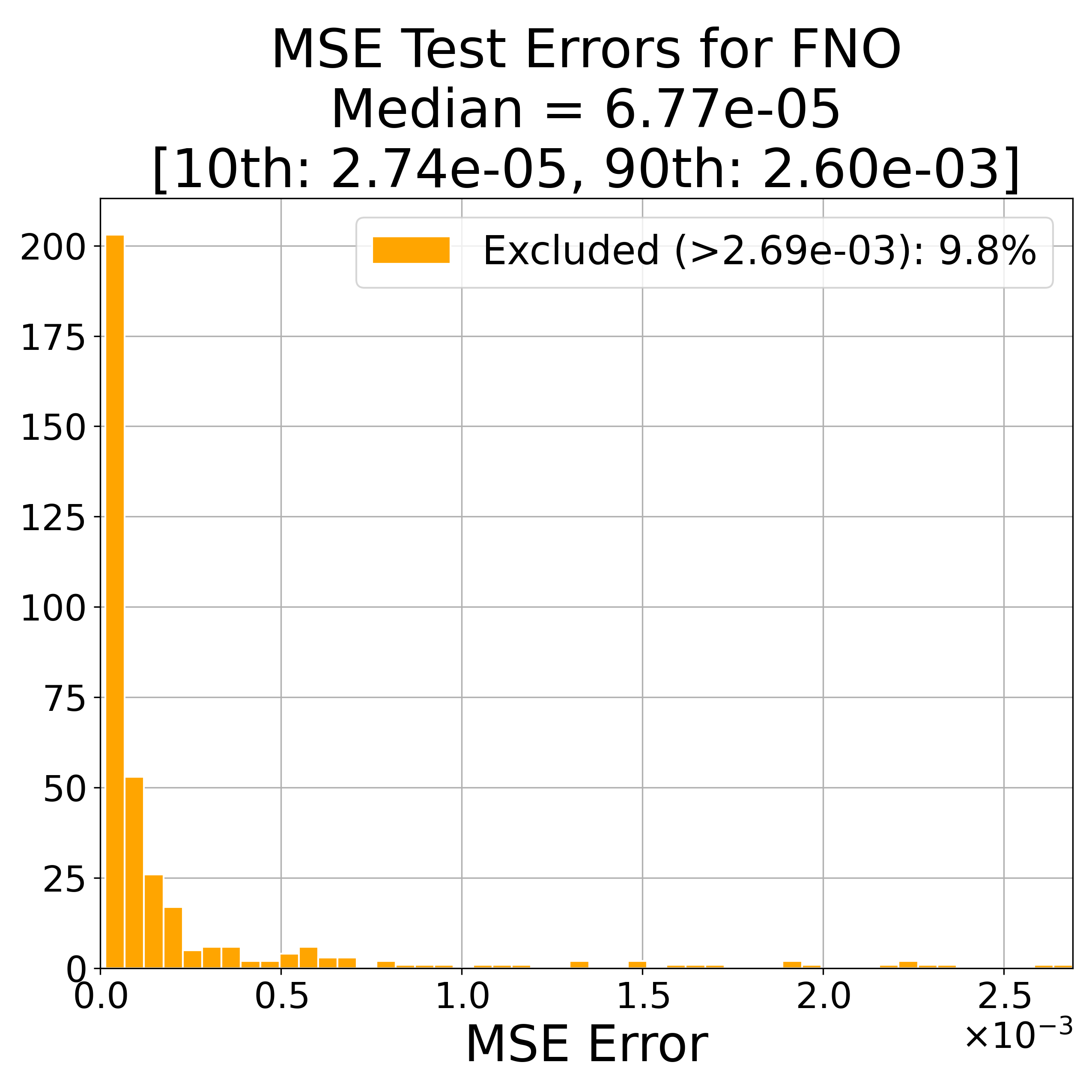}
        \caption{$N=300$}
    \end{subfigure}
    \hfill
    \begin{subfigure}[b]{0.29\textwidth}
        \centering
        \includegraphics[width=\textwidth]{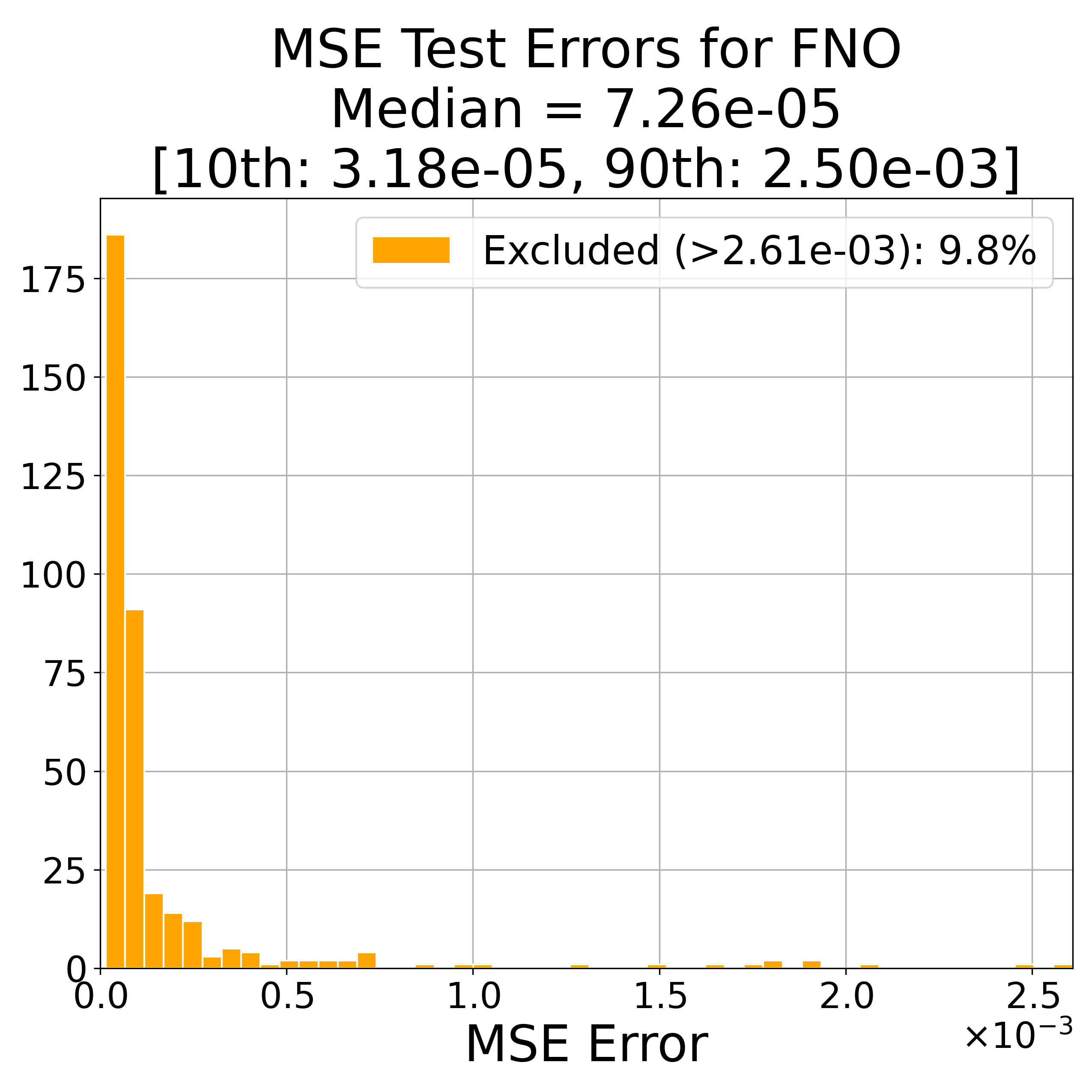}
        \caption{$N=400$}
    \end{subfigure}
    \hfill
    \begin{subfigure}[b]{0.29\textwidth}
        \centering
        \includegraphics[width=\textwidth]{BA_Network_FNO_L2histoFNO_S=42_N=500.png}
        \caption{$N=500$}
    \end{subfigure}
    
    \vspace{0.1cm}
    
    \begin{subfigure}[b]{0.29\textwidth}
        \centering
        \includegraphics[width=\textwidth]{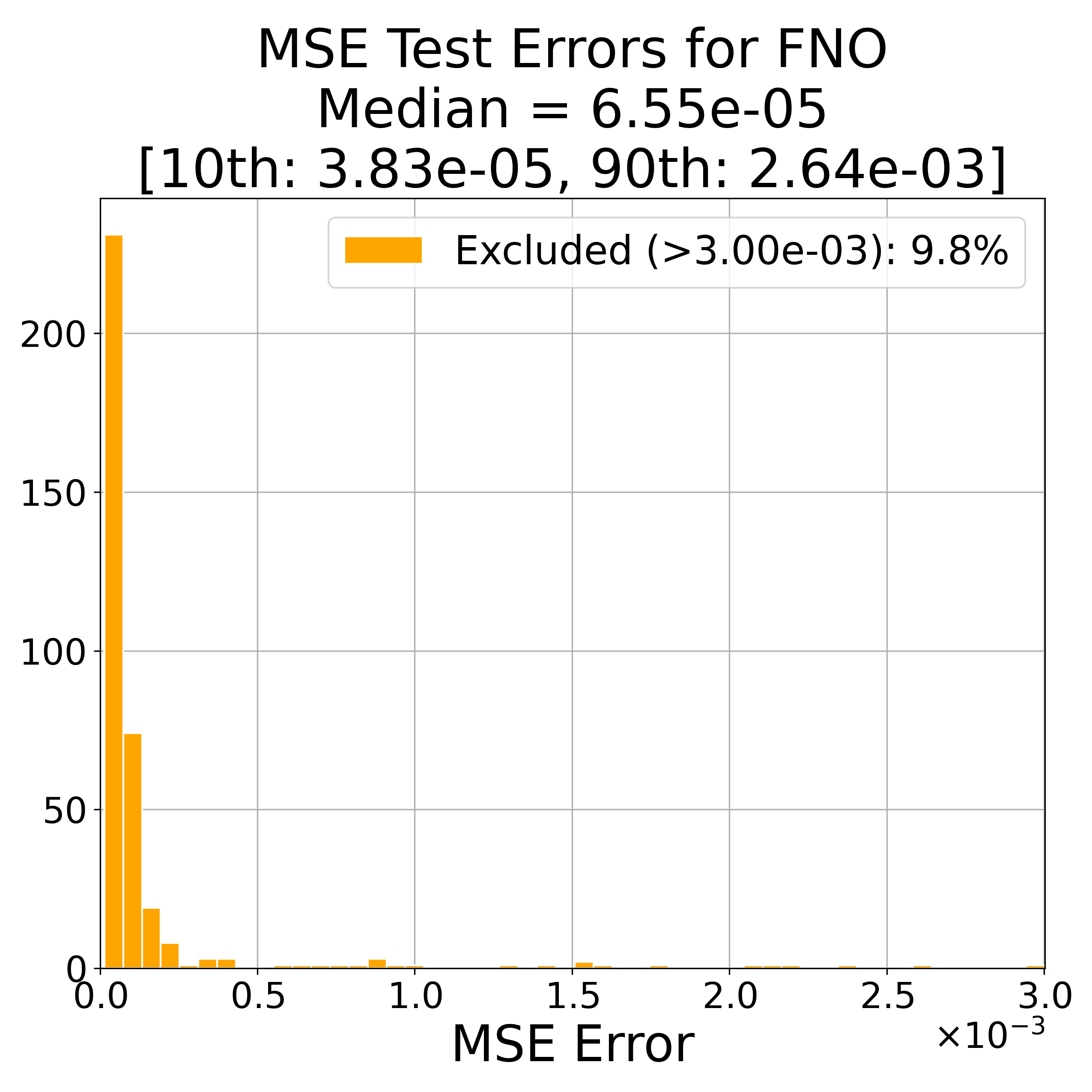}
        \caption{$N=600$}
    \end{subfigure}
    \hfill
    \begin{subfigure}[b]{0.29\textwidth}
        \centering
        \includegraphics[width=\textwidth]{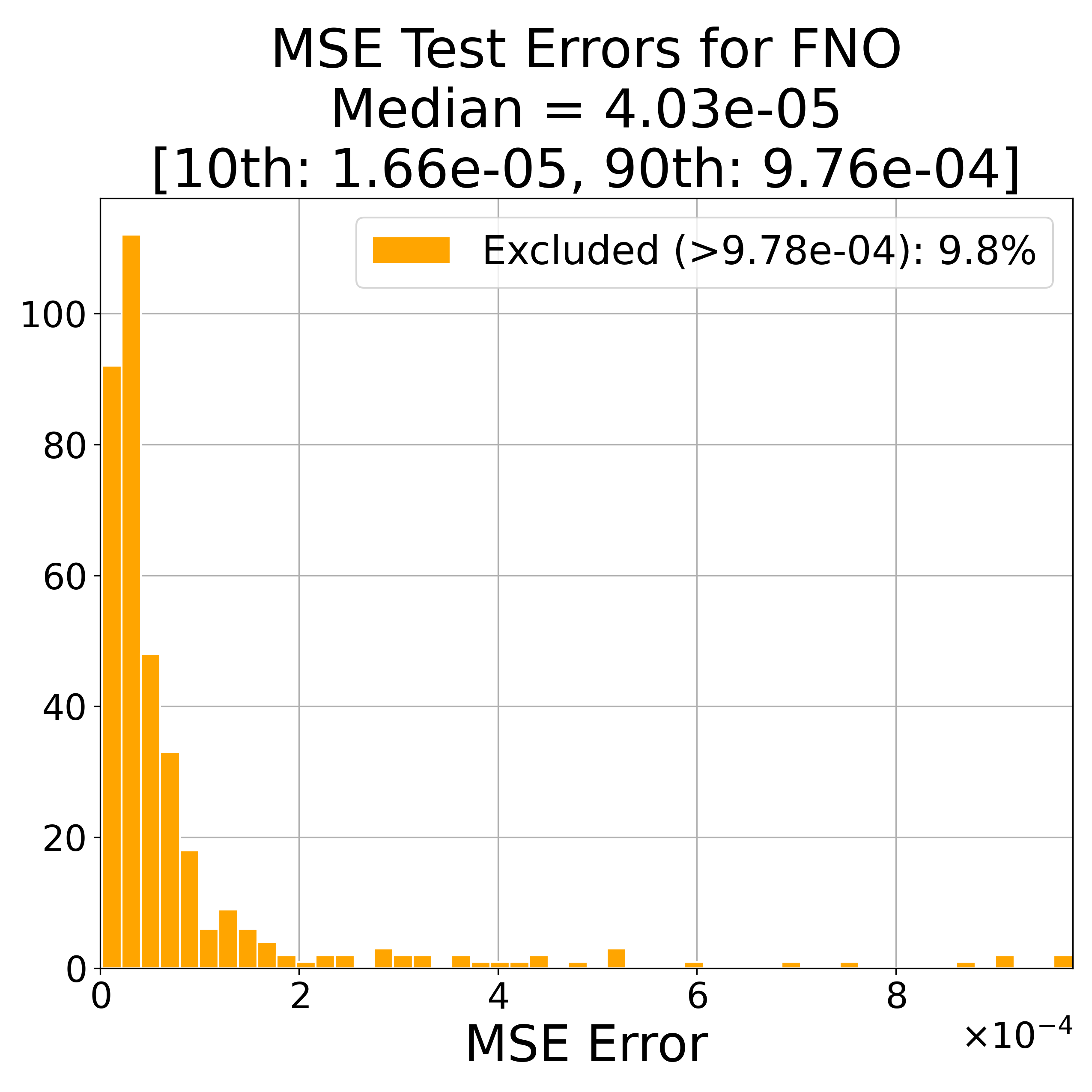}
        \caption{$N=700$}
    \end{subfigure}
    \hfill
    \begin{subfigure}[b]{0.29\textwidth}
        \centering
        \includegraphics[width=\textwidth]{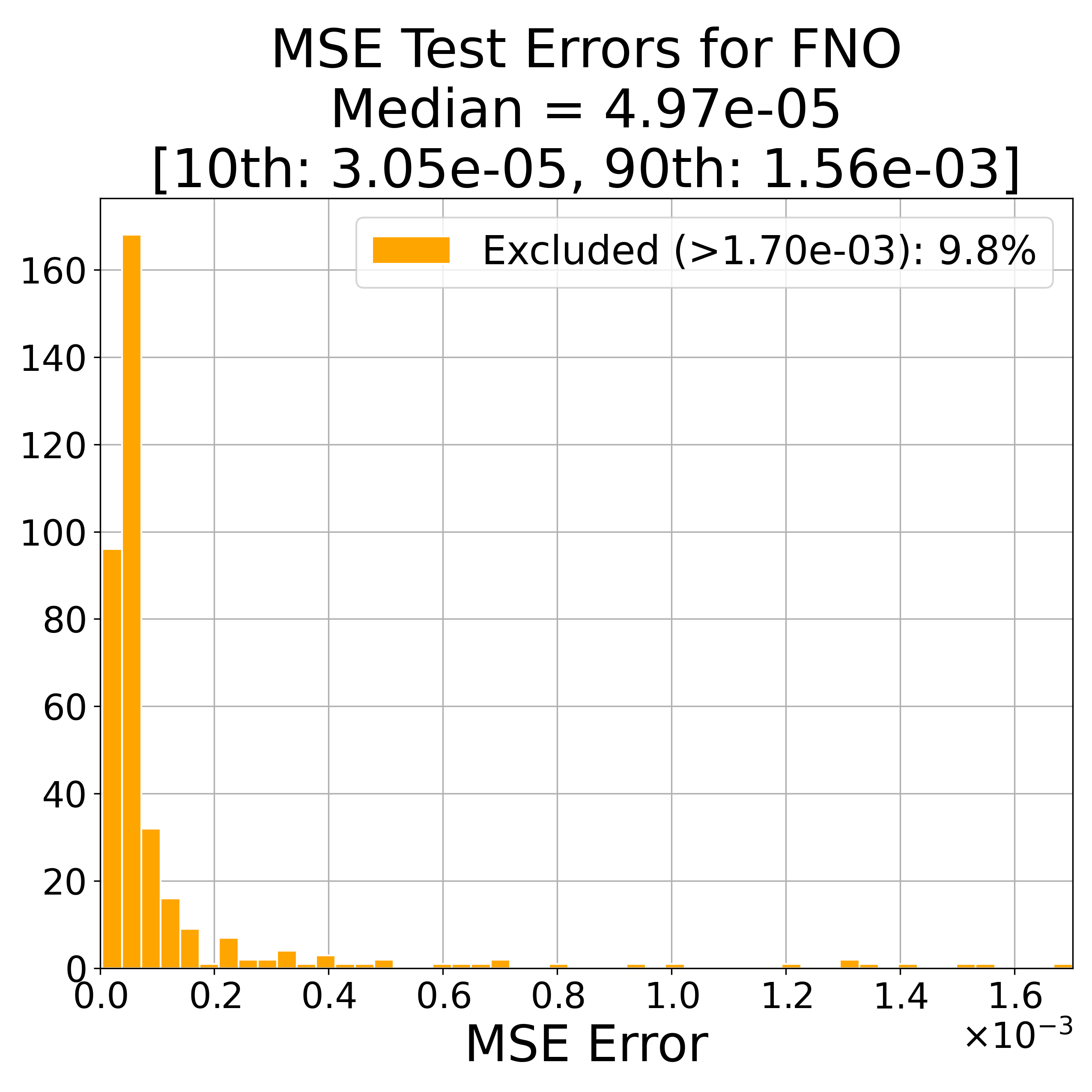}
        \caption{$N=800$}
    \end{subfigure}
    
    \vspace{0.1cm}
    
    \begin{subfigure}[b]{0.29\textwidth}
        \centering
        \includegraphics[width=\textwidth]{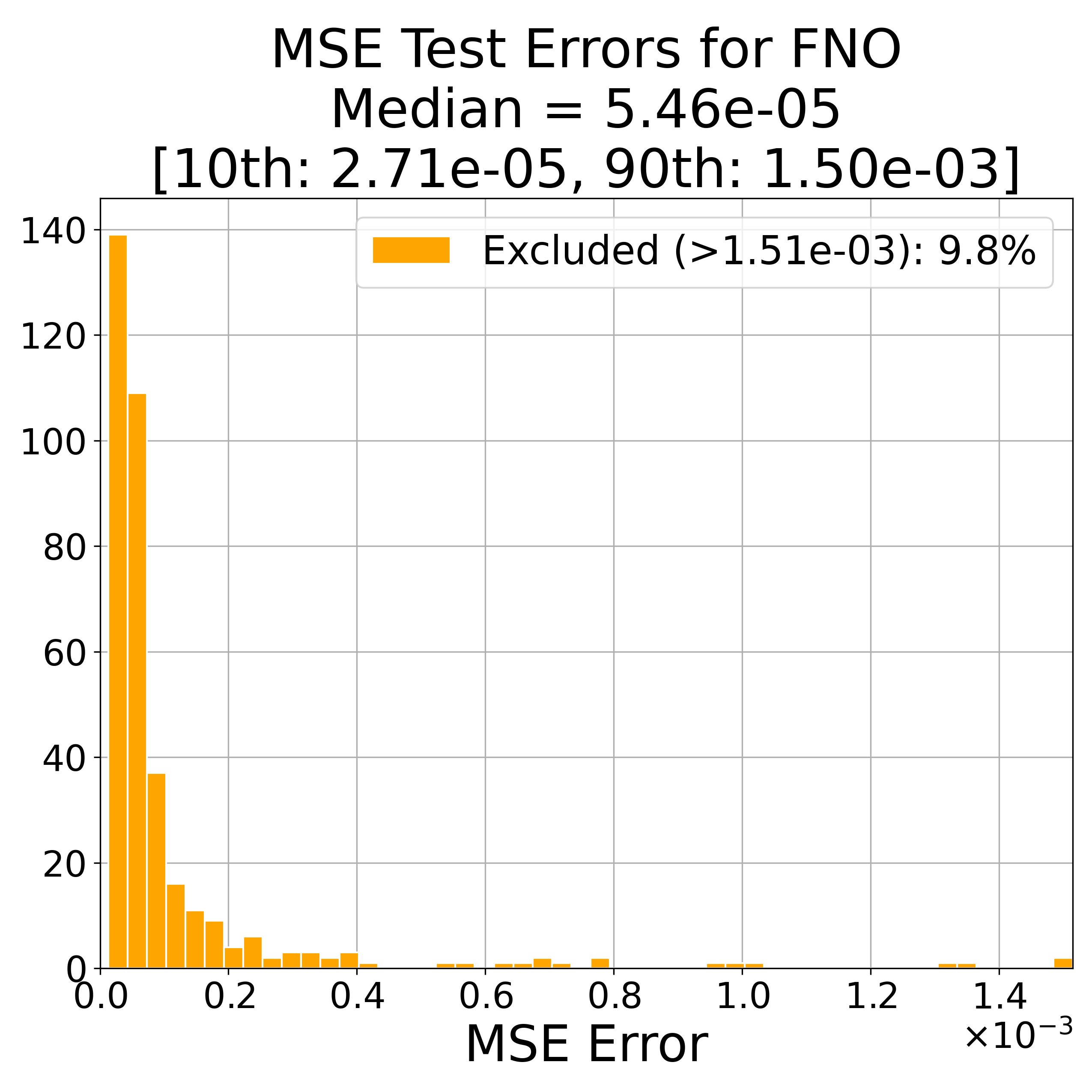}
        \caption{$N=900$}
    \end{subfigure}
    \hfill
    \begin{subfigure}[b]{0.29\textwidth}
        \centering
        \includegraphics[width=\textwidth]{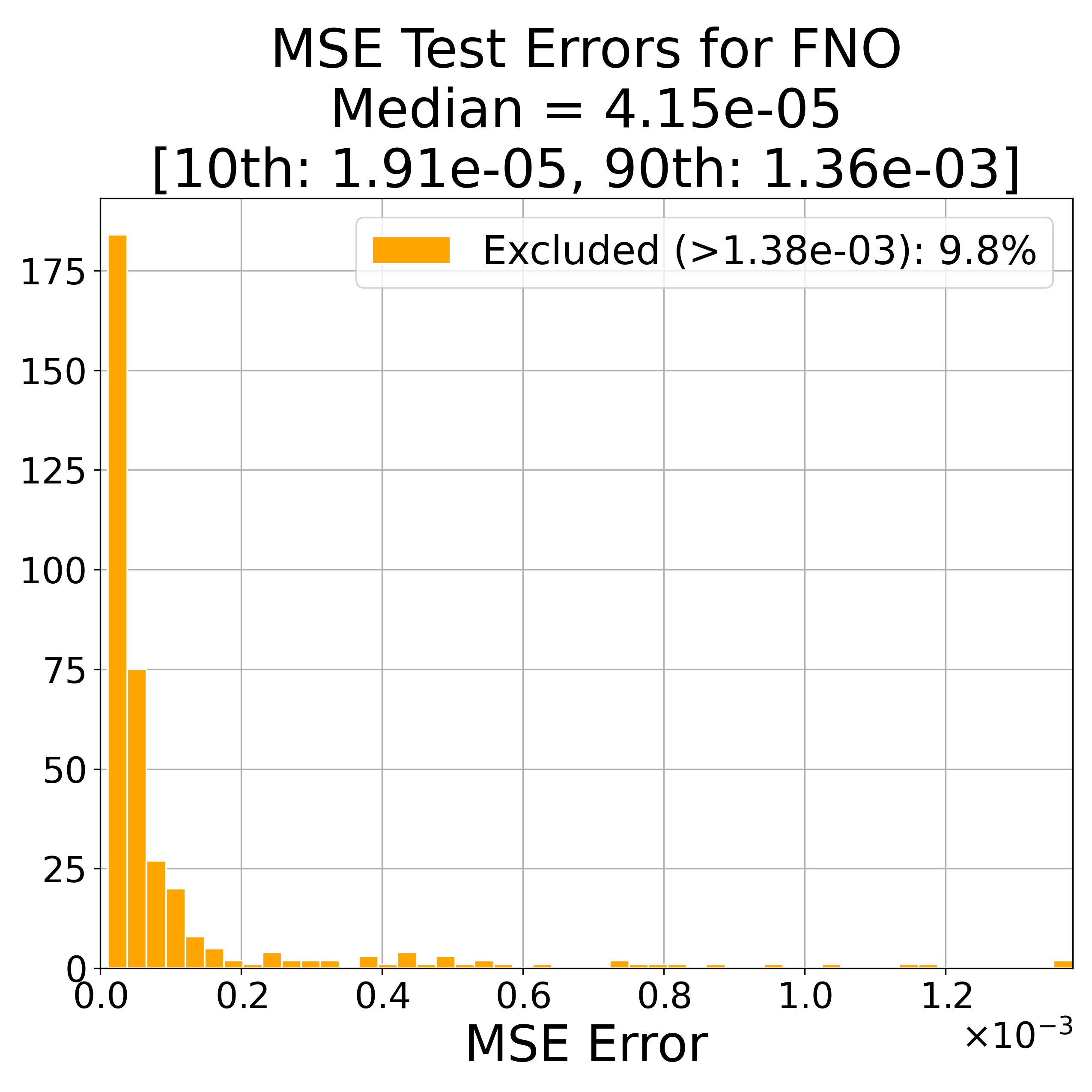}
        \caption{$N=1000$}
    \end{subfigure}
    \hfill
    \begin{subfigure}[b]{0.29\textwidth}
        \centering
        \includegraphics[width=\textwidth]{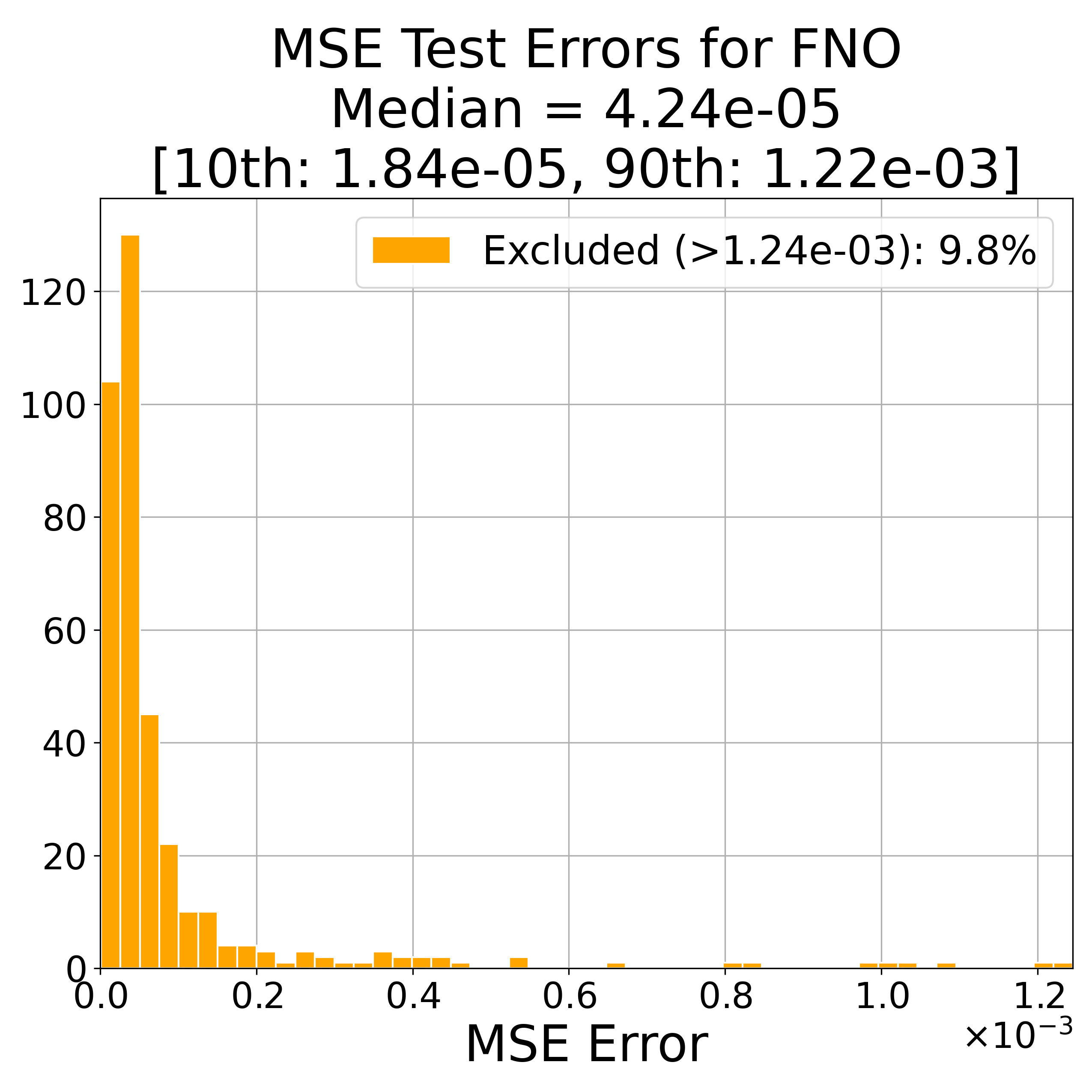}
        \caption{$N=2000$}
    \end{subfigure}
    
    \caption{MSE distributions across the test set for different Barab\'asi-Albert network sizes ($N$).}
    \label{fig:ba_histo_grid}
\end{figure}

\begin{figure}[htbp]
    \centering
    \begin{subfigure}[b]{0.32\textwidth}
        \centering
        \includegraphics[width=\textwidth]{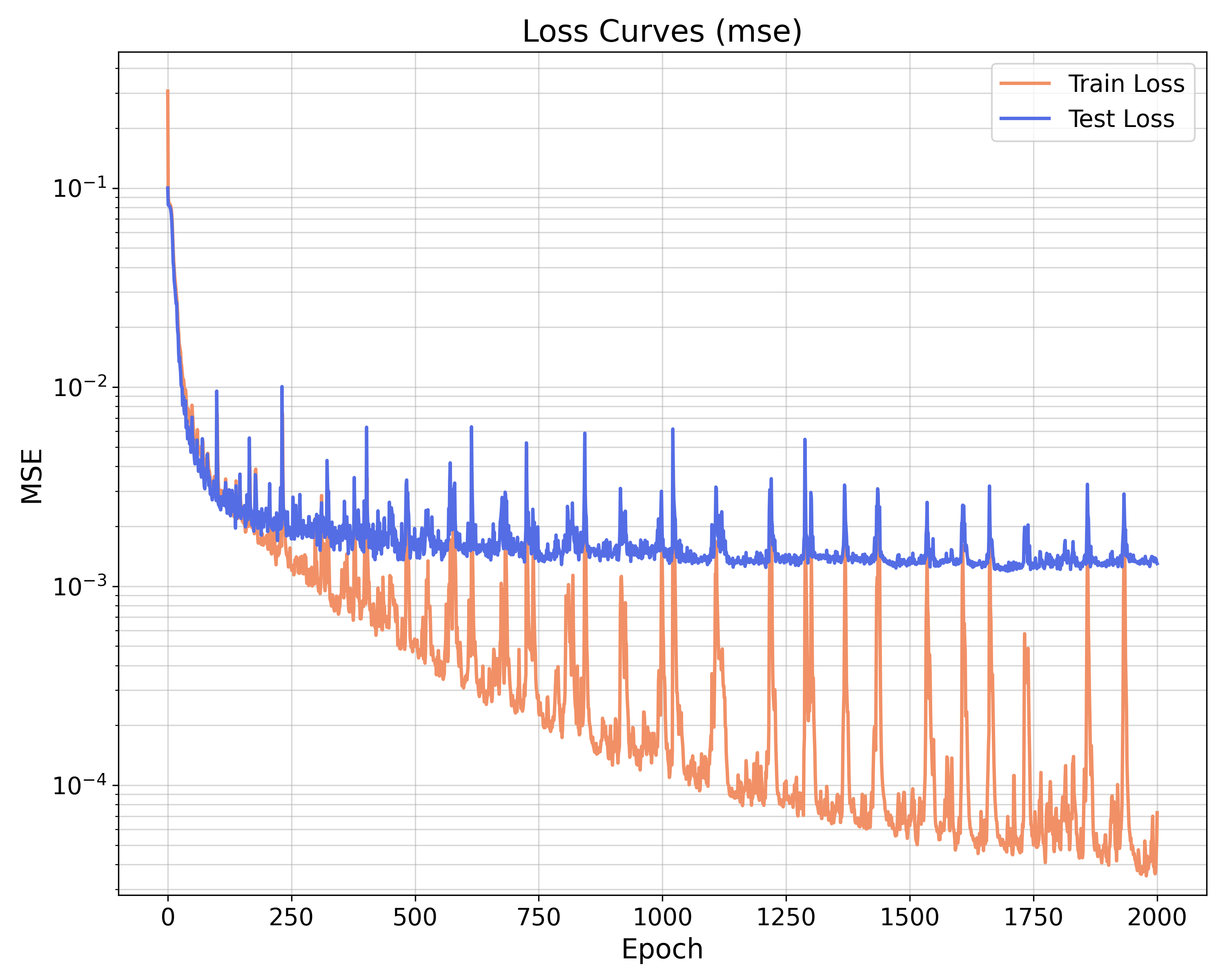}
        \caption{$N=10$}
    \end{subfigure}
    \hfill
    \begin{subfigure}[b]{0.32\textwidth}
        \centering
        \includegraphics[width=\textwidth]{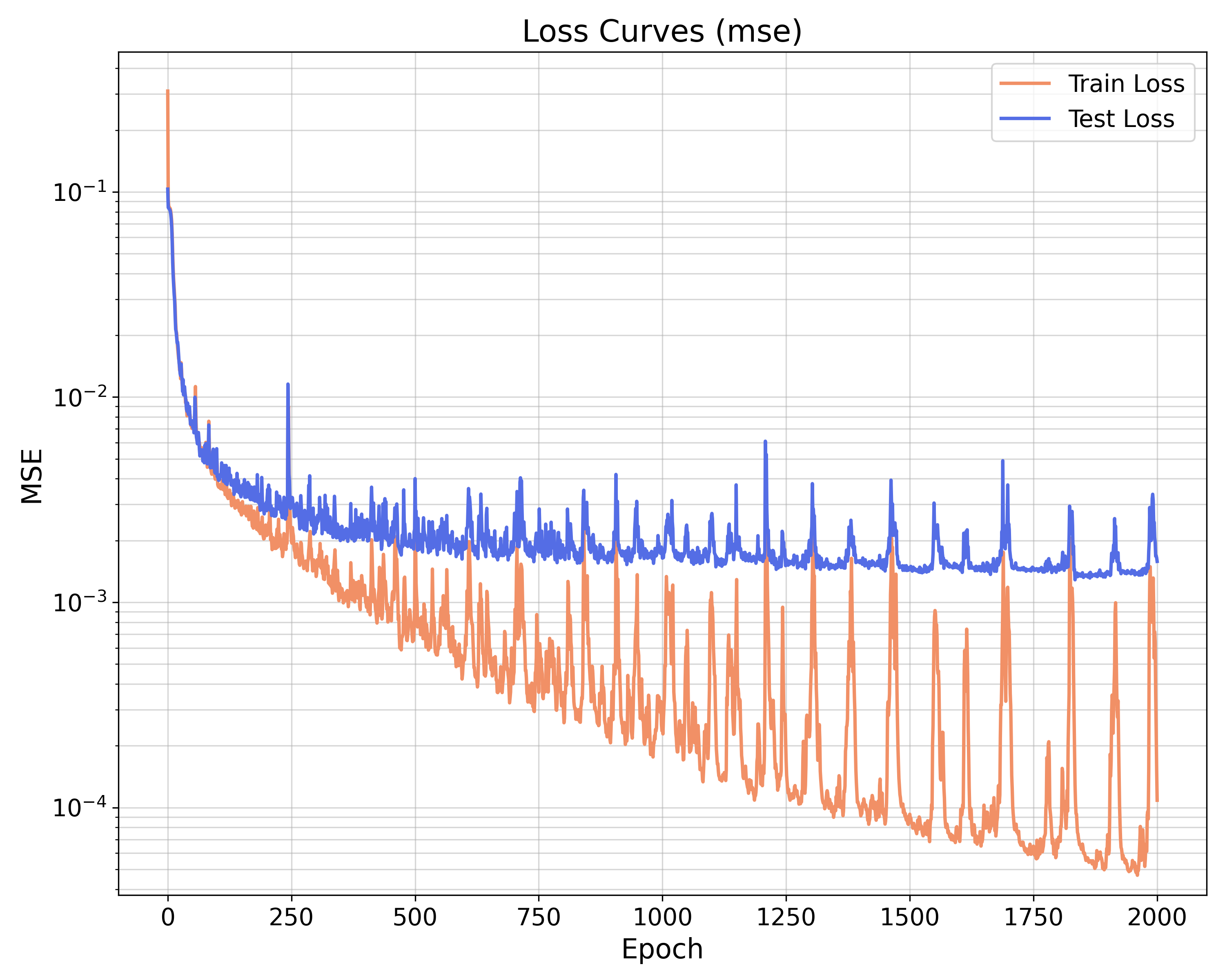}
        \caption{$N=100$}
    \end{subfigure}
    \hfill
    \begin{subfigure}[b]{0.32\textwidth}
        \centering
        \includegraphics[width=\textwidth]{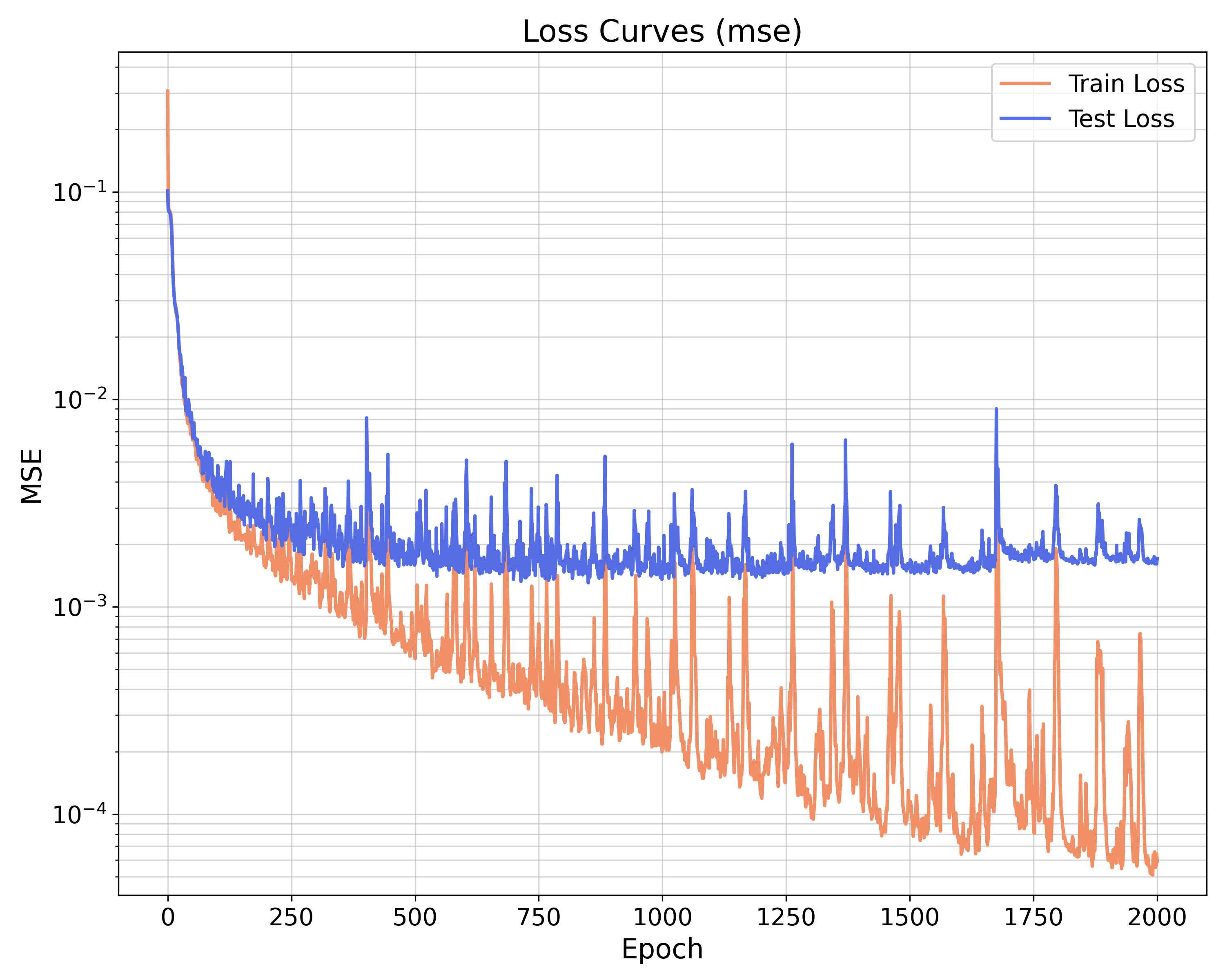}
        \caption{$N=200$}
    \end{subfigure}
    
    \vspace{0.4cm}
    
    \begin{subfigure}[b]{0.32\textwidth}
        \centering
        \includegraphics[width=\textwidth]{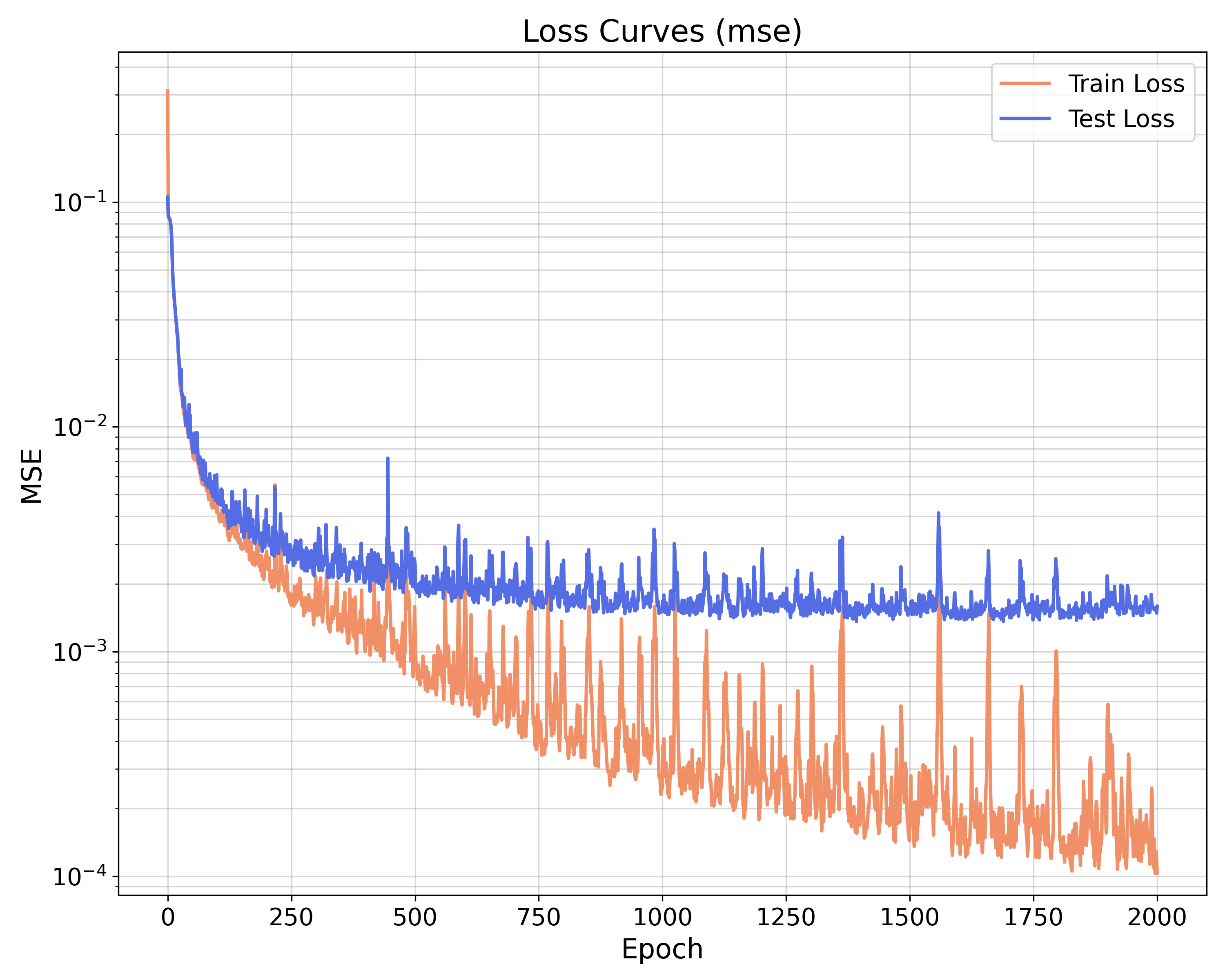}
        \caption{$N=300$}
    \end{subfigure}
    \hfill
    \begin{subfigure}[b]{0.32\textwidth}
        \centering
        \includegraphics[width=\textwidth]{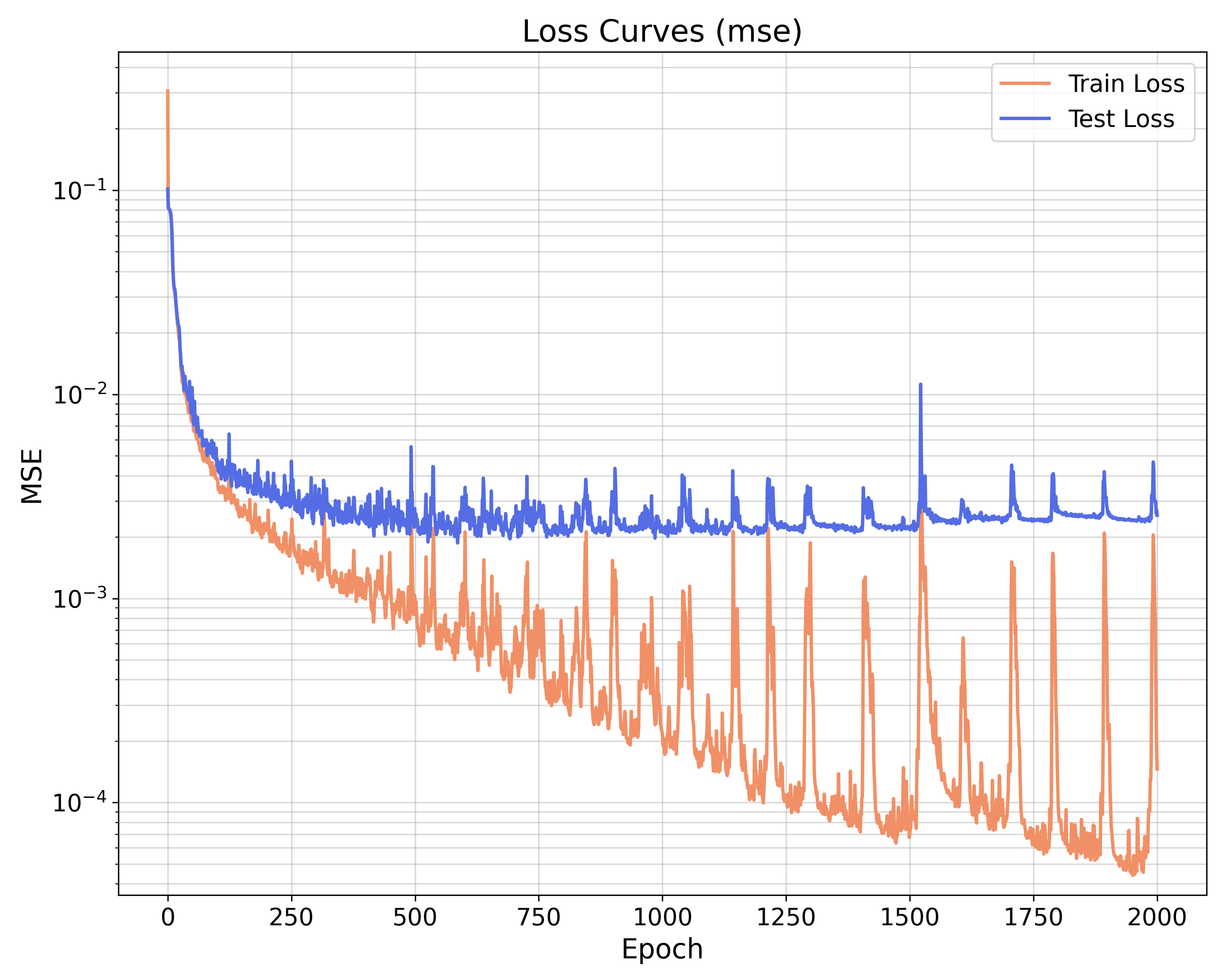}
        \caption{$N=400$}
    \end{subfigure}
    \hfill
    \begin{subfigure}[b]{0.32\textwidth}
        \centering
        \includegraphics[width=\textwidth]{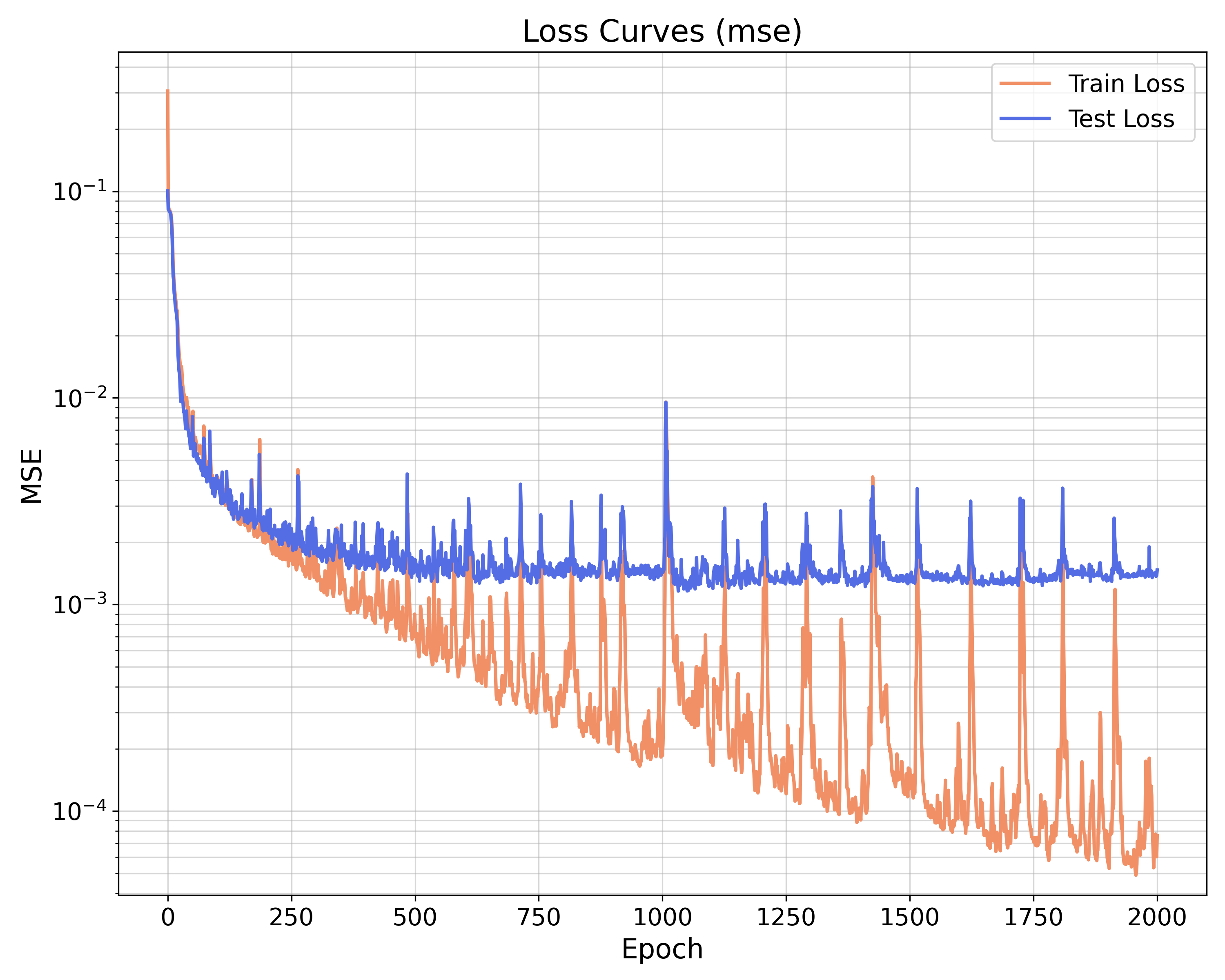}
        \caption{$N=500$}
    \end{subfigure}
    
    \vspace{0.4cm}
    
    \begin{subfigure}[b]{0.32\textwidth}
        \centering
        \includegraphics[width=\textwidth]{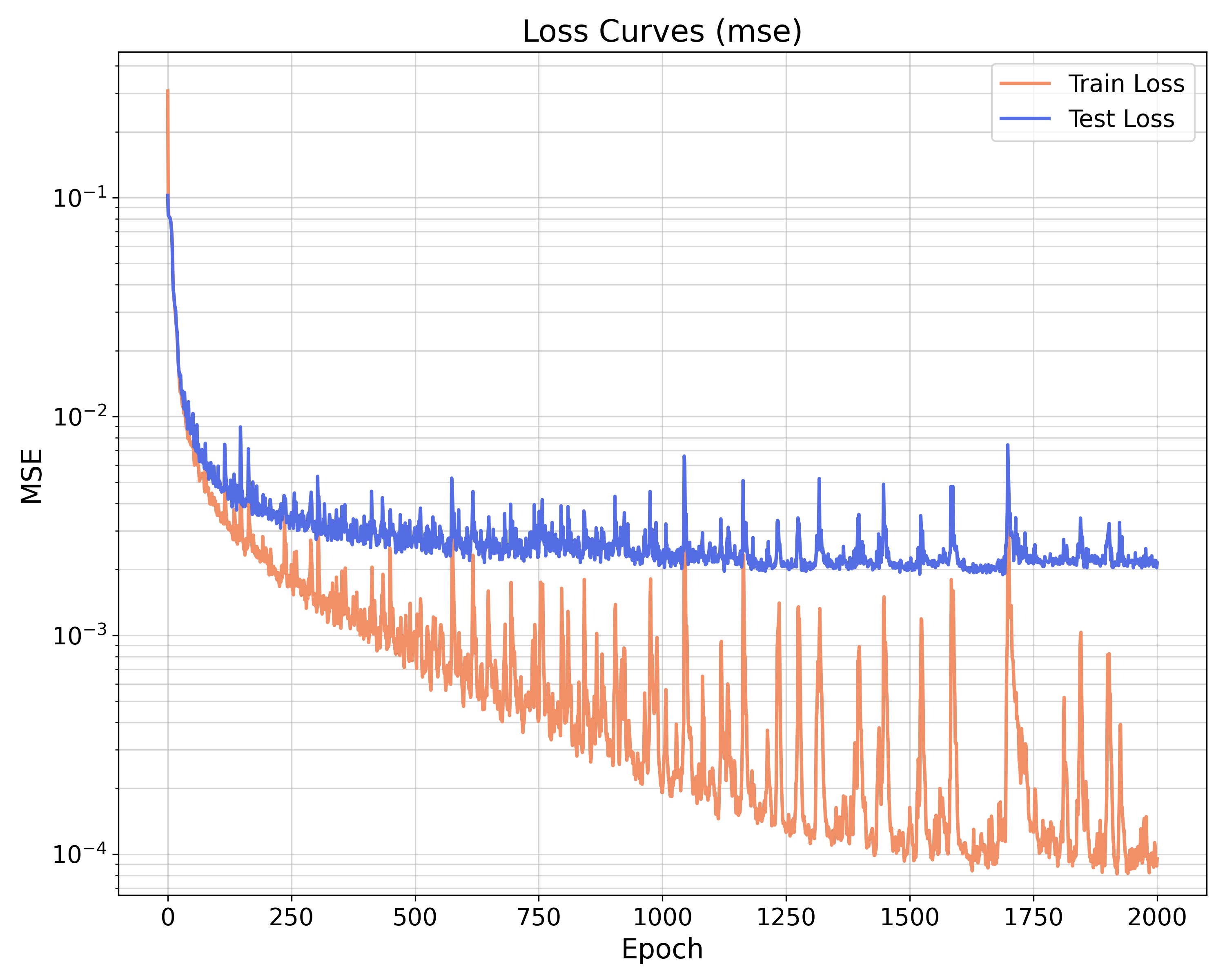}
        \caption{$N=600$}
    \end{subfigure}
    \hfill
    \begin{subfigure}[b]{0.32\textwidth}
        \centering
        \includegraphics[width=\textwidth]{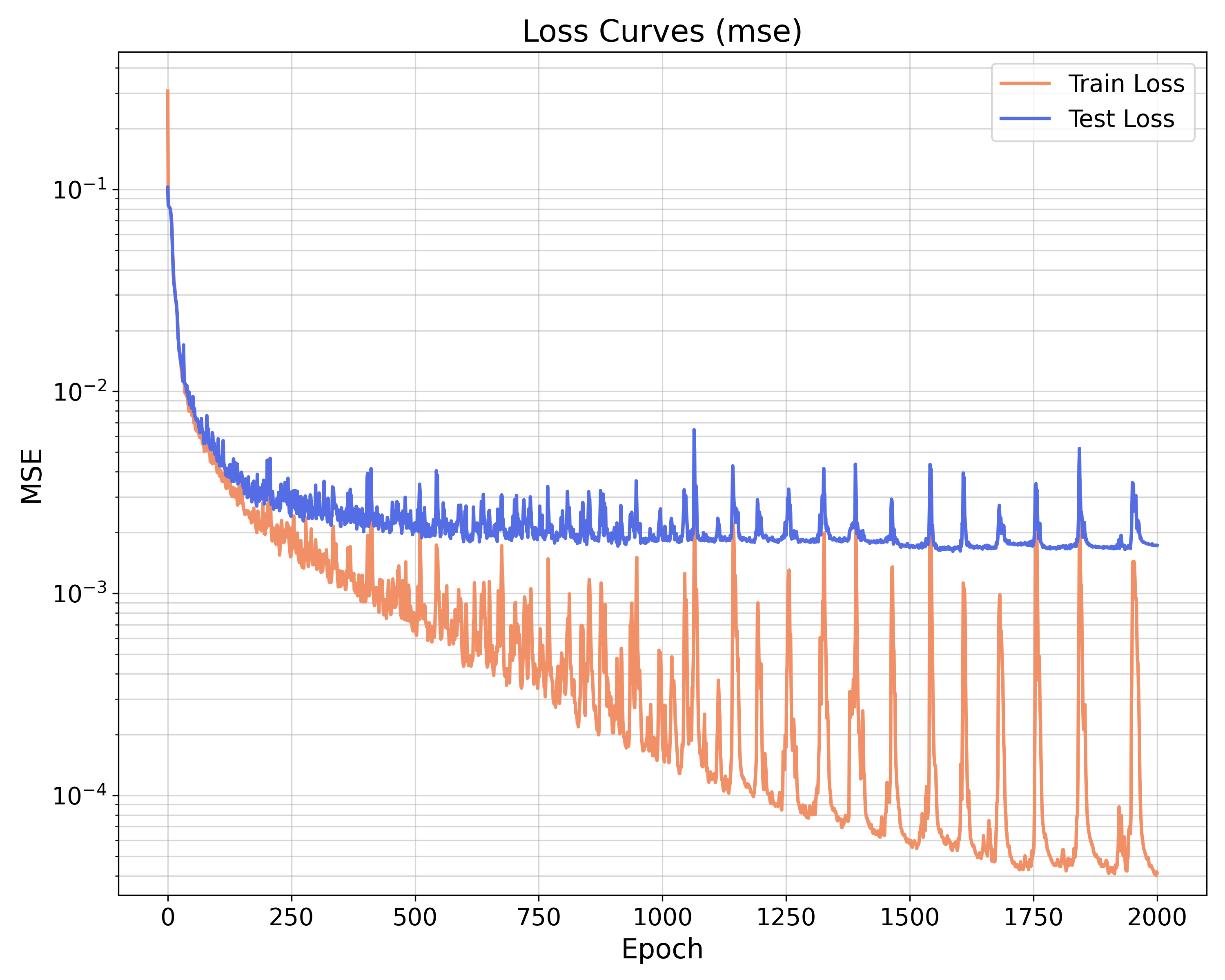}
        \caption{$N=700$}
    \end{subfigure}
    \hfill
    \begin{subfigure}[b]{0.32\textwidth}
        \centering
        \includegraphics[width=\textwidth]{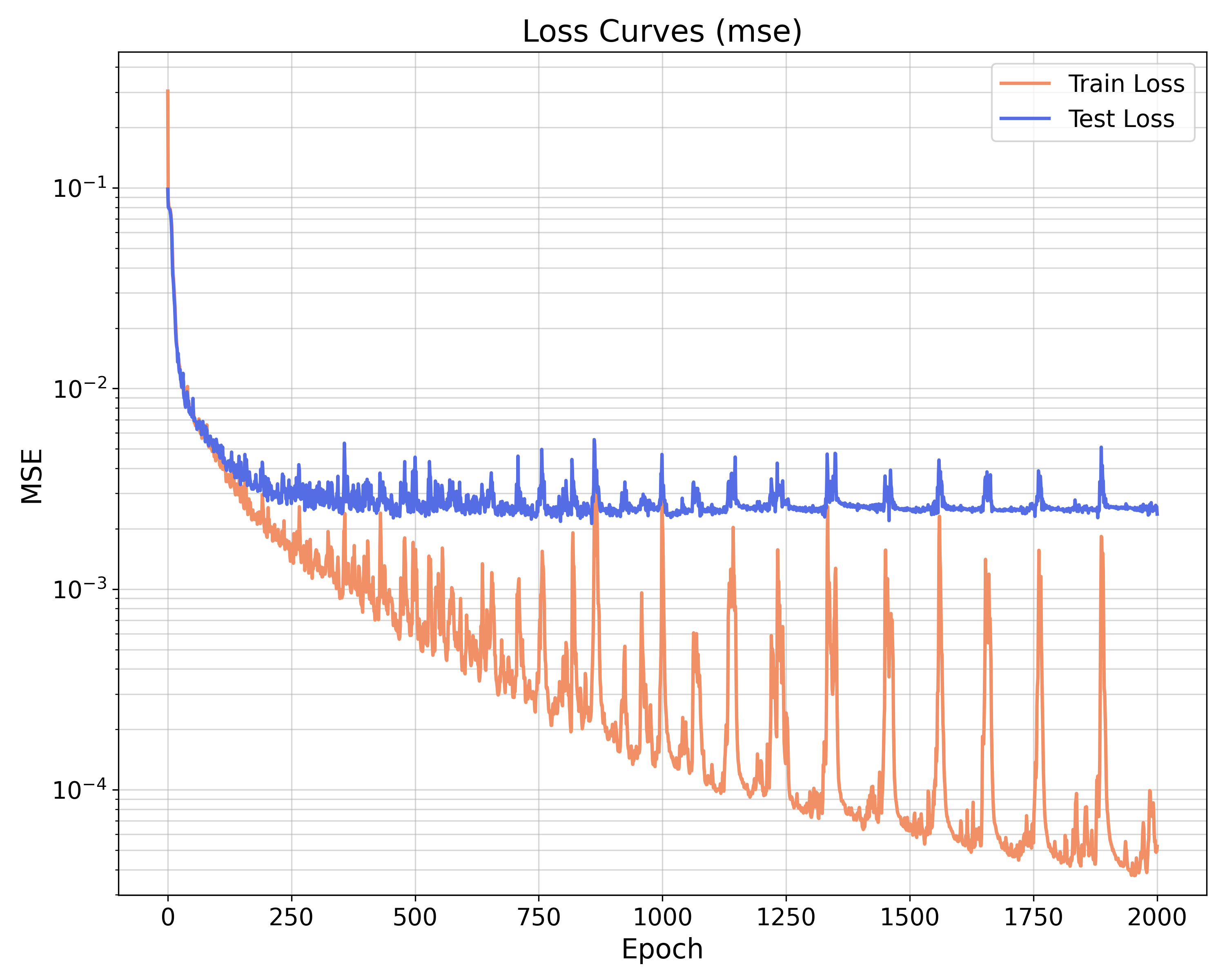}
        \caption{$N=800$}
    \end{subfigure}
    
    \vspace{0.4cm}
    
    \begin{subfigure}[b]{0.32\textwidth}
        \centering
        \includegraphics[width=\textwidth]{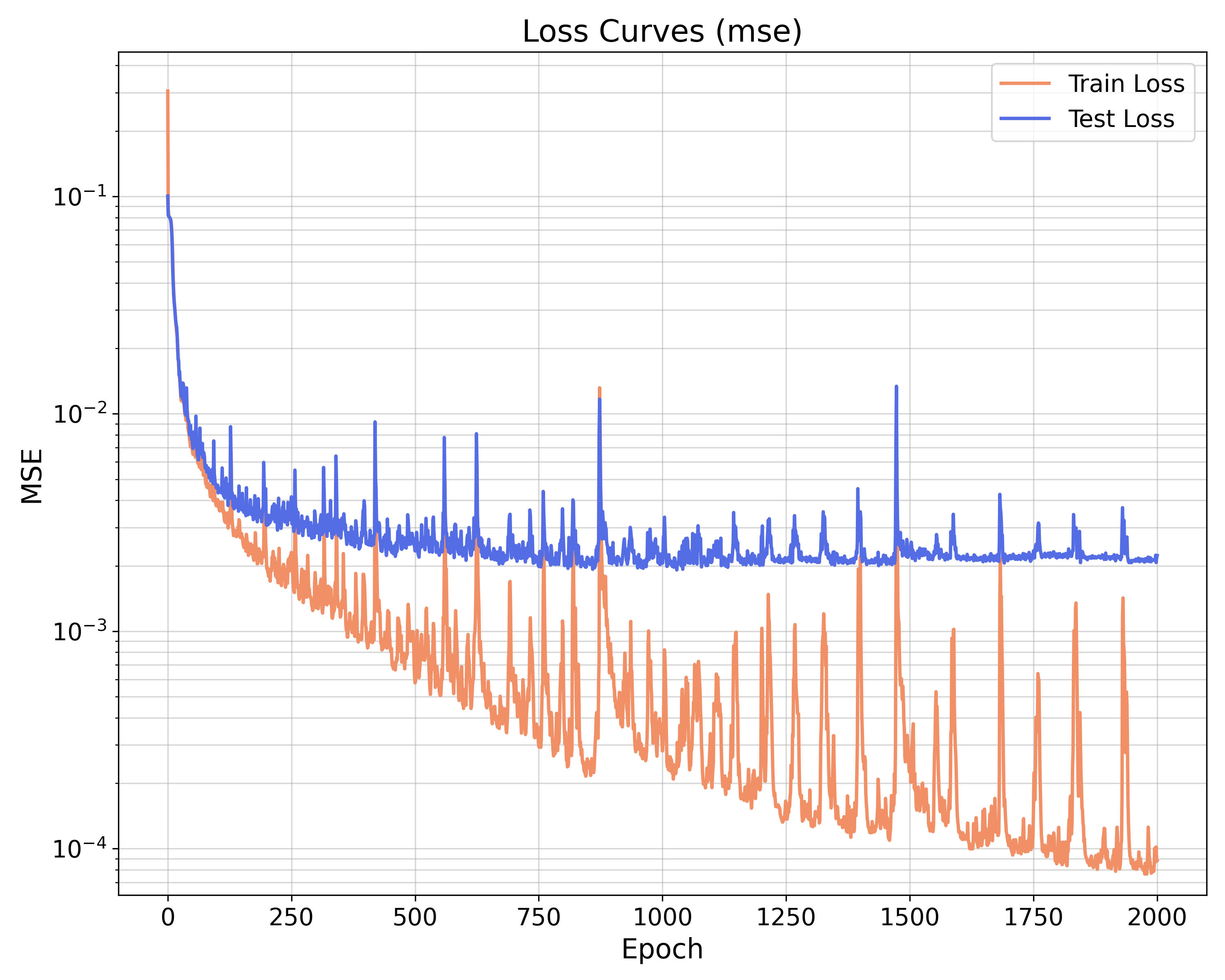}
        \caption{$N=900$}
    \end{subfigure}
    \hfill
    \begin{subfigure}[b]{0.32\textwidth}
        \centering
        \includegraphics[width=\textwidth]{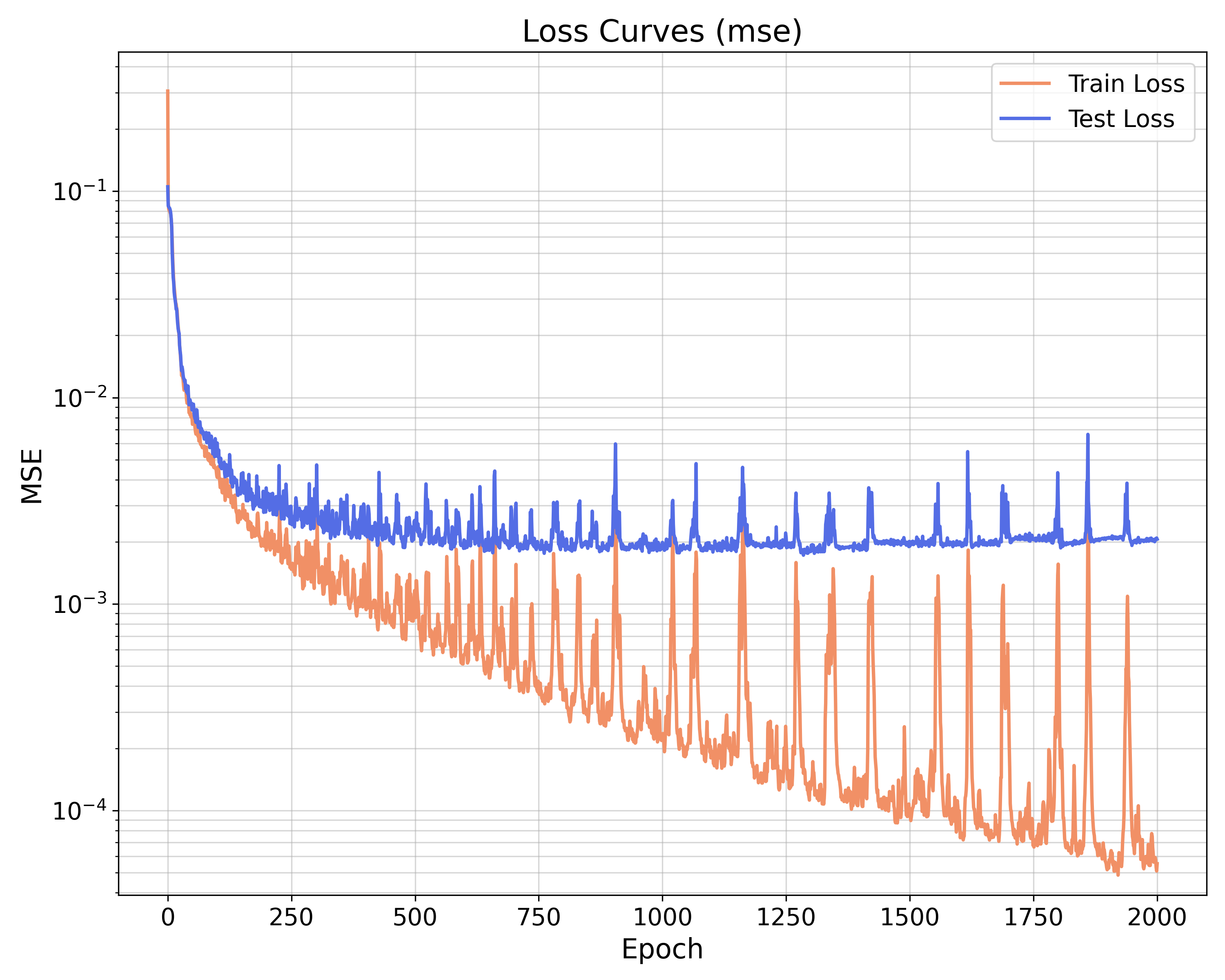}
        \caption{$N=1000$}
    \end{subfigure}
    \hfill
    \begin{subfigure}[b]{0.32\textwidth}
        \centering
        \includegraphics[width=\textwidth]{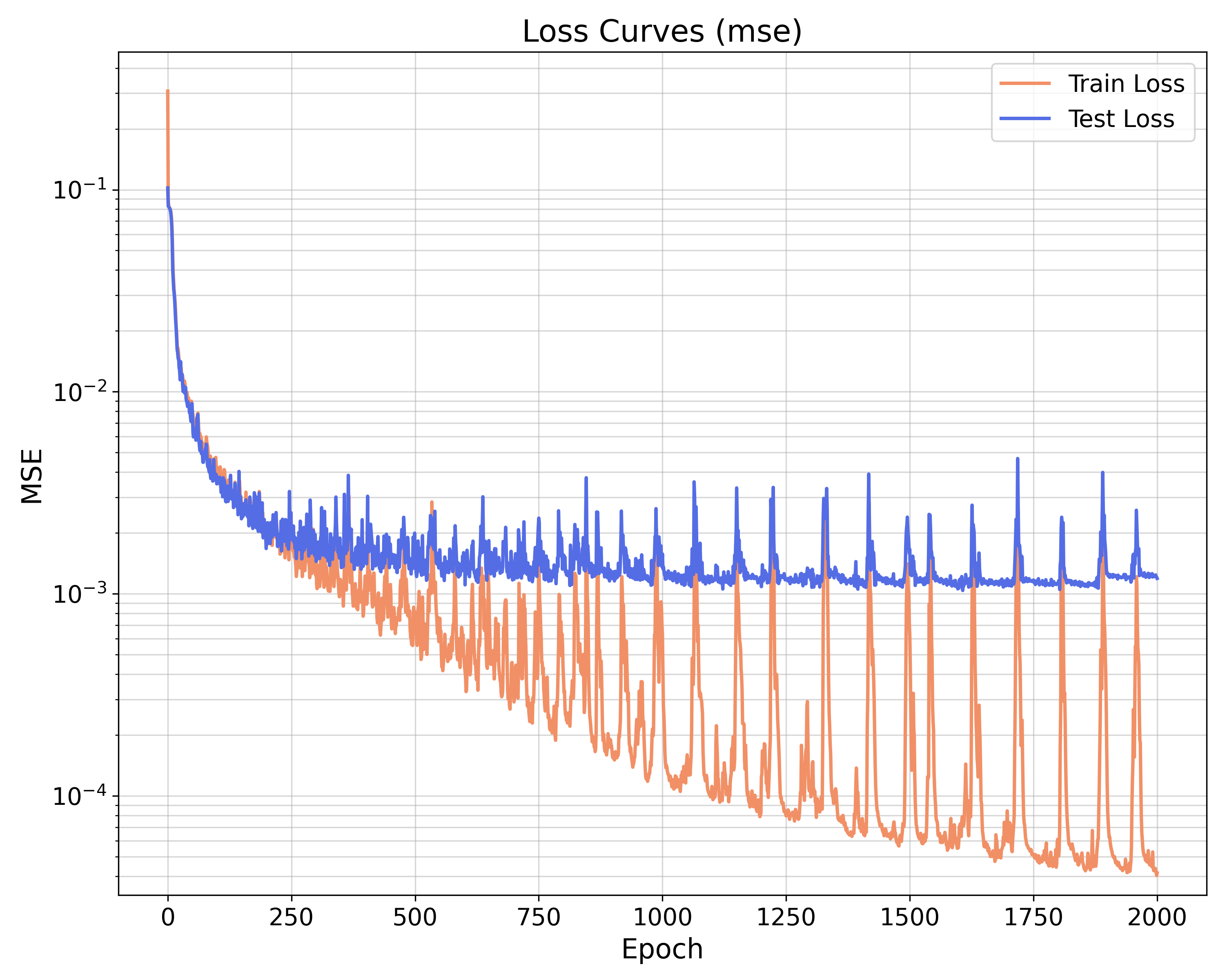}
        \caption{$N=2000$}
    \end{subfigure}
    
    \caption{Training and test loss convergence for different Barab\'asi-Albert network sizes ($N$).}
    \label{fig:ba_loss_grid}
\end{figure}


\clearpage
\subsection{Real Data}
We will here report additional tests against the real dataset introduced in Section \ref{Chp:5}. These include the usage of a bidirectional LSTM to further test the robustness of the predictions of the FNO. We here recall that a bidirectional LSTM is an extension of standard Long Short-Term Memory (LSTM) models that processes sequential data in both forward and backward directions simultaneously.

\begin{figure}[H]
    \centering
    \includegraphics[width=\textwidth, height=0.7\textheight, keepaspectratio]{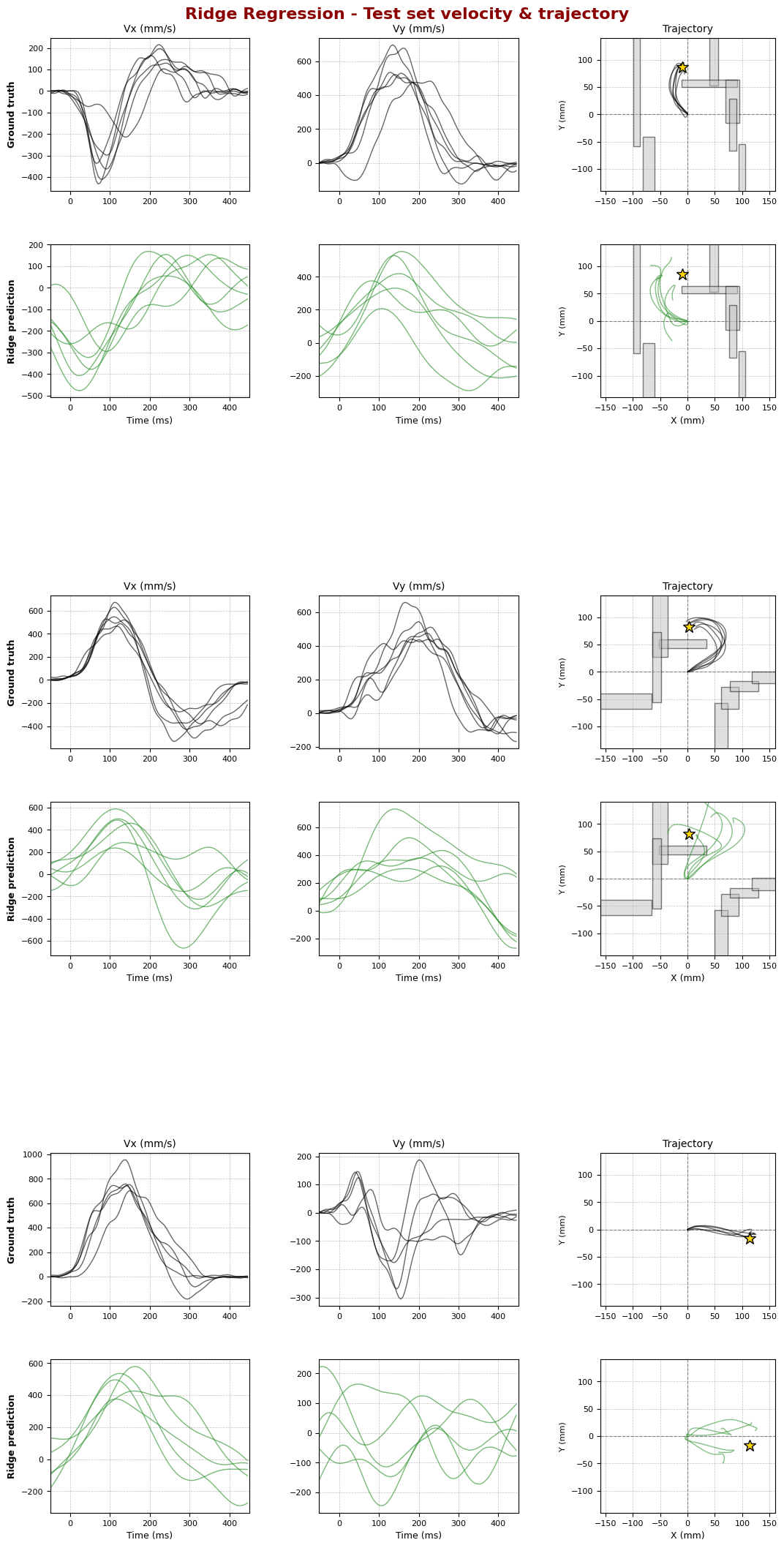}
    \caption{Ridge Regression: velocity components $V_x$, $V_y$ and reconstructed hand trajectories for the test trials of three maze conditions. In each block, the top row shows the recorded data (ground truth) and the bottom row the regression predictions; the trajectories are obtained by numerically integrating the velocities. Grey rectangles are the virtual barriers and the star marks the reached target.}
    \label{fig:qual_ridge}
\end{figure}

\begin{figure}[p]
    \centering
    \includegraphics[width=\textwidth, height=0.9\textheight, keepaspectratio]{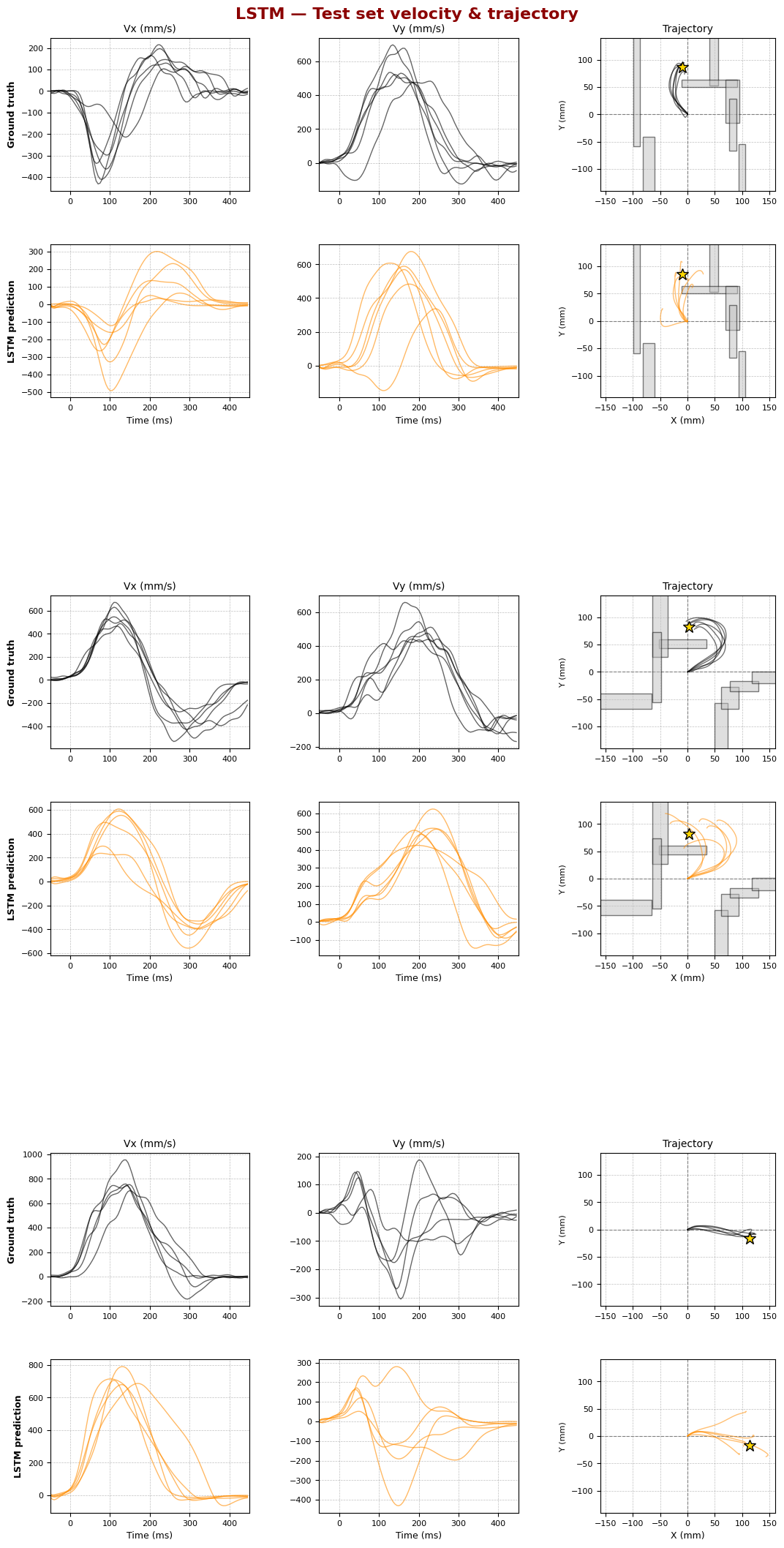}
    \caption{LSTM: velocity components $V_x$, $V_y$ and reconstructed hand trajectories for the test trials of three maze conditions. In each block, the top row shows the recorded data (ground truth) and the bottom row the LSTM predictions; the trajectories are obtained by numerically integrating the velocities. Grey rectangles are the virtual barriers and the star marks the reached target.}
    \label{fig:qual_lstm}
\end{figure}

\begin{figure}[p]
    \centering
    \includegraphics[width=\textwidth, height=0.9\textheight, keepaspectratio]{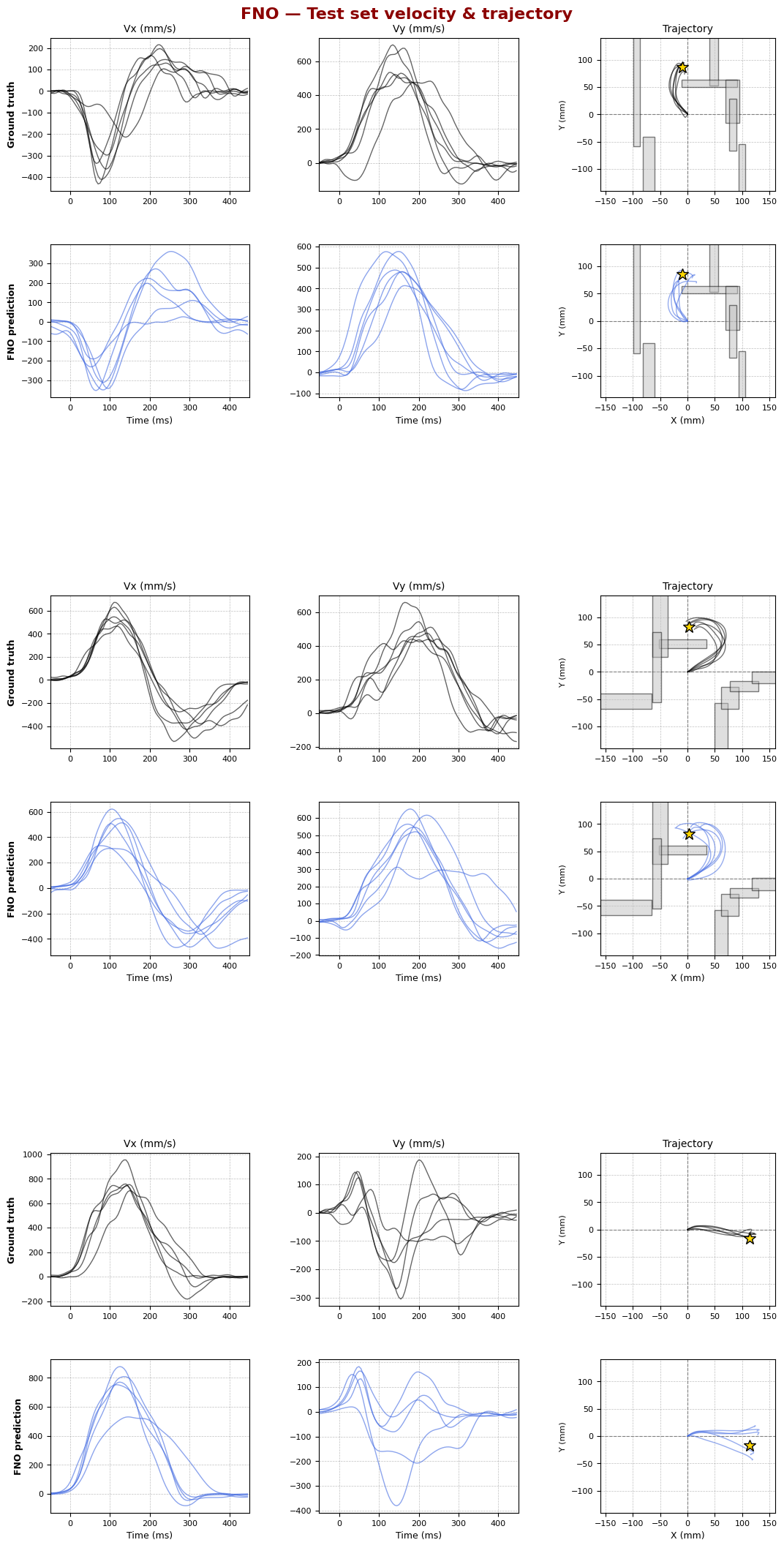}
    \caption{FNO: velocity components $V_x$, $V_y$ and reconstructed hand trajectories for the test trials of three maze conditions. In each block, the top row shows the recorded data (ground truth) and the bottom row the FNO predictions; the trajectories are obtained by numerically integrating the velocities. Grey rectangles are the virtual barriers and the star marks the reached target.}
    \label{fig:qual_fno}
\end{figure}

\begin{figure}[p]
    \centering
    \includegraphics[width=\textwidth, height=0.9\textheight, keepaspectratio]{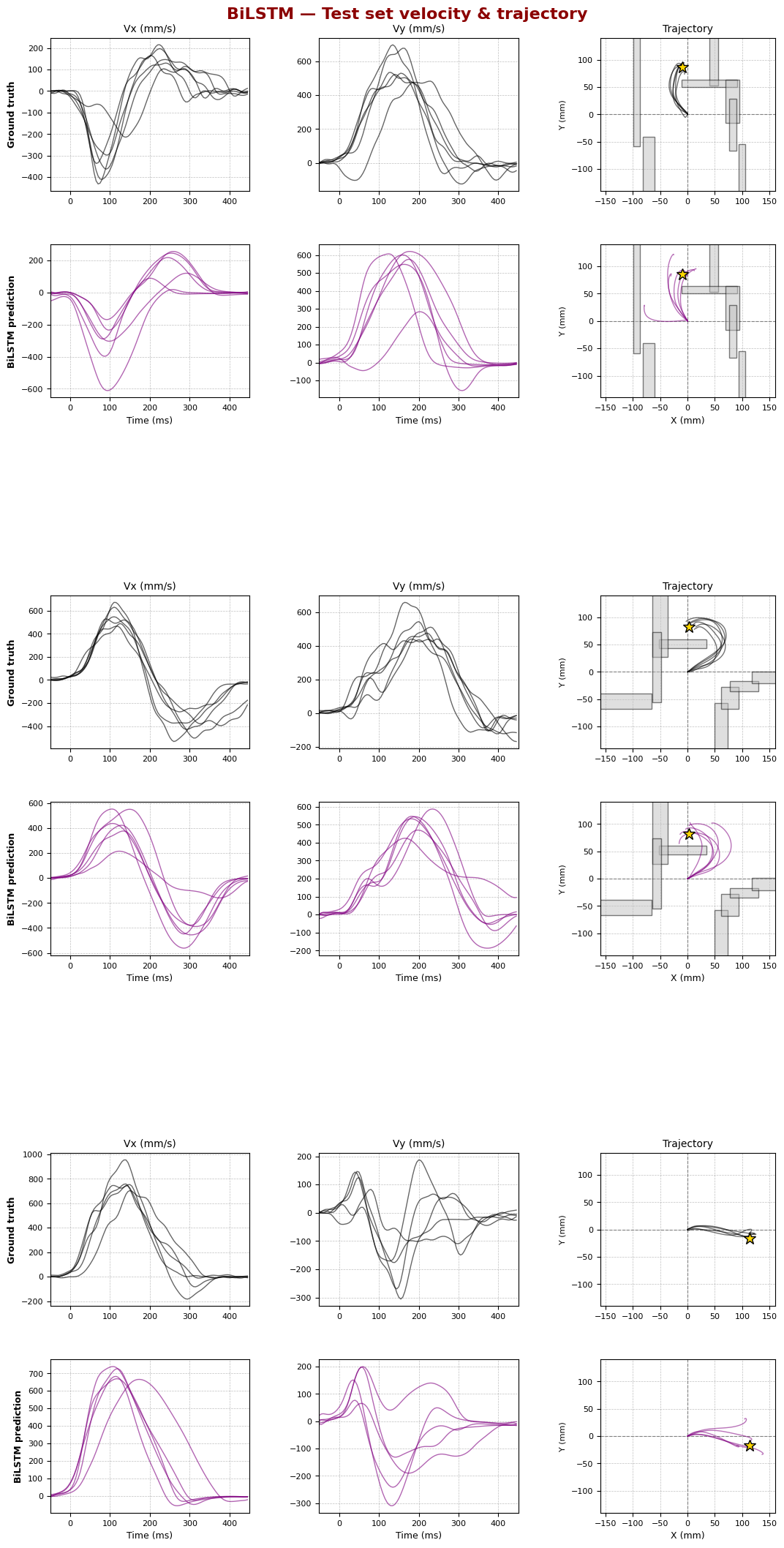}
    \caption{Bidirectional LSTM: velocity components $V_x$, $V_y$ and reconstructed hand trajectories for the test trials of three maze conditions. In each block, the top row shows the recorded data (ground truth) and the bottom row the BiLSTM predictions; the trajectories are obtained by numerically integrating the velocities. Grey rectangles are the virtual barriers and the star marks the reached target.}
    \label{fig:qual_bilstm}
\end{figure}

\begin{figure}[htbp]
    \centering
    \begin{subfigure}[b]{0.42\textwidth}
        \centering
        \includegraphics[width=\textwidth]{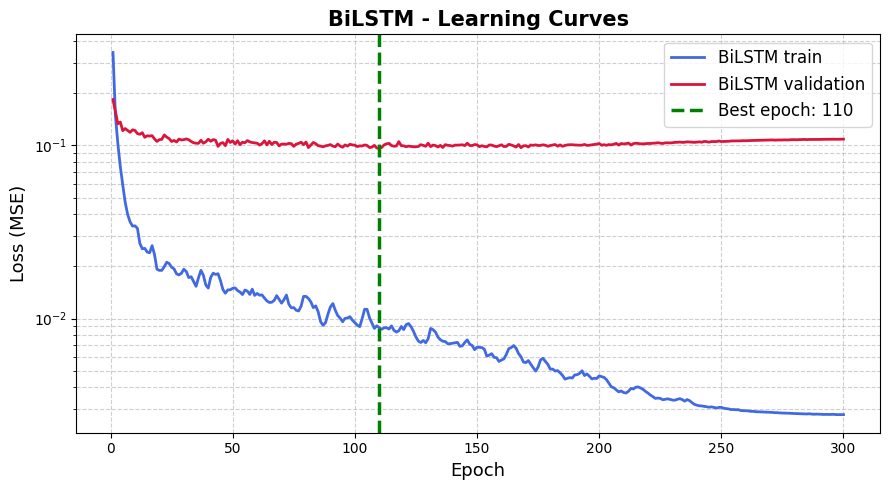}
        \caption{Training dynamics.}
        \label{fig:bilstm_loss}
    \end{subfigure}
    \hfill
    \begin{subfigure}[b]{0.56\textwidth}
        \centering
        \includegraphics[width=\textwidth]{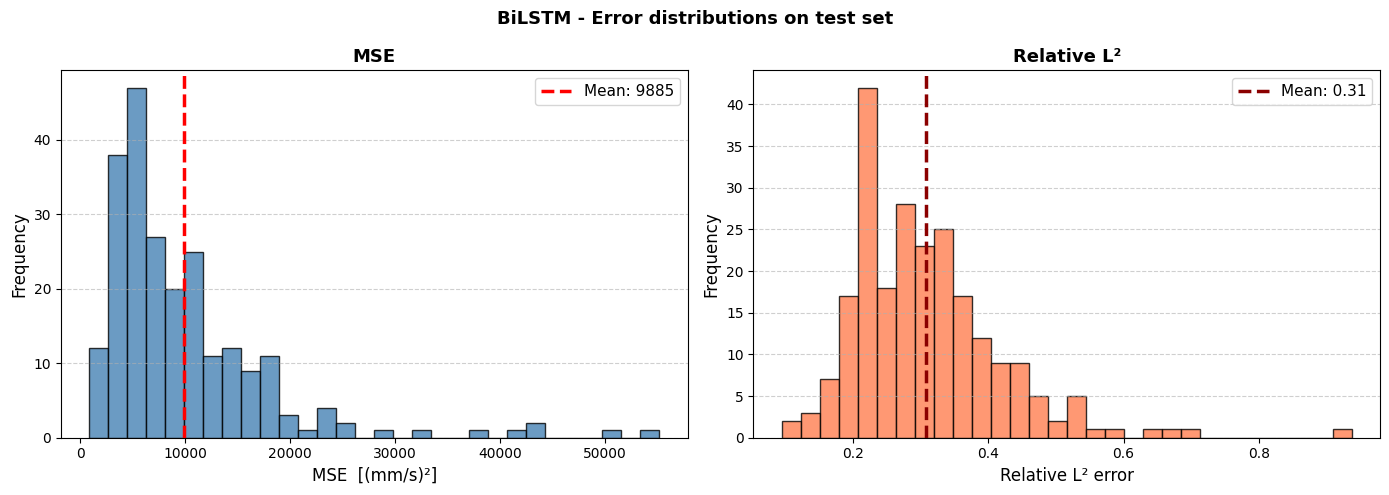}
        \caption{Error distributions on the test set.}
        \label{fig:bilstm_hist}
    \end{subfigure}
    \caption{Bidirectional LSTM. \textbf{(a)} Training and validation loss (MSE on standardized velocities) over 300 epochs; the dashed line marks the epoch with the lowest validation loss, whose weights were retained. \textbf{(b)} Histograms of the per-trial MSE (left) and relative $L_2$ error (right) on the 230 test trials; dashed lines indicate the means.}
    \label{fig:bilstm_training}
\end{figure}

\begin{figure}[p]
    \centering
    \includegraphics[width=\textwidth, height=0.9\textheight, keepaspectratio]{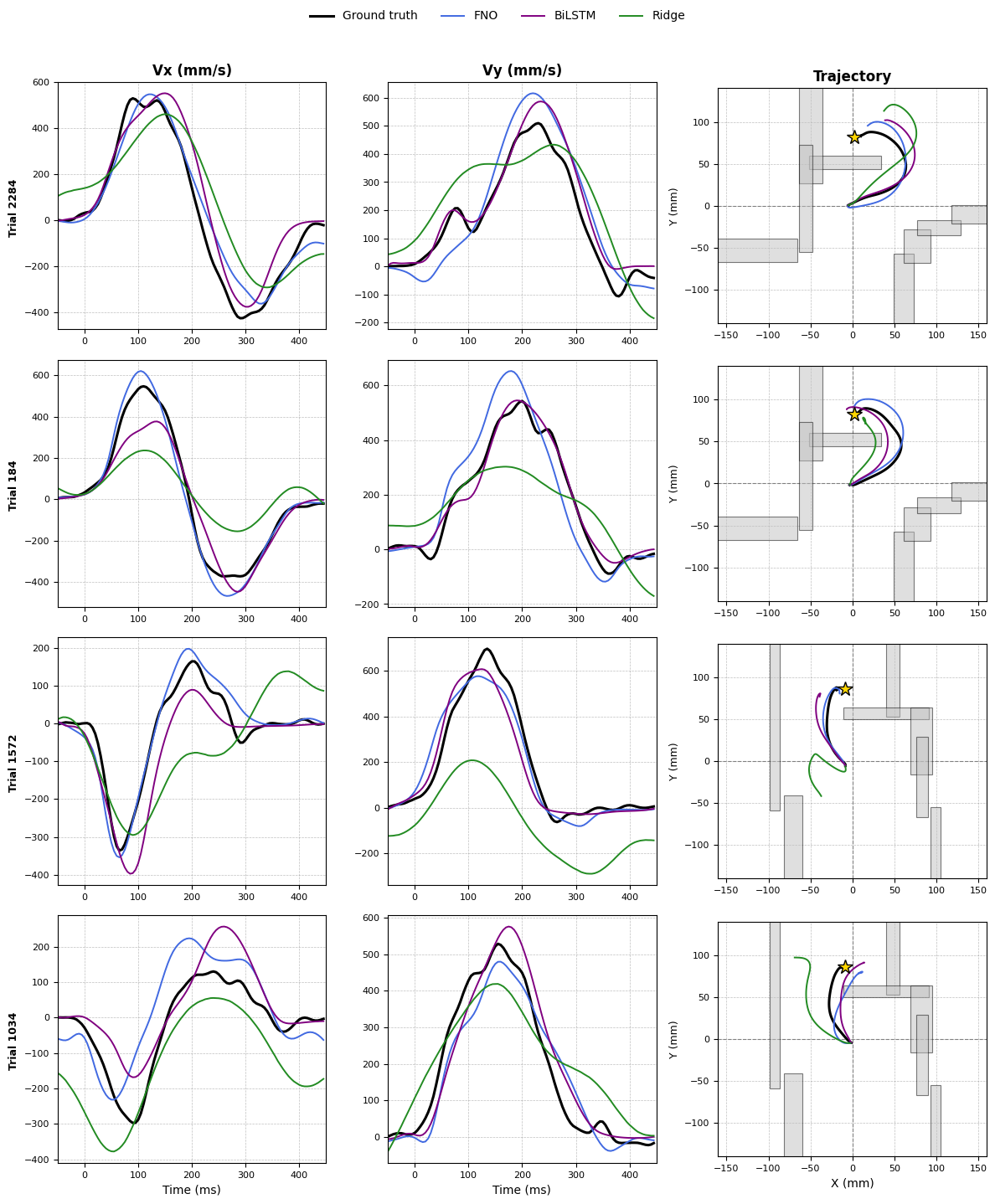}
    \caption{Comparison of the FNO, the bidirectional LSTM and Ridge Regression on four test trials. Each row is one trial; columns show $V_x$, $V_y$ and the reconstructed trajectory, with the ground truth in black. Grey rectangles are the virtual barriers, the star marks the reached target.}
    \label{fig:comparison_bilstm}
\end{figure}
\clearpage

\end{appendices}
\clearpage

\bibliographystyle{unsrt}
\bibliography{bibliography}

\end{document}